\documentclass[fleqn,usenatbib,useAMS]{mnras}

\usepackage[section]{placeins}
\usepackage{graphicx}	% Including figure files
\usepackage{grffile}
\usepackage{amsmath}	% Advanced maths commands
\usepackage{amssymb}	% Extra maths symbols
\usepackage{multicol}        % Multi-column entries in tables
\usepackage{bm}		% Bold maths symbols, including upright Greek
\usepackage{pdflscape}	% Landscape pages
\usepackage{nccmath}
\usepackage[section]{placeins}
\usepackage{caption}
\usepackage{multirow}
\usepackage{subcaption}
\defcitealias{osul18}{OS18}

\usepackage[T1]{fontenc}
\usepackage{ae,aecompl}

\usepackage{newtxtext,newtxmath}

\title[Polarization properties of radio galaxies]{A low frequency study of linear polarization in radio galaxies}

\author[V. H. Mahatma]{V. H. Mahatma$^{1,2}$\thanks{Contact e-mail: \href{mailto:vmahat@tls-tautenburg.de}{vmahat@tls-tautenburg.de}}, M. J. Hardcastle$^{1}$, J. Harwood$^{1}$, 
S. P. O'Sullivan$^{3}$,
G. Heald$^{4}$,\newauthor
C. Horellou$^{5}$
and D.J.B. Smith$^{1}$
\\
$^{1}$Centre for Astrophysics Research, School of Physics, Astronomy
\& Mathematics, University of Hertfordshire, College Lane, Hatfield
AL10 9AB, UK\\
$^{2}$ Thüringer Landessternwarte, Sternwarte 5, 07778 Tautenburg, Germany \\
$^{3}$ School of Physical Sciences and Centre for Astrophysics \& Relativity, Dublin City University, Glasnevin, D09 W6Y4, Ireland.	\\
$^{4}$CSIRO Astronomy and Space Science, PO Box 1130, Bentley, WA 6102, Australia\\
$^{5}$Chalmers University of Technology, Dept of Space, Earth and Environment, Onsala Space Observatory, SE-43992 Onsala, Sweden	
}

\date{Accepted 2020 December 22. Received 2020 December 11; in original form 2020 August 19}

\pubyear{2020}

\begin{document}
\label{firstpage}
\pagerange{\pageref{firstpage}--\pageref{lastpage}}
\maketitle

\begin{abstract}
Radio galaxies are linearly polarized -- an important property that allows us to infer the properties of the magnetic field of the source and its environment. However at low frequencies, Faraday rotation substantially depolarizes the emission, meaning that comparatively few polarized radio galaxies are known at low frequencies. Using the LOFAR Two Metre Sky Survey at 150 MHz and at 20 arcsec resolution, we select 342 radio galaxies brighter than 50 mJy and larger than 100 arcsec in angular size, of which 67 are polarized (18 per cent detection fraction). These are predominantly Fanaroff Riley type II (FR-II) sources. The detection fraction increases with total flux density, and exceeds 50 per cent for sources brighter than 1 Jy. We compare the sources in our sample detected by LOFAR to those also detected in NVSS at 1400 MHz, and find that our selection bias toward bright radio galaxies drives a tendency for sources depolarized between 1400 and 150 MHz to have flatter spectra over that frequency range than those that remain polarized at 150 MHz. By comparing observed rotation measures with an analytic model we find that we are preferentially sensitive to sources in low mass environments. We also infer that sources with one polarized hotspot are inclined by a small angle to the line of sight, while sources with hotspots in both lobes lie in the plane of the sky. We conclude that low frequency polarization in radio galaxies is related to a combination of environment, flux density and jet orientation.
\end{abstract}

\begin{keywords}
polarization -- galaxies: active -- galaxies: jets -- techniques: polarimetric -- radiation mechanisms: non-thermal
\end{keywords}

%%%%%%%%%%%%%%%%%%%%%%%%%%%%%%%%%%%%%%%%%%%%%%%%%%

%%%%%%%%%%%%%%%%% BODY OF PAPER %%%%%%%%%%%%%%%%%%

% The MNRAS class isn't designed to include a table of contents, but for this document one is useful.
% I therefore have to do some kludging to make it work without masses of blank space.

\section{Introduction}
\subsection{Polarized emission of radio-loud AGN}\label{sect:polarisation_intro}
The synchrotron radiation by which we observe the jets and lobes of radio-loud active galactic nuclei (RLAGN) arises from relativistic electrons gyrating in magnetic fields. As a consequence of this process, the radiation is intrinsically linearly polarized. RLAGN, which can have a maximum degree of polarization of up to $\sim70$ per cent \citep{pach70}, are strong sources of polarized radiation which can be observed by radio telescopes \citep[see reviews by][]{saik88,wiel12}. Information on the polarized intensity from RLAGN is important to obtain for the following reasons:
\begin{itemize}
    \item Since the position angle of the electric field vector of the radiation we observe is perpendicular to the projected magnetic field direction in the plane of the sky, polarization observations, if calibrated correctly, can directly give information on the structure of magnetic fields in the plane of the sky. This has led to studies of the magnetic field structure in the jets, lobes and hotspots of RLAGN, and their surrounding environment, on parsec scales \citep[e.g.][]{gabu92,homa05} to kiloparsec scales (e.g. \citealt{lain80,hard97,lain08,osul09,guid11} and see review by \citealt{brid84}). The hotspots of RLAGN, which are thought to contain compressed and ordered magnetic fields \citep{lain80,hugh89}, are expected to be prime locations for polarized emission, so observations of polarization may enhance our understanding of particle acceleration processes.
    \item A lack of detectable polarization for high surface brightness objects gives evidence for substantial \textit{depolarization} -- a combination of factors such as the finite telescope beam, inhomogeneous magnetic field structures in the lobes or in their surrounding environment and Faraday rotation will reduce the observed polarization (as described below). Measurements of depolarization in the lobes can, in principle, trace their magnetoionic properties and their thermal particle content, or that of their environment \citep{dreh87,osul17,osul19,knue19}.
\end{itemize}
Effects caused by Faraday rotation results in frequency-dependent depolarization: as linearly polarized emission travels through birefringent magnetised media (the intergalactic medium, for example), a difference in the phase velocity occurs for the right and left circular polarization constituents of the linear polarization. This manifests as a wavelength-dependent rotation of the polarization angle as
\begin{ceqn} 
\begin{equation}\label{equation:polangle}
\chi = \chi_0 + RM\lambda^2 \text{\hspace{5mm} [rad]}
\end{equation}
\end{ceqn}
where $\chi$ is the observed polarization angle (in radians), $\chi_0$ is the intrinsic polarization angle, $RM$ is the rotation measure (in rad m$^{-2}$) and $\lambda$ is the wavelength (in m). The $RM$ is related to the properties of the line-of-sight magnetised media by 
\begin{ceqn}
\begin{align}\label{eq:rm}
RM = 0.81\int^{\text{telescope}}_{\text{source}} n_e \vec{\bm{B}}_\parallel \cdot \text{ \textbf{d}}\vec{\bm{l}} \text{\hspace{2mm} [rad m$^{-2}$]}
\end{align}
\end{ceqn}
where $n_e$ is the electron density (in cm$^{-3}$), $\vec{\bm{B}_{\parallel}}$ is the line of sight magnetic field strength (in $\mu$Gauss) and the integral is taken with respect to the path lengths $\text{ \textbf{d}}\vec{\bm{l}}$ (in parsecs) through all intervening material between the source and the telescope. Differential Faraday rotation, and/or inhomogeneous magnetic field structures in the source, lead to different polarization angles across the telescope beam, which are then vector-averaged and lead to depolarization. Further depolarization can occur when these effects apply significantly within the observing bandwidth, since Faraday rotation is frequency-dependent.  

Depolarization\footnote{It should be noted that, except in the case of differential Faraday rotation across the band, the magnitude of the $RM$ and depolarization of a source are not strictly related, rather the latter is associated with the dispersion in $RM$ across the telescope beam or along the line of sight.} of linearly polarized emission from RLAGN confirms the presence of magnetic fields in thermal plasma along lines of sight through their environments. Significant depolarization is generally attributed to environments local to the source that cause large $RM$s (e.g. in the interstellar or intracluster medium; \citealt{hard03} and \citealt{cari02}, respectively, or in the shocked gas surrounding radio lobes; \citealt{hard12}), and hence $RM$s are useful in inferring properties of the environment which are otherwise difficult to obtain. For RLAGN environments well described by hot ($T \sim 10^{7}$ K) plasma that radiates in X-rays due to thermal bremsstrahlung, sensitive X-ray maps allow a measure of the gas density $n_e$ \citep[e.g.][]{cros08,hick13,maha20}, but may be expensive to obtain for large samples of radio galaxies. $RM$ maps can give valuable (but indirect) information on the surrounding environment of RLAGN, while giving constraints on the structure of magnetic fields. 

In general, the interpretation of observed $RM$s is difficult, as they are in general a superposition of Faraday rotation from; the Earth ionosphere, the magnetized plasma in the Galaxy, the intergalactic medium, the intracluster/intragroup medium and within the source itself. Radio lobes in particular carry entrained thermal gas from their surroundings \citep{bick84}, and source-intrinsic Faraday rotation can also be a significant contributor to observed $RM$s when the Galactic contributions have been subtracted. Precise $RM$ information can help to disentangle effects from different line of sight contributions if their respective $RM$s are found. In order to accurately quantify the $RM$ and polarization properties of RLAGN in general, large-sample statistics are needed. 
\subsection{Low frequency polarization}
Low frequency ($\sim 100$ MHz) linearly polarized source detections, particularly in a statistical study, are scarce. Due to the $\lambda^2$ factor in Equation \ref{equation:polangle}, Faraday rotation, and by extension depolarization, is much more important at low frequencies. RLAGN samples with polarization information exist in surveys at 1.4 GHz or greater \citep[e.g.][]{tayl07,hale14,rudn14}, where depolarization is less significant in general, although this is also due to the fact that there are many more completed large-area radio surveys at GHz frequencies. \cite{tayl09} produced an $RM$ map of the sky using the NRAO VLA Sky Survey \citep[NVSS;][]{cond98}, a 1400 MHz survey, detecting 37,543 polarized radio sources at declinations $>-40^{\circ}$. However at lower frequencies the total flux density for any steep-spectrum RLAGN ($\alpha\leqslant-0.7$ where $S\propto\nu^{\alpha}$) is much higher\footnote{Down until the self-absorption regime which affects the radio continuum at frequencies lower than $\sim$100 MHz \citep[e.g.][]{sche68}}, which may give adequate polarized signal to noise for low surface brightness regions that are undetected at higher frequencies. Moreover, and more crucially, since the $RM$ precision depends on the interval in $\lambda^2$, low frequency instruments can out-perform centimetre-wave instruments by a few orders of magnitude in $RM$ precision. Larger source counts at these frequencies will test the robustness of the previously mentioned polarization studies, after understanding the detection statistics and possible selection biases of large samples at low frequencies. In sampling this new low-frequency parameter space, it is crucial to have radio telescopes with the ability to perform wide-area surveys of the sky combined with the required sensitivity to observe large samples of RLAGN -- past surveys such as 3CRR \citep{3crr} are severely biased towards the most luminous sources such as Fanaroff-Riley type-II objects \citep[FR-II;][]{FR74}. Recently, \cite{rise20} presented the POlarised GaLactic and Extragalactic All-Sky MWA Survey-the POlarised GLEAM Survey (POGS) in the frequency range 169-231 MHz at a resolution of 3 arcmin (at the highest), with 484 polarized RLAGN detected in the entire southern sky. However, instruments with the capability to perform sub-arcmin resolution surveys are more ideal in resolving the different components of RLAGN (core and lobes) and also mitigate the effects of beam depolarization in small angular size sources.

The LOw Frequency ARray \citep[LOFAR;][]{vanh13} is one such instrument, giving an angular resolution of 6 arcsec at 150 MHz. LOFAR is able to obtain a Faraday depth resolution (ability to resolve structures in Faraday depth space, where Faraday depth is the more generalised form of $RM$ in Equation \ref{equation:polangle}) of $\sim1$ rad m$^{-2}$ at 150 MHz, significantly better than that obtained by higher-frequency instruments (a factor of 200 better than the upcoming VLA Sky Survey, VLASS; \citealt{vlass}). Additionally, LOFAR's mixture of long and short baselines and its sensitivity to large extended structures (such as the lobes of nearby FR-I and FR-II radio galaxies) enables straightforward selection of RLAGN. The LOFAR Two-Metre Sky Survey \citep[LoTSS;][]{shim19}, an on-going  survey of the northern hemisphere enables large samples of RLAGN to be obtained for  polarization studies. The first data release (DR1; \citealt{shim19}) covered the area of the Hobby-Eberly Telescope Dark Energy eXperiment (HETDEX: \citealt{hetdex}) Spring field; over 420 square degrees on the sky within $161^{\circ}<$ RA $<231^{\circ}$ and $45.5^{\circ}<$ DEC $<57^{\circ}$, observed at 6 arcsec resolution with a median sensitivity of $\sim 70  \mu$Jy beam$^{-1}$. With a large low-frequency sample of polarized RLAGN, in combination with measurements at higher frequencies, we may start to answer questions about the main driver of observed wavelength-dependent depolarization. Moreover, we may test whether polarized emission is seen as a result of the physical effects of the `Faraday screen' (i.e. an external magnetoionic medium), or whether different AGN properties drive different levels of polarized emission for a population of sources. A statistical study of the observational nature of polarized emission from RLAGN will also be a vital prerequisite for upcoming radio surveys (Square Kilometre Array, VLASS), for which broad-band radio polarimetry is a scientific goal \citep{heal20}.

The polarization data in LoTSS have already been analysed by \cite{vane18}, \cite{vane19} and \cite{osul18} (hereafter \citetalias{osul18}), using the $RM$ synthesis technique (see Section \ref{sect:detection}). In the former two studies, the authors searched for polarized point sources and diffuse sources, respectively, within the HETDEX region at an angular resolution of 4.3 arcmin, with the study of \cite{vane18} producing a catalogue of 92 polarized point sources. \citetalias{osul18} studied the sources in this catalogue in the LoTSS DR1 area ($\sim$80 per cent of which have optical identifications) at a higher resolution of 20 arcsec. These sources have radio luminosities consistent with being RLAGN, and while the sample includes a mixture of extended radio galaxies and blazars, the majority of detections came from the hotspots of large FR-II radio galaxies. \cite{stua20} presented a polarization study of giant radio galaxies, which are $>0.7$ Mpc in size, in order to infer the physical properties of the intergalactic medium. Their study selects polarized sources in the LoTSS survey and cross-matches their sources with the giant radio galaxy catalogue of \cite{dabh19}. However, a low frequency study that determines the statistical polarization properties of radio galaxies as a population is required. Such a study requires an independent selection of radio galaxies without physical selection effects before searching for polarized emission. 

In this paper, we utilise data from LoTSS DR2 to select a flux-complete sample of extended radio galaxies, forming a parent sample to search for polarized emission at 150 MHz. We use $RM$ synthesis \citep{bren05} to produce polarization and $RM$ maps of all sources and compare the bulk observational and physical properties between the detected and non-detected sources, with the aim of inferring the primary driver of observed polarized emission in radio galaxies. In Section \ref{sect:observations} we describe the selection of our parent sample and our polarized detection criteria. In Section \ref{sect:analysis} we present our analysis and results on detectability, host galaxies, observed and predicted $RM$s using an analytic model. We summarise our results and conclude in Section \ref{sect:conclusions}.

Throughout this paper we define the spectral index $\alpha$ in the sense $S\propto\nu^{\alpha}$. We use a $\Lambda$CDM cosmology in which $H_0 = 71$ kms$^{-1}$Mpc$^{-1}$, $\Omega_m$ = 0.27 and $\Omega_{\Lambda}$ = 0.73.
\section{Observations}
\label{sect:observations}
\subsection{RLAGN sample}
LoTSS DR2 (scheduled for public release in early 2021) will have a northern sky coverage of over 5700 deg$^2$, including the DR1 area which covered 424 deg$^2$. While DR1 does not contain polarization information, DR2 contains Stokes QU cubes (at 20 arcsec angular resolution) as data products, enabling polarization information to be extracted in this sky area. However, DR2 does not contain optical IDs for radio sources at the time of writing. For the purposes of our study we required a sample of radio galaxies with physical information such as radio luminosities. Hence, we use the DR2 polarization products to find polarized radio galaxies in the DR1 catalogue, which is publicly available and includes a value added catalogue with optical identifications \citep{shim19,dunc19,will19}. This is particularly important since a radio galaxy catalogue from the DR1 sources has been made \citep{hard19}, which we use to select sources for this study. 

The DR1 radio galaxy selection details are given by \cite{hard19}, but we briefly describe them here. Sources were selected as having an optical ID from either Pan-STARRs \citep{panstarrs} or the Wide Infrared Survey Explorer \citep[WISE;][]{wise}, and either a spectroscopic redshift or a photometric redshift with a fractional error < 10\%. From this sample of 71,955 sources, star-forming galaxies (SFG) were identified using the MPA-JHU catalogue\footnote{\url{https://www.sdss.org/dr15/data_access/value-added-catalogs/?vac_id=mpa-jhu-stellar-masses}} and were removed. Objects were further removed if their WISE colours were consistent with those of SFG colours unless either; they are classed as AGN in the MPA-JHU catalogue, their total radio luminosity $L_{150}>10^{25}$ W Hz$^{-1}$ and their host galaxy $K_s$-band absolute magnitude $M_{K_s}$>-25, or $M_{K_s}$>-25 and $\log_{10}(L_{150})>25.3-0.06(25+K_s)$, resulting in a sample of 23,344 RLAGN. Given the nature of the selection criteria applied, it is likely that some RLAGN have been missed from the survey, particularly if their hosts are strongly star-forming galaxies, unless the radio luminosity $L_{150}>10^{25}$ W Hz$^{-1}$ (which selects radio-loud quasars). For the purposes of our study we require extended and bright double-lobed radio galaxies and so this  sample adequately describes the population of radio galaxies detected in DR1. 

In order to create a sample with a polarization detection fraction high enough for a statistical study, we selected sources that are both bright and large -- this also removes compact objects such as blazars that are not of interest to this study. From the RLAGN catalogue of \cite{hard19}, we selected sources with total flux density $S_{144}\geqslant 50$ mJy (as are all sources detected in polarization in the study by \citealt{stua20}) and with angular size $L \geqslant 100$ arcsec\footnote{Various other values for these criteria were tested, with the result that lower cut-off values resulted in a large number of undetected and unresolved sources in polarization for the purposes of this study.}. These criteria resulted in a total of 382 sources in the DR1 area of 424 deg$^2$, from which we study the bulk polarization properties in the rest of the paper. 
\subsection{Polarized emission detection}
\label{sect:detection}
To produce polarization and $RM$ maps of our sample, we utilised the $RM$ synthesis technique \citep{bren05}, using \textsc{pyrmsynth}\footnote{\url{https://github.com/sabourke/pyrmsynth_lite}}, a Python script developed primarily for LOFAR Stokes Q and U cubes. The complex polarization ($P=Q+iU$) can be written as 
\begin{ceqn}
\begin{equation} \label{equation:polvector}
    P(\lambda^2) = \int^{+\infty}_{-\infty}F\left(\phi\right)e^{2i\phi\lambda^2}d\phi
\end{equation}
\end{ceqn}
\citep{burn66}, where $P(\lambda^2)$ is the polarized intensity as a function of wavelength ($\lambda$) squared and $F\left(\phi\right)$ is the Faraday spectrum (polarized intensity as a function of the Faraday depth $\phi$). $RM$ synthesis transforms the cubes from frequency space into Faraday depth space by inverting Equation \ref{equation:polvector} so that (an approximated reconstruction of) the Faraday spectrum is calculated. The reconstructed Faraday spectrum ($\bar{F}[\phi]$) is then used to measure the peak polarized signal for each pixel in the image. The value of $\phi$ at which a peak in the Faraday spectrum is found is taken as the $RM$ of each pixel\footnote{Note that this value for the $RM$ applies in the case of a delta function for the Faraday spectrum.}. This technique has already been extensively applied to LOFAR data \citep{vane18,vane19,osul18,osul19,stua20,osul20}. We extracted the (unCLEANed) Stokes Q and U cubes of all sources in our sample, which were spatially masked with a 3$\sigma$ cut-off based on the Stokes I image of the source at 20 arcsec resolution. We inputted the QU cubes for each source into \textsc{pyrmsynth}, using the \textsc{rmclean} tool \citep{heal09} to deconvolve the Faraday spectrum using a maximum of 1000 iterations, fitting a Gaussian to the peak of the reconstructed CLEANed Faraday spectrum, resulting in linearly polarized intensity and $RM$ maps. Note that this procedure implicitly assumes only one peak in the Faraday spectrum of each source, but we verified that multiple and equally strong peaks were not present (except in the case of leakage -- see below) by inspecting the spectra. We limited the Faraday depth range to -150 $\leqslant \phi$ (rad m$^{-2}$) $\leqslant$ 150 to search for polarized emission, with increments of $\delta\phi =0.3$ rad m$^{-2}$. Though the Faraday depth magnitude can be up to 450 rad m$^{-2}$ for LOFAR, with initial analysis we found that we do not detect peaks in the Faraday spectra outside the range stated above. With this spectral setup of RM synthesis we are sensitive to scales $\leqslant$ 1 rad m$^{-2}$ in Faraday space. As no corrections were made for leakage signal, which can dominate near Faraday depths close to zero, we further exclude the range $-3\leqslant\phi\leqslant1.5$ rad m$^{-2}$ from the fitting of the peak in the Faraday spectrum. Hence, we are only Faraday depth-complete outside this range. The linear polarization and $RM$ maps of six sources in our sample are shown in Figure \ref{fig:pol_fr1} and Figure \ref{fig:pol_fr2}. It should be noted that the typical Galactic $RM$ in the HETDEX region is $0\leqslant RM_{\text{Galactic}}$ (rad m$^{-2}$) $\leqslant +23$ \citep{oppe15}.
\begin{figure*} % "[t!]" placement specifier just for this example
\captionsetup{type=figure}\addtocounter{figure}{-1}
\begin{subfigure}{0.46\textwidth}
\includegraphics[scale=0.32]{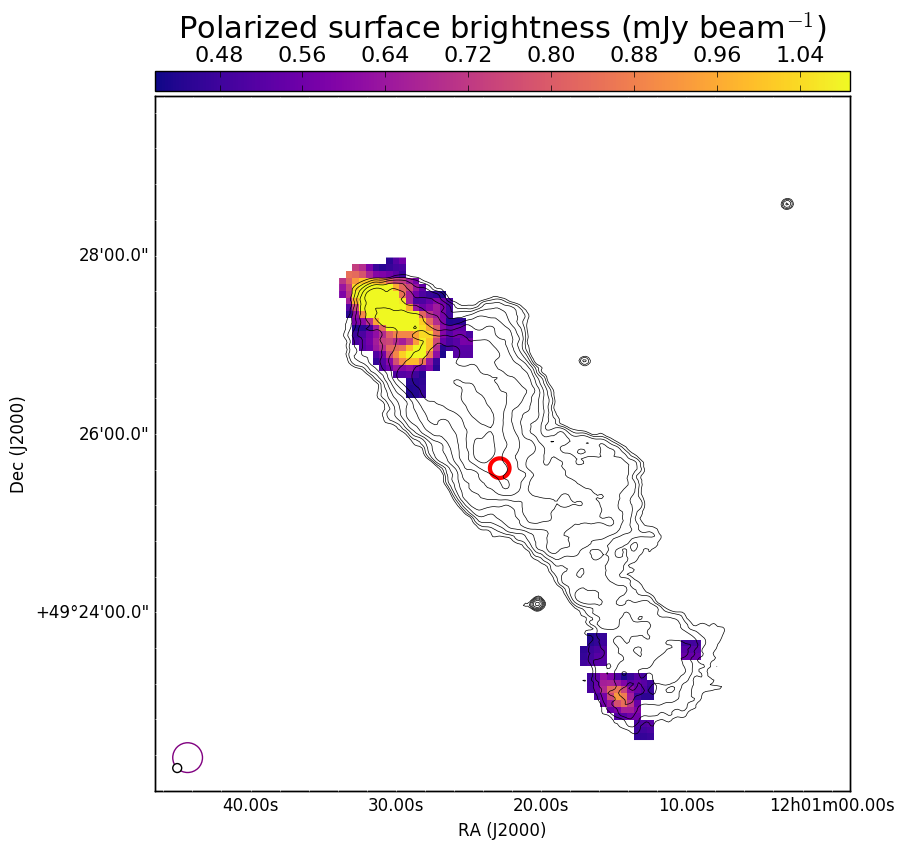}
\caption{ILTJ120122.67+492554.0} \label{fig:a}
\end{subfigure}\hspace*{0.1cm}
\begin{subfigure}{0.46\textwidth}
\includegraphics[scale=0.32]{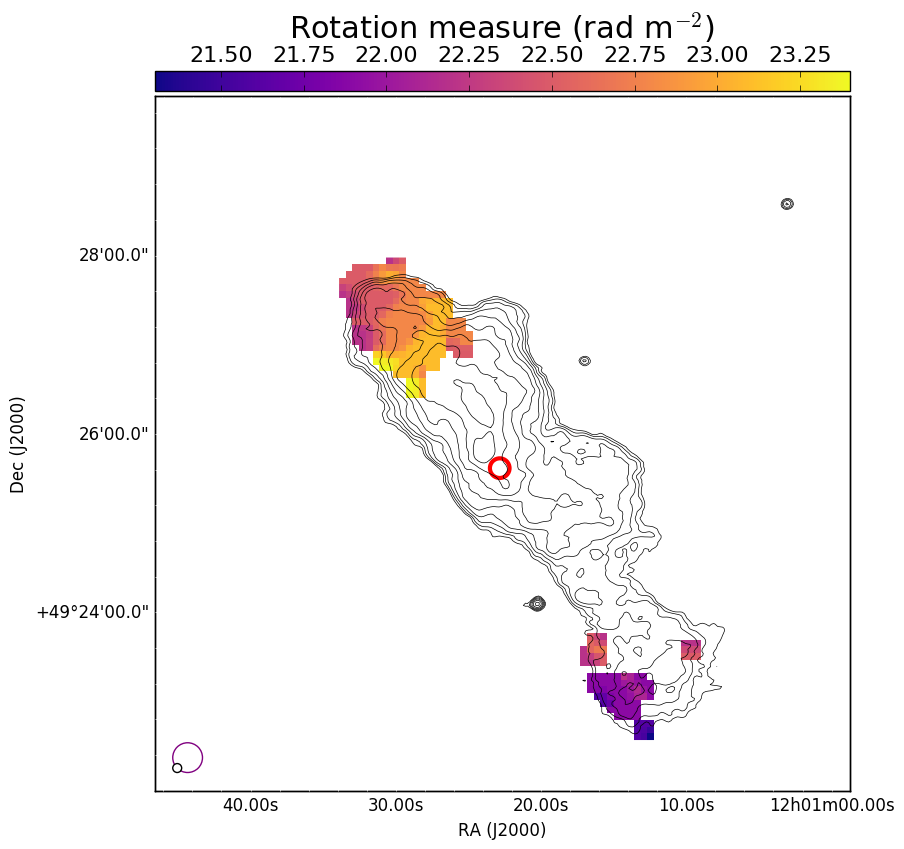}
\caption{ILTJ120122.67+492554.0} \label{fig:b}
\end{subfigure}

\medskip
\begin{subfigure}{0.46\textwidth}
\includegraphics[scale=0.32]{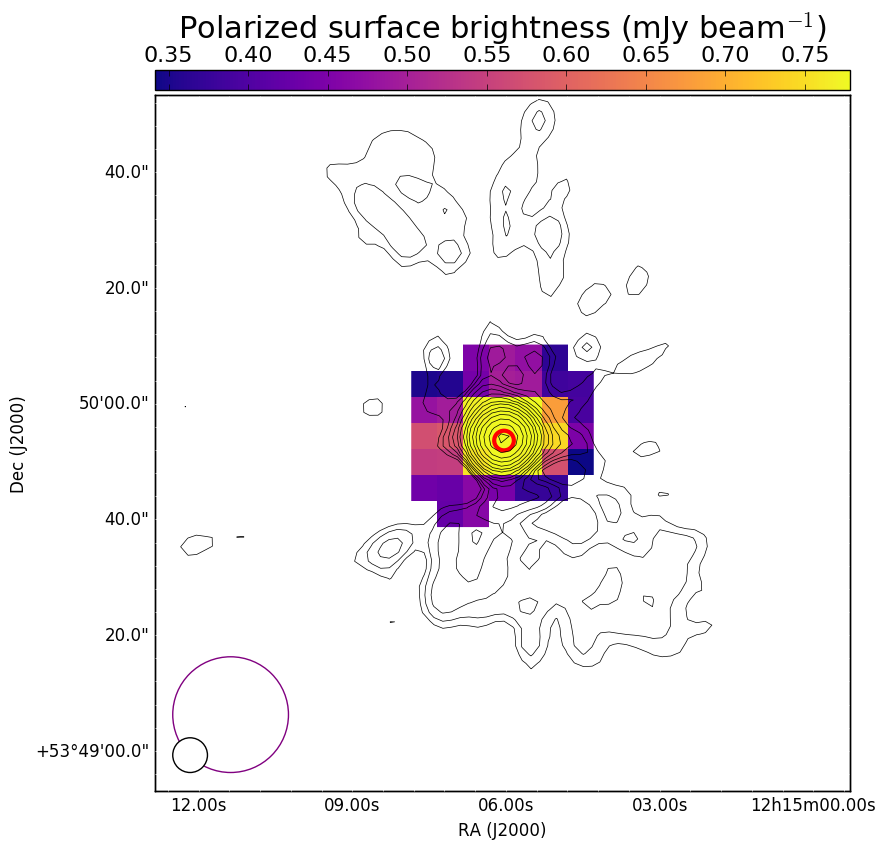}
\caption{ILTJ121506.07+534953.2} \label{fig:c}
\end{subfigure}\hspace*{0.1cm}
\begin{subfigure}{0.46\textwidth}
\includegraphics[scale=0.32]{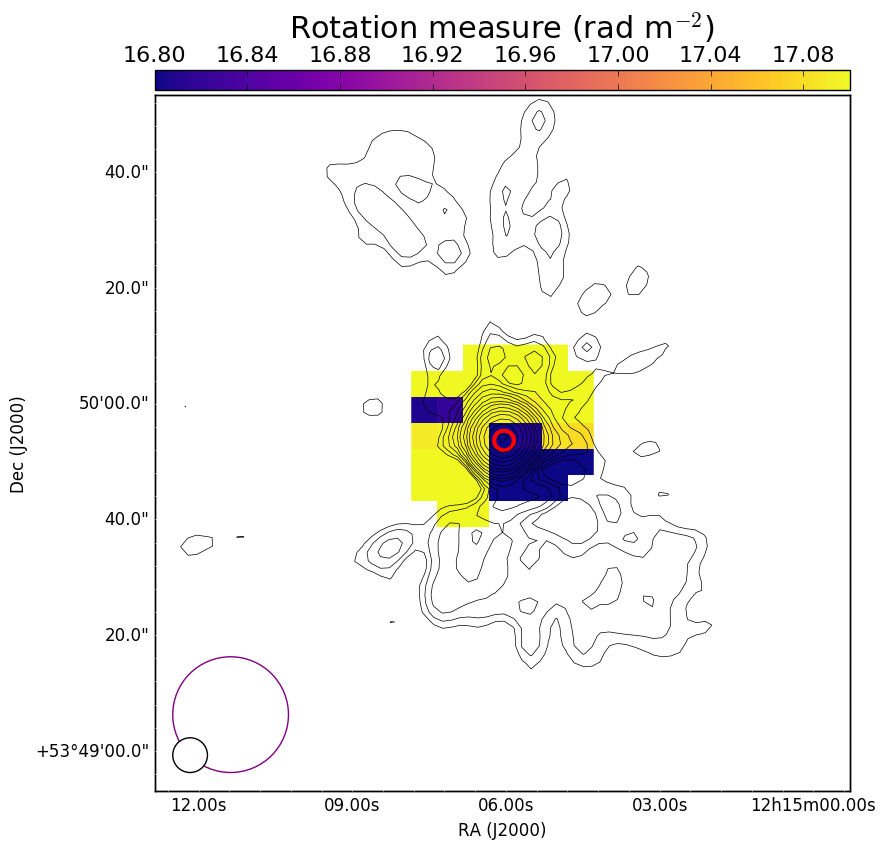}
\caption{ILTJ121506.07+534953.2} \label{fig:d}
\end{subfigure}

\medskip
\begin{subfigure}{0.46\textwidth}
\includegraphics[scale=0.32]{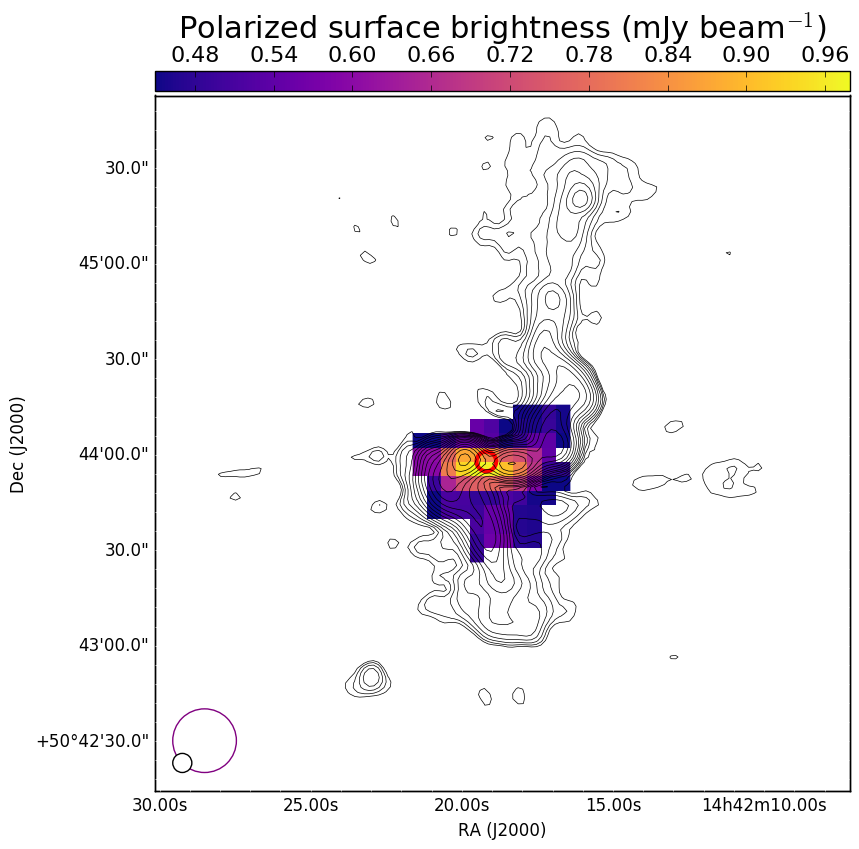}
\caption{ILTJ144218.66+504403.7} \label{fig:e}
\end{subfigure}\hspace*{0.1cm}
\begin{subfigure}{0.46\textwidth}
\includegraphics[scale=0.32]{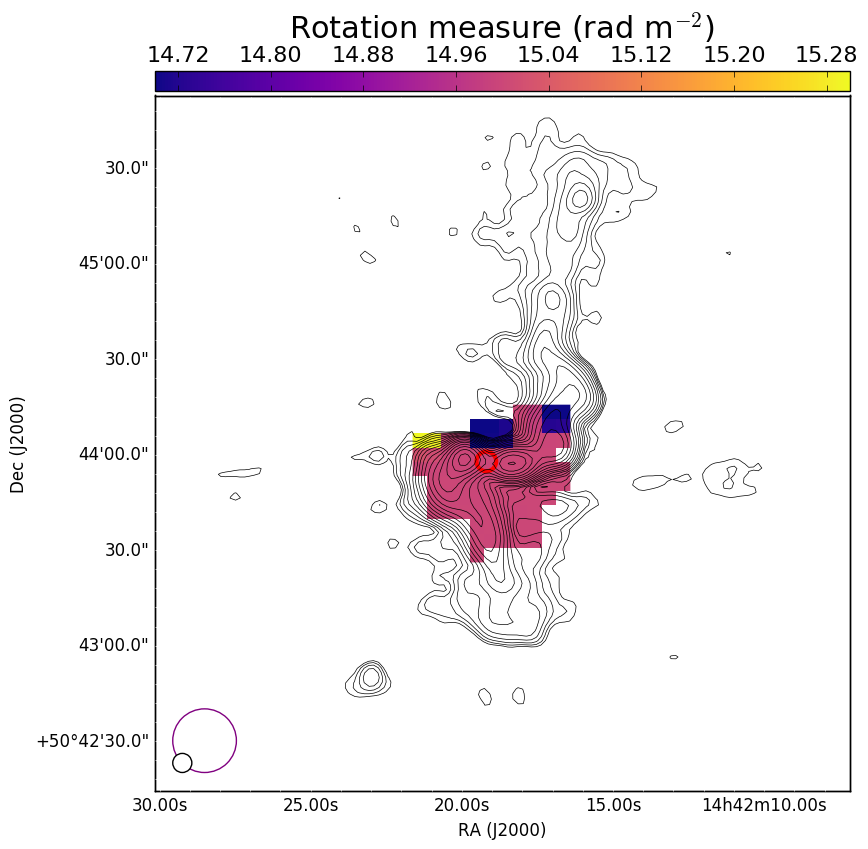}
\caption{ILTJ144218.66+504403.7} \label{fig:f}
\end{subfigure}

\captionof{figure}{150 MHz polarized FR-Is from our RLAGN sample. Coloured pixels represent 20 arcsec polarization detections: left panels are the polarized intensity and the right panels are the $RM$s, while black contours display the 6 arcsec Stokes I emission (beam sizes shown as magenta and black circles in the lower left of each image, respectively). Red circles show the position of the catalogued host galaxy in LoTSS DR1.} \label{fig:pol_fr1}
\end{figure*}
\begin{figure*} % "[t!]" placement specifier just for this example
%\captionsetup{type=figure}\addtocounter{figure}{-1}
\begin{subfigure}{0.46\textwidth}
\includegraphics[scale=0.33]{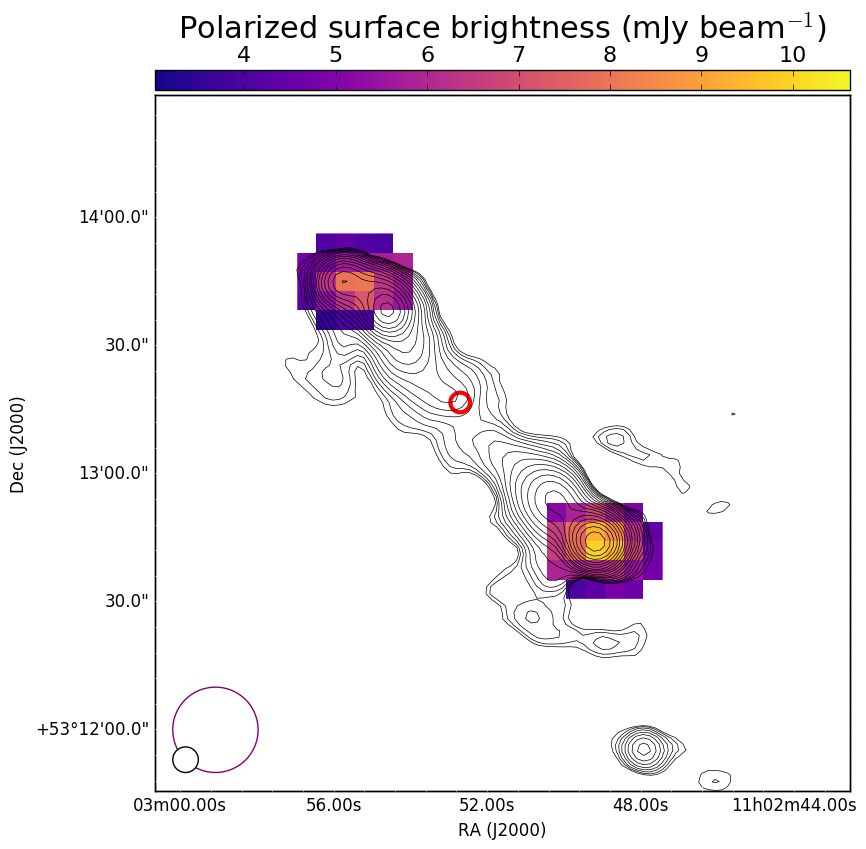}
\caption{ILTJ110251.61+531307.1} \label{fig:a}
\end{subfigure}\hspace*{\fill}
\begin{subfigure}{0.48\textwidth}
\includegraphics[scale=0.33]{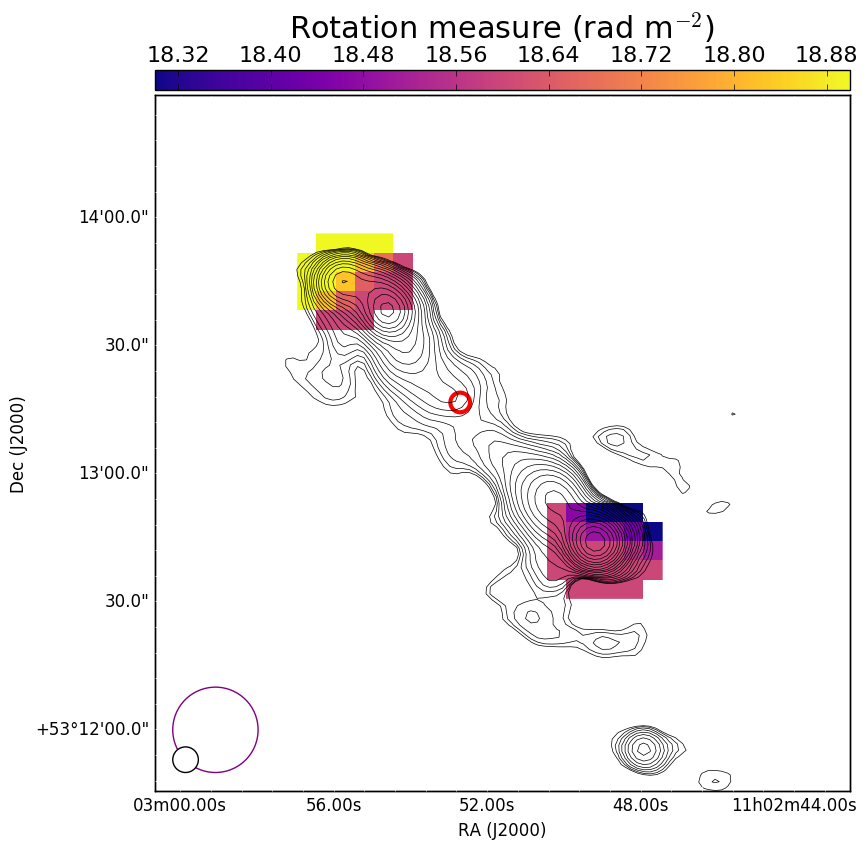}
\caption{ILTJ110251.61+531307.1} \label{fig:b}
\end{subfigure}

\medskip
\begin{subfigure}{0.48\textwidth}
\includegraphics[scale=0.33]{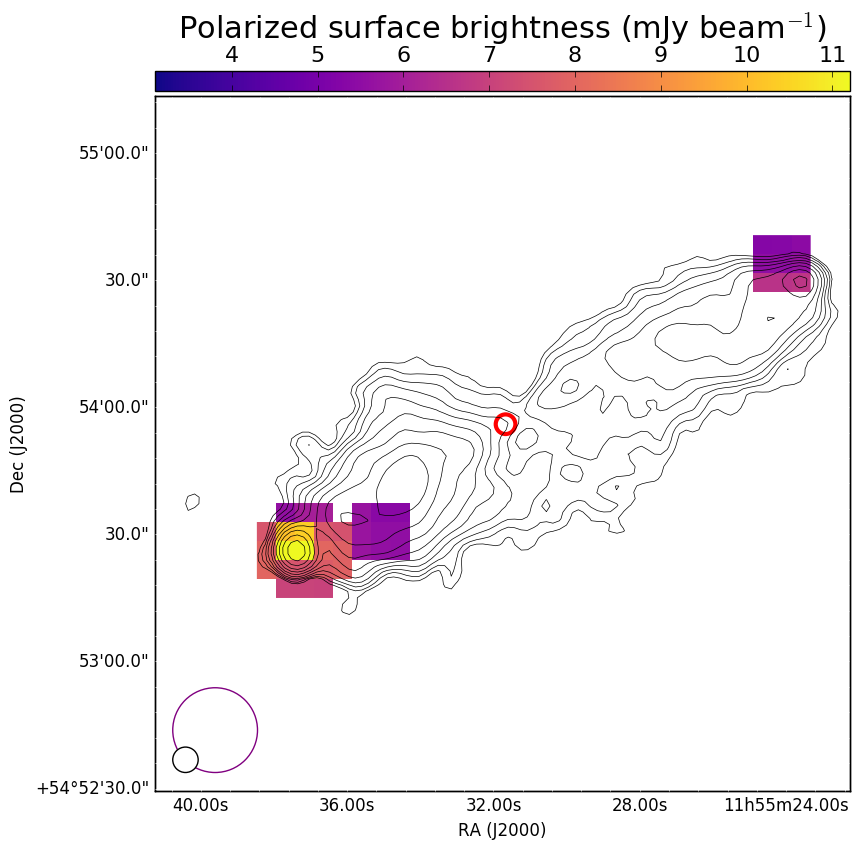}
\caption{ILTJ115531.76+545351.5} \label{fig:c}
\end{subfigure}\hspace*{\fill}
\begin{subfigure}{0.48\textwidth}
\includegraphics[scale=0.33]{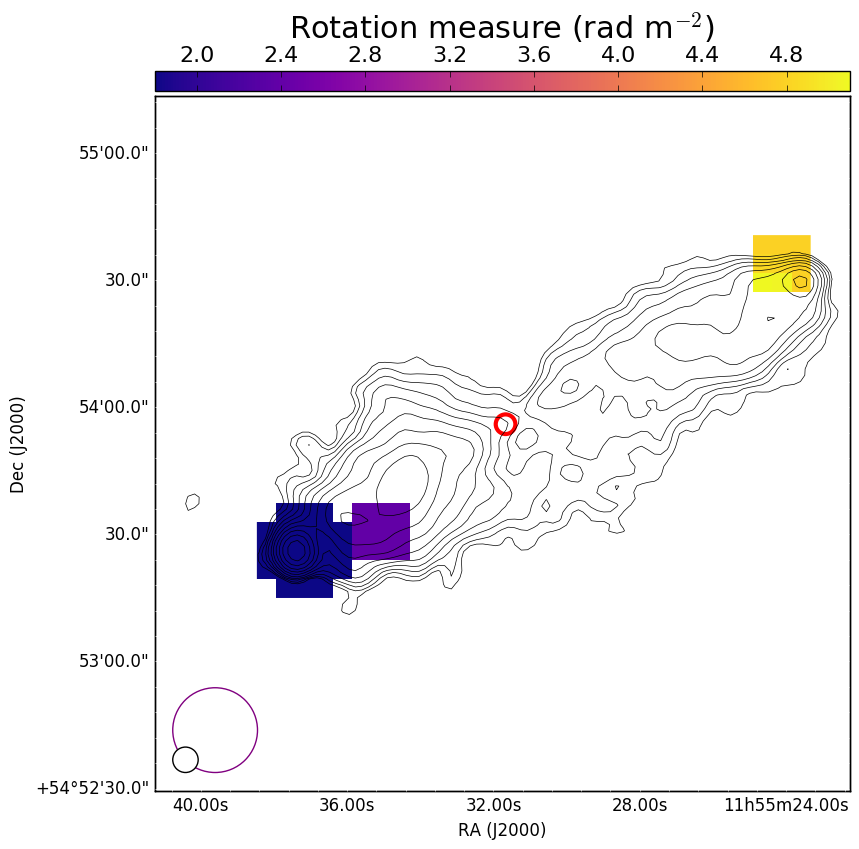}
\caption{ILTJ115531.76+545351.5} \label{fig:d}
\end{subfigure}

\medskip
\begin{subfigure}{0.48\textwidth}
\includegraphics[scale=0.33]{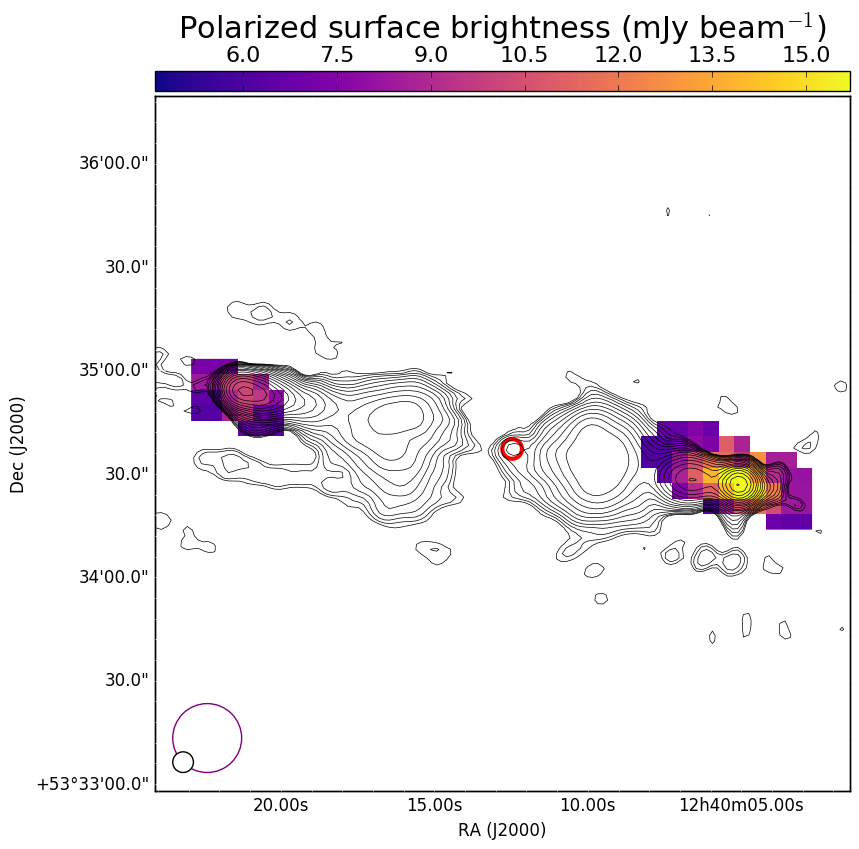}
\caption{ILTJ124012.79+533438.9} \label{fig:e}
\end{subfigure}\hspace*{\fill}
\begin{subfigure}{0.48\textwidth}
\includegraphics[scale=0.33]{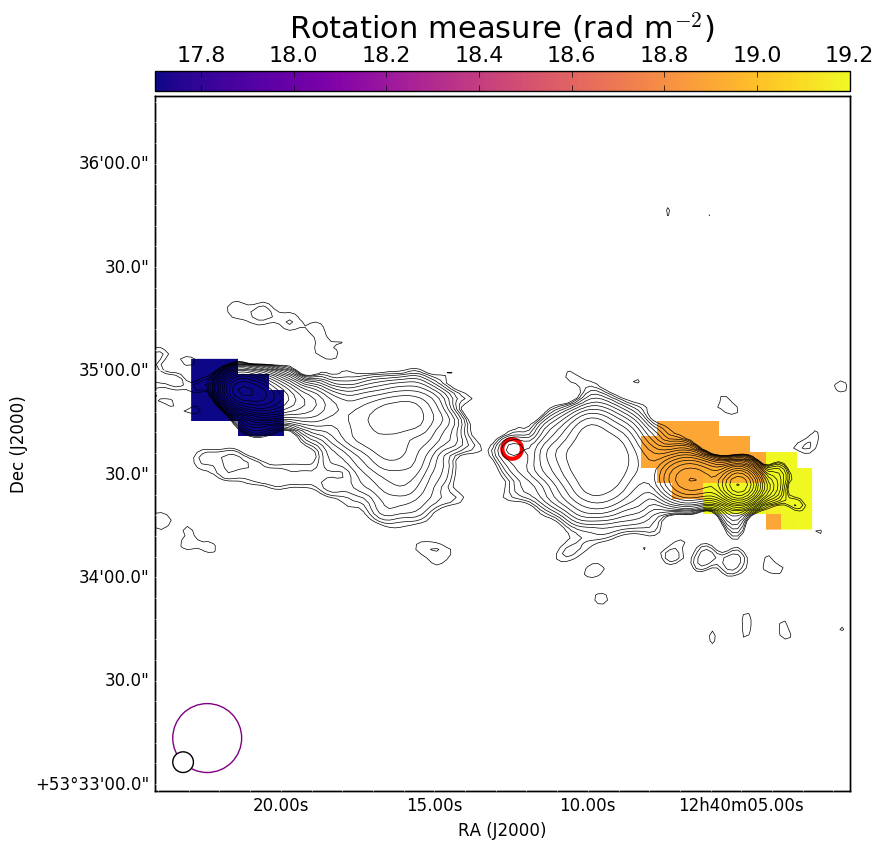}
\caption{ILTJ124012.79+533438.9} \label{fig:f}
\end{subfigure}

\caption{150 MHz polarized FR-IIs from our RLAGN sample. See Figure \ref{fig:pol_fr1} for image descriptions.} \label{fig:pol_fr2}
\end{figure*}

To determine which sources have detections of polarized emission, given the relatively low resolution of the maps and the expectancy of low S/N polarization, we use a simple island-finding method by masking non-detected pixels\footnote{Each map has a pixel size of 4.5 arcsec}. First, we remove background noise pixels by masking pixels which have surface brightnesses less than the mean pixel value in the image plus 3$\sigma_b$, where $\sigma_b$ is the standard deviation of the pixel brightnesses in the image. For some sources that were clearly not polarized (sporadic regions of high pixel intensity, mostly off-source) but still had unmasked regions, particularly those with large angular size, we manually disallowed a detection. Detections were also manually disallowed which still had unmasked regions of nearby polarized sources in the map unrelated to the source (i.e. background quasars). For some sources generally with small angular sizes and hence a low number of background pixels present in the maps (as the cubes were constructed near the source to only contain on-source pixels), mean pixel surface brightnesses and $\sigma_b$ were generally overestimated as the $3\sigma_b$ threshold prevented detections of clearly polarized hotspots. We therefore implemented a procedure to iteratively mask >5$\sigma_b$ regions in the polarized intensity maps and re-calculate $\sigma_b$, until the fractional difference between $\sigma_b$ in the current and last iteration became $\left(\sigma_{b,\text{last}}-\sigma_{b,\text{current}}\right)$/$\sigma_{b,\text{current}}$ $<1\times10^{-6}$. For some source images still containing too few background pixels, where the image is dominated by bright emission, we reduce the masking criterion to pixels >$3\sigma_b$.

We then label as detections of polarized emission where groups of unmasked pixels are in a $2\times2$ configuration or larger, so as to prevent single pixel detections which we regard as insufficient as all of our sources are resolved at 20 arcsec in Stokes I. We measured polarized flux densities as the sum of the detected pixel surface brightnesses divided by the beam area. The uncertainty on the measured polarized flux densities are quoted as 3$\sigma_{\text{QU}}$, where $\sigma_{\text{QU}}$ is given as the mean of the detected pixels in the linearly polarized rms map output from $RM$ synthesis\footnote{The rms is given by the average spread of the `wings' of the Faraday spectrum over Stokes Q and U, limited by our Faraday depth range of -150 $\leqslant \phi$ (rad m$^{-2}$) $\leqslant$ 150.}. For the $RM$ of each source, we take a weighted mean pixel value from the $RM$ map as
\begin{ceqn}
\begin{equation}
\langle RM\rangle = \frac{\sum\limits_{i}w_i RM_i}{\sum\limits_{i} w_i}
\end{equation}
\end{ceqn}
where $w_i$ is the normalised pixel brightness of pixel $i$ of all detected pixels in the polarized intensity map. The $RM$ error is given by
\begin{ceqn}
\begin{equation}
\sigma_{RM}=\frac{\text{RMSF FWHM}}{2\times S/N}
\label{equation:rm_err}
\end{equation}
\end{ceqn}
\citep{bren05}, where the RMSF FWHM (Full Width Half Maximum) is $1.16$ rad m$^{-2}$ for LoTSS pointings, and the signal to noise ratio $S/N$ is the peak pixel brightness over the rms value in that pixel. An additional, more dominant error arises from the systematic error of the ionospheric $RM$ correction included in $RM$ synthesis, and results in uncertainties of 0.1-0.3 rad m$^{-2}$ \citep{soto13}. We give conservative error estimates by adding in quadrature the error from Equation \ref{equation:rm_err} and a maximum systematic error of 0.3 rad m$^{-2}$ for each source. We also correct our polarized intensities for Ricean bias, which can be significant at low signal to noise, using Equation 5 of \cite{geor12}. No corrections have been made for the dependence of the derived Faraday spectrum on spectral index, but it does not affect the peak of the Faraday spectrum and hence the corrections are minimal \citep{bren05}.

We then identified sources where we find evidence of leakage signal being manifested as detections -- the Faraday depth range we excluded from $RM$ synthesis  ($-3\leqslant\phi\leqslant1.5$ rad m$^{-2}$) is not precisely centred on zero due to the ionospheric $RM$ correction(s) shifting the leakage signals. It is possible for this exclusion range to not capture all sources with leakage signal and some of our detected sources which have $RM$s below -3 rad m$^{-2}$ and above +1.5 rad m$^{-2}$ may show strong leakage, particularly at the locations of the sidelobes of the leakage signal. Since leakage signals will generally have low fractional polarization ($\Pi\leqslant 1$\%), we removed sources in our sample with $\Pi\leqslant 1$\% that were detected at $-3.5\leqslant\phi\leqslant-3$ rad m$^{-2}$ and $1.5\leqslant\phi\leqslant2$ rad m$^{-2}$ in their Faraday spectra. This resulted in the removal of seven sources. Faraday spectra of three detected sources in our sample that were not excluded are shown in Figure \ref{fig:fd_spectra}.

Applying our methods we finally obtain a reliable polarization detection in 67 out of 382 sources -- a detection fraction of 18 per cent at 150 MHz. We regard the non-detected sources as depolarized. This represents a polarized radio galaxy surface density within the HETDEX field, for sources brighter than 50 mJy and larger than 100 arcsec in angular size, of 0.16$\pm0.02$ deg$^{-2}$ (errors quoted here represent Poisson statistics). As a comparison with recent studies from the LoTSS data in the same area, \cite{vane18} found 92 point sources at a surface density of 0.16$\pm 0.02$ deg$^{-2}$ at 4.3 arcmin resolution, while \cite{mulc14} and \cite{neld18} found 6 securely-detected polarized sources in a single LOFAR pointing at the same resolution as ours, finding a surface density of 0.30$\pm0.12$ deg$^{-2}$, demonstrating consistency between LOFAR studies and confirming that RLAGN are the predominant source of polarization at 150 MHz. Subsets of our detected sources are presented in Figure \ref{fig:pol_fr1} and Figure \ref{fig:pol_fr2}.
\subsection{Caveats affecting our polarized sample}
The most important limitation to our analysis is the use of unCLEANED Stokes QU cubes that are used to produce polarized intensity (dirty) maps. We cannot reliably CLEAN the cubes prior to $RM$ synthesis due to the low signal to noise in each channel of the cubes. The effect of this is higher noise and artefacts in the resulting polarized intensity maps, particularly around bright point sources, than in the case of CLEANed maps. Since the aims of our study do not rely on accurate astrometry and analysis of spatial structure, the low image fidelity due to the use of dirty QU cubes does not affect our analysis or results.

We also note the issue of using $RM$ synthesis and the associated rms map in polarized intensity to determine detection rates: imperfect imaging and calibration (as is the case here) can lead to non-Gaussian tails in the Stokes Q and U Faraday spectra, whence the rms is calculated. While this will overestimate errors, it may also cause high false detection rates in low signal to noise pixels \citep{geor12} where relatively low detection thresholds are used (3$\sigma_{\text{QU}}$, as used here). While \cite{geor12} propose 8$\sigma_{\text{QU}}$ to serve as a detection threshold, we note that our detection method involved visual inspection, and in the case of polarized emission not associated with a core, lobe or hotspot (i.e structures of high intensity in the Stokes I image), sources were discarded as false detections. A 5$\sigma_{\text{QU}}$ threshold was tested, giving the result that low signal to noise sources that had clearly polarized emission (e.g. in hotspots) were undetected with our method. 
\section{Analysis}
\label{sect:analysis}
\begin{figure*}
    \centering
    \includegraphics[scale=0.43, trim={1.2cm 0 2cm 0}, clip]{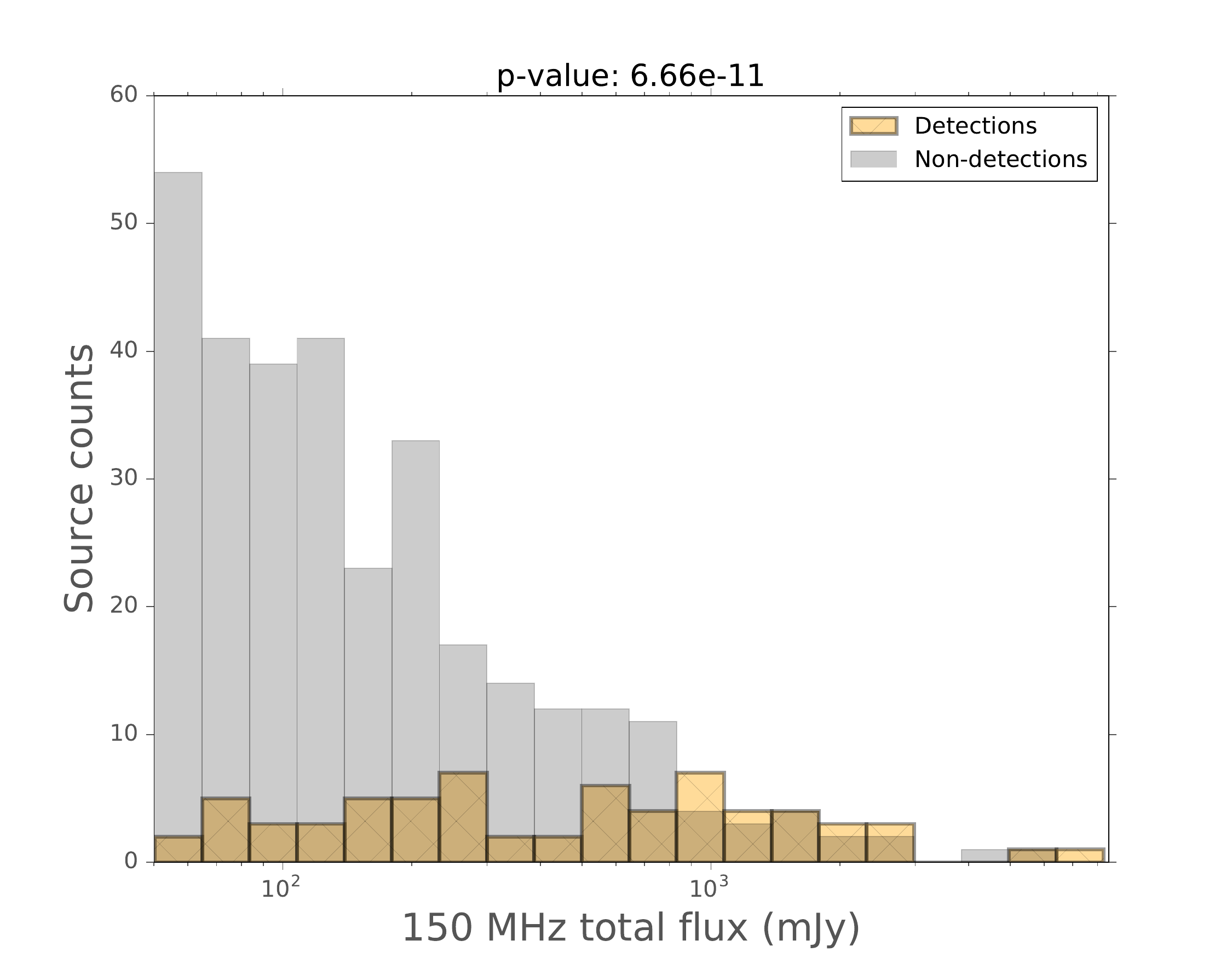} \hspace{0cm}
    \includegraphics[scale=0.43, trim={1.2cm 0 2cm 0}, clip]{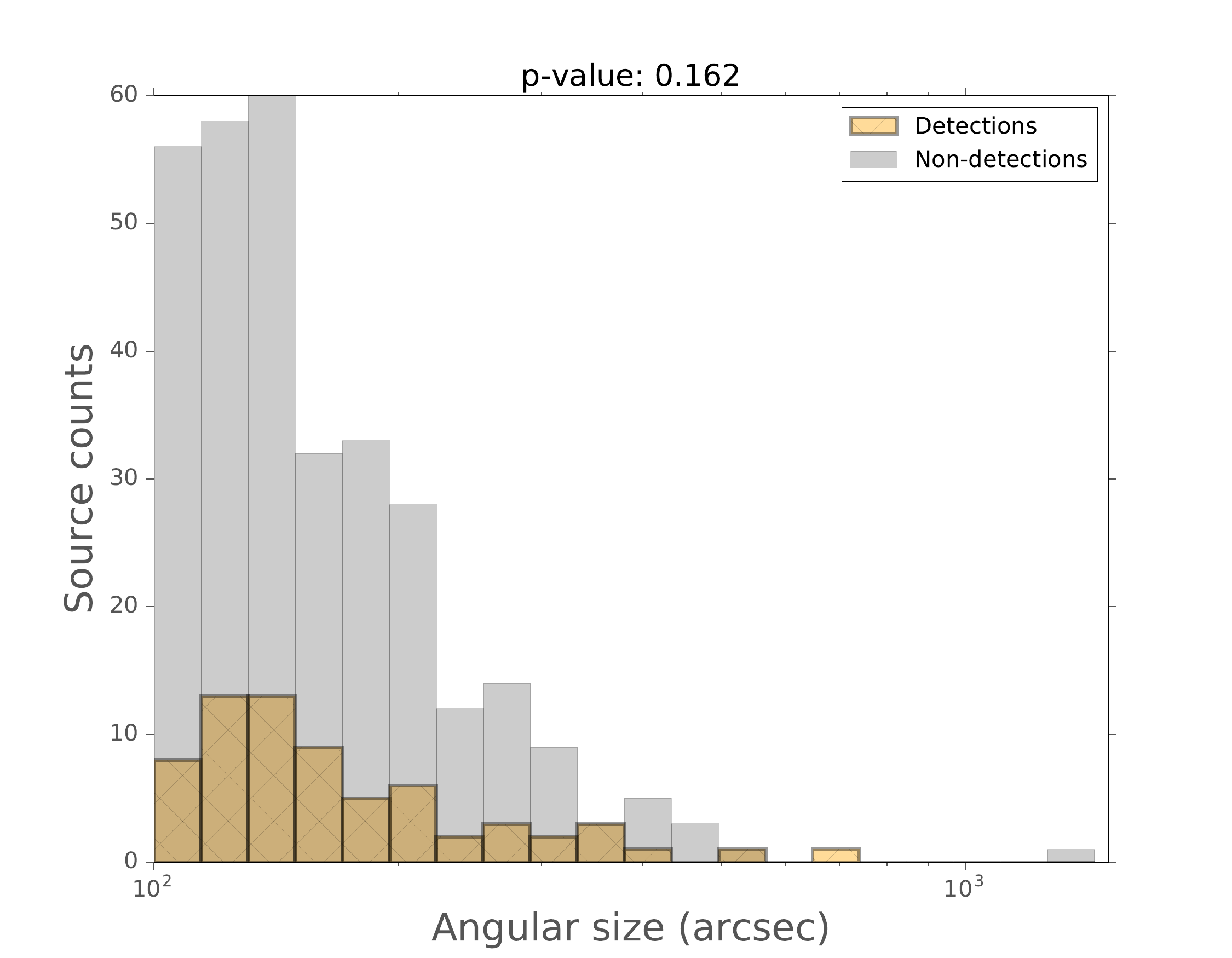}\\
    \includegraphics[scale=0.43, trim={1.2cm 0 2cm 0}, clip]{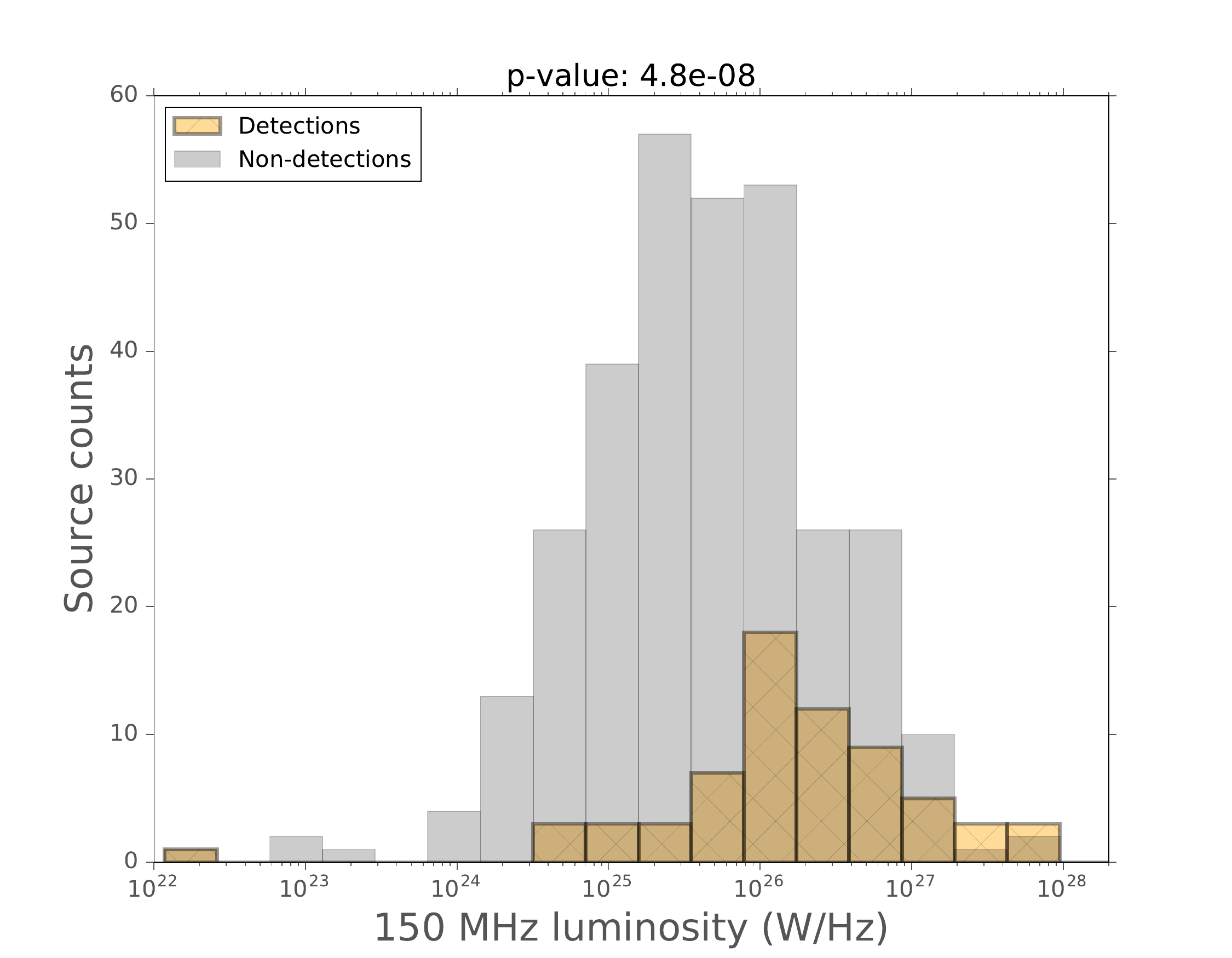}
    \includegraphics[scale=0.43, trim={1.2cm 0 2cm 0}, clip]{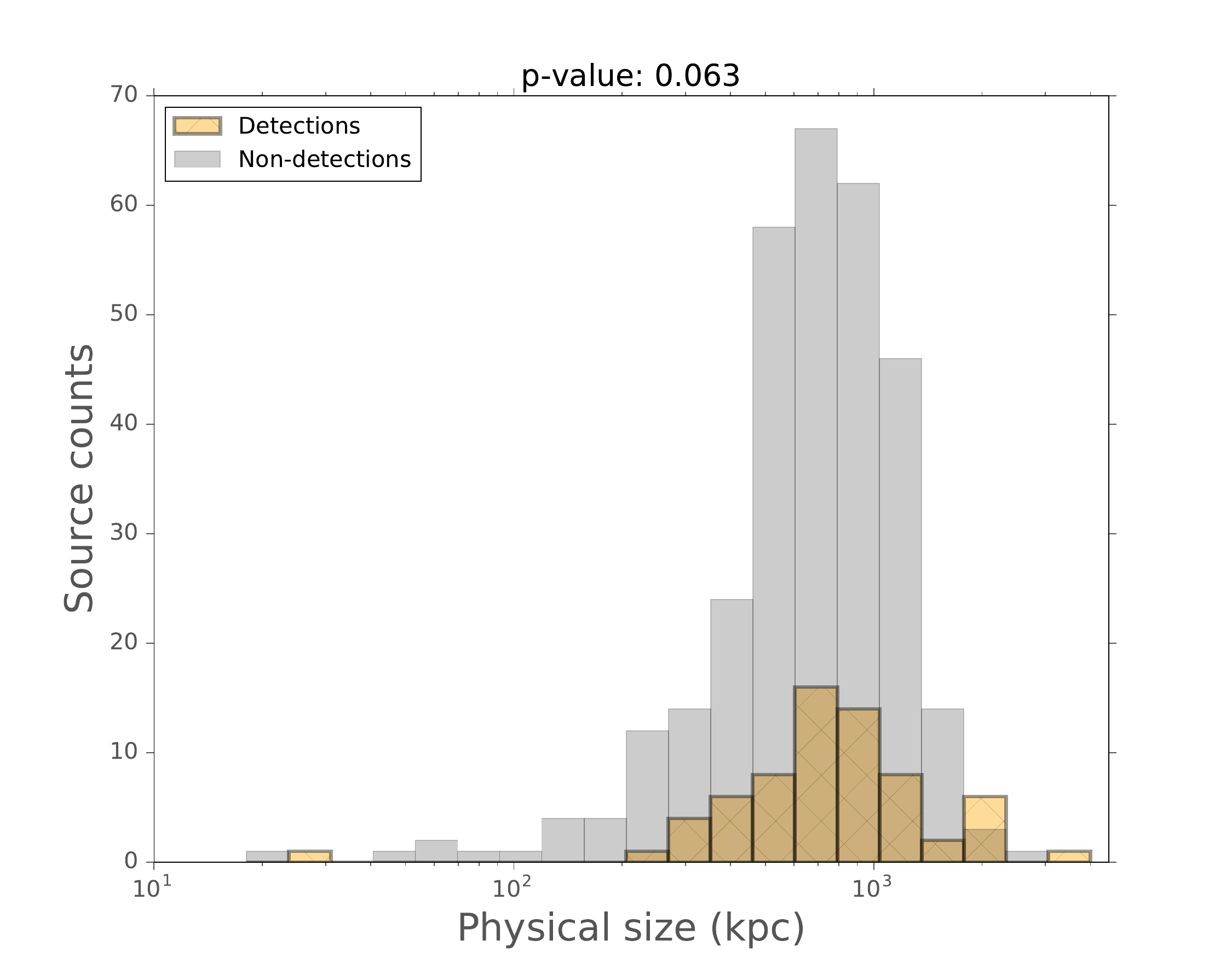}
    \caption{Distributions of observational and physical radio properties of our RLAGN sample; 150 MHz flux density (top left), angular size (top right), radio luminosity (bottom left) and physical size (bottom right). Grey and hatched beige bars represent our non-detected and detected sources in polarization, respectively.}
    \label{fig:radio_props_hist}
\end{figure*}
In this section we present a statistical analysis of the polarization properties of our RLAGN sample. Unless otherwise stated, to distinguish the characteristics of two distributions we quote the $p$-value from a Wilcoxon-Mann-Whitney test \citep{mann47}, using a 95 per cent confidence level (i.e. a p-value $<0.05$ means we can reject the null hypothesis that the two samples have identical median values). 
\subsection{Observational properties}
In the top panel of Figure \ref{fig:radio_props_hist} we plot the distributions in 150 MHz flux density and angular size of our detected and non-detected sources. We see that, statistically, our polarized sources (hatched beige) are significantly brighter compared to those that are depolarized (grey), as expected. Further, the detection fraction is greater than 50 per cent for sources brighter than 1 Jy. On the other hand, there are statistically similar medians in angular size between detected and non-detected sources, meaning that the detectability of polarization amongst RLAGN in our sample is driven primarily by flux density rather than, or in addition to, angular size. Similar statistics are seen with physical properties as shown in the bottom panel of Figure \ref{fig:radio_props_hist}, where the polarized sources are more luminous but similar in physical size. This is consistent with the idea that brighter and more luminous RLAGN have a preference for detectable polarization.

\begin{figure*}
    \centering
    \includegraphics[scale=0.41, trim={0.7cm 0 1cm 0}, clip]{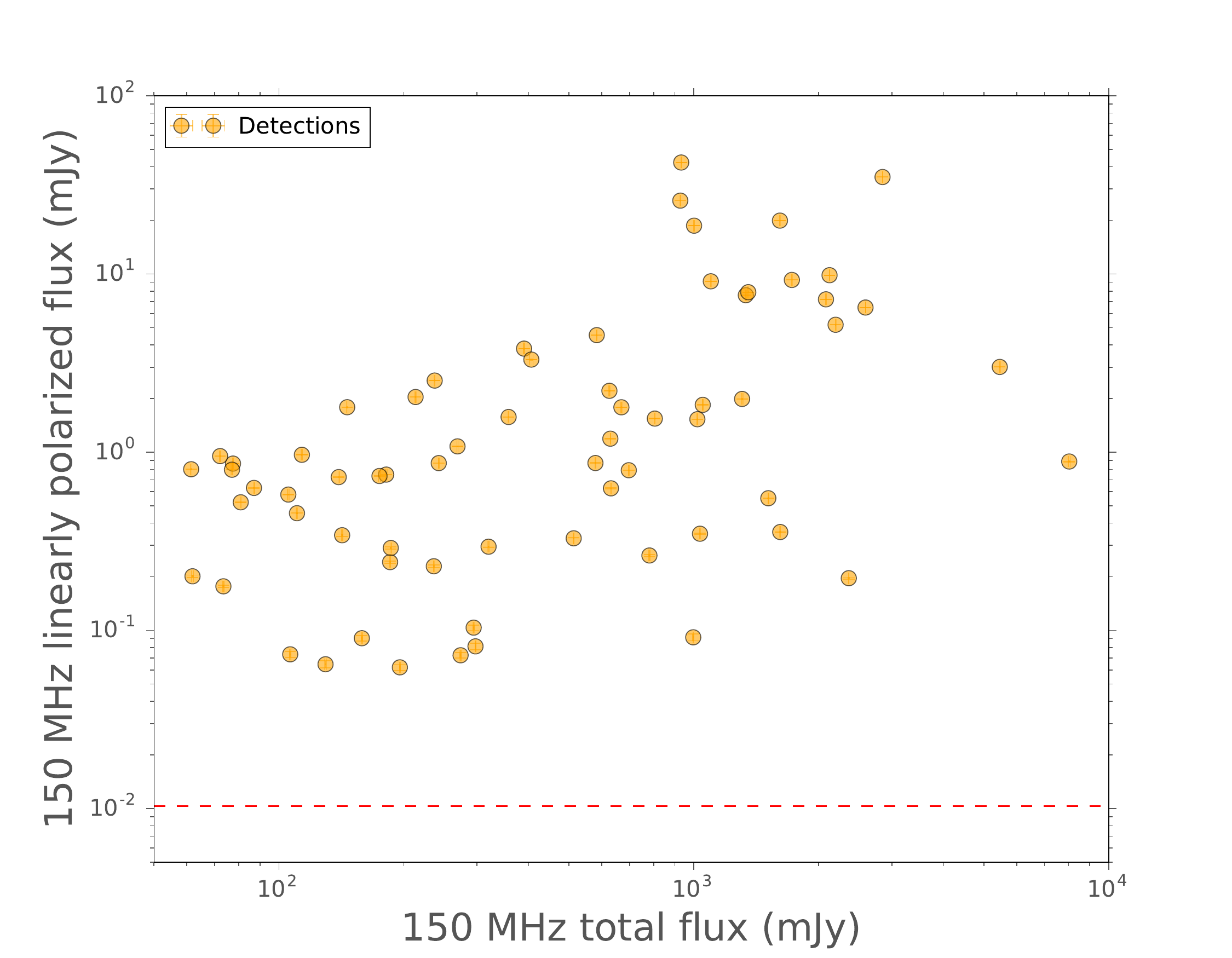}
    \includegraphics[scale=0.41, trim={0.7cm 0 1cm 0}, clip]{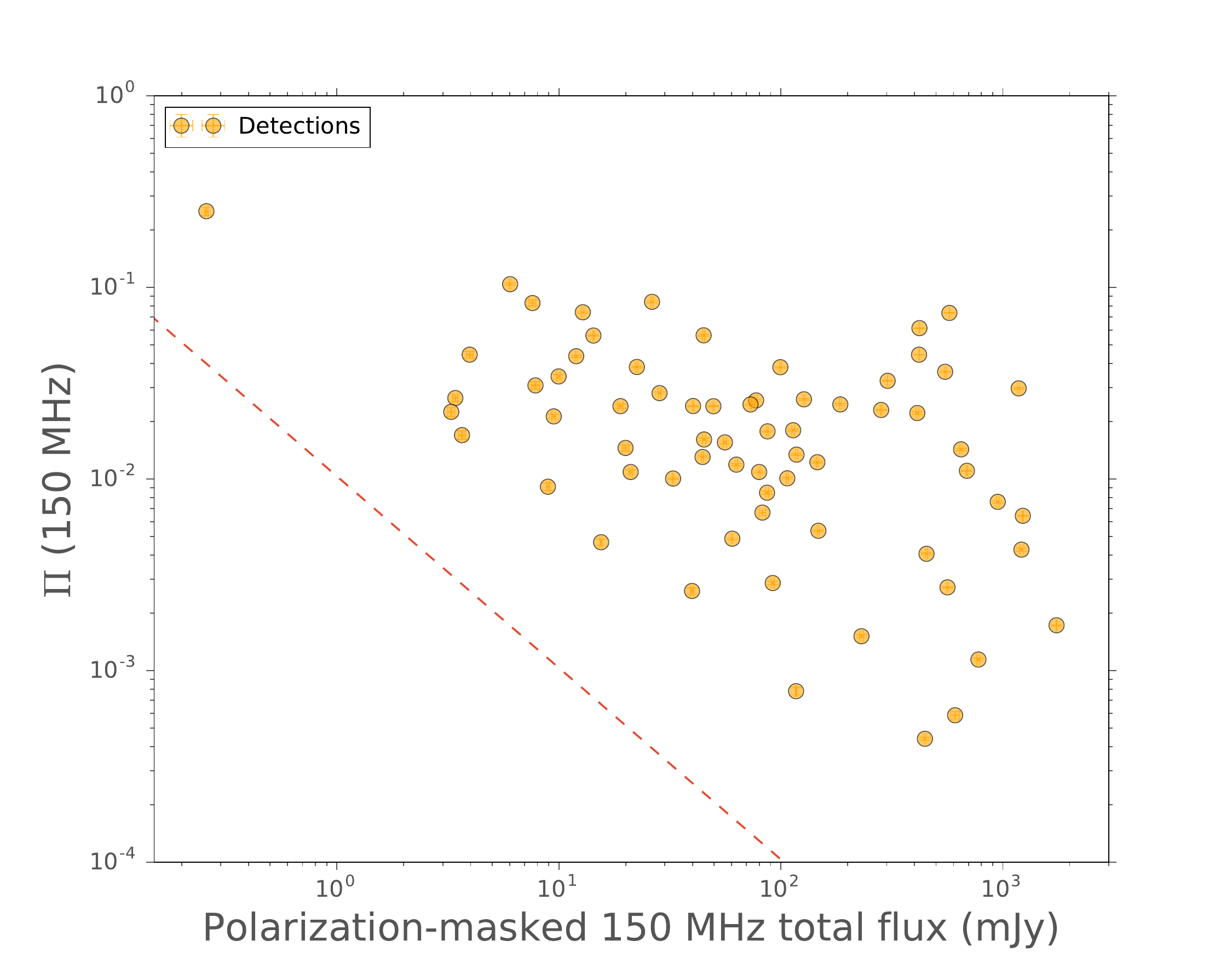}
    \caption{150 MHz linearly polarized flux density (left) and fractional polarization (right) against total flux density for our detected sources. Note that for the right panel we only plot the total flux density for the regions of the source that are polarized. Uncertainties are shown as 3$\sigma$ error bars, but are not visible due to the logarithmic scale.Dashed lines give an indicative estimate of our sensitivity to polarized emission, plotted as $3\sigma_{\text{mean},QU}$, where $\sigma_{\text{mean},QU}$ is the average rms error of polarized fluxes in our sample.}
    \label{fig:flux/size_polflux}
\end{figure*}

In the left panel of Figure \ref{fig:flux/size_polflux} we plot the 6 arcsec total flux density against the 20 arcsec polarized flux density for our polarized sources. We do not see any clear correlation, meaning that although the polarized detection fraction increases with flux density (as seen in Figure \ref{fig:radio_props_hist}), the \textit{amount} of polarized emission is not entirely driven by total flux density\footnote{We note that the flux density cut in selecting sources in our sample introduces a selection bias, and our results do not necessarily apply for sources with $S_{150}<50$ mJy.}. Rather, a range of polarized emission is detected in RLAGN for a large range in total flux density at 150 MHz. This is consistent with a model in which the level of polarized emission seen in RLAGN is strongly related to the characteristics of the associated Faraday screen and the attributed depolarization, some components of which are unrelated to the source (e.g. the foreground IGM). We also plot the fractional polarization $\Pi$ (ratio of polarized emission to total emission in the polarization-detected pixels) against the total flux density measured in the 20 arcsec Stokes I image, in the right panel of Figure \ref{fig:flux/size_polflux} (the red dashed line indicates the approximate sensitivity to polarization by taking an average of $3\sigma_{QU}$ from all sources). As expected, since the latter observable is the denominator of the former, the fractional polarization decreases with increasing total flux density. However there is a large scatter, particularly for $>1$ Jy sources, which is likely driven by a combination of different jet properties and environmental Faraday depolarization in the line of sight at 150 MHz.
\begin{figure*}
    \centering
    \includegraphics[scale=0.41, trim={0.7cm 0 1cm 0}, clip]{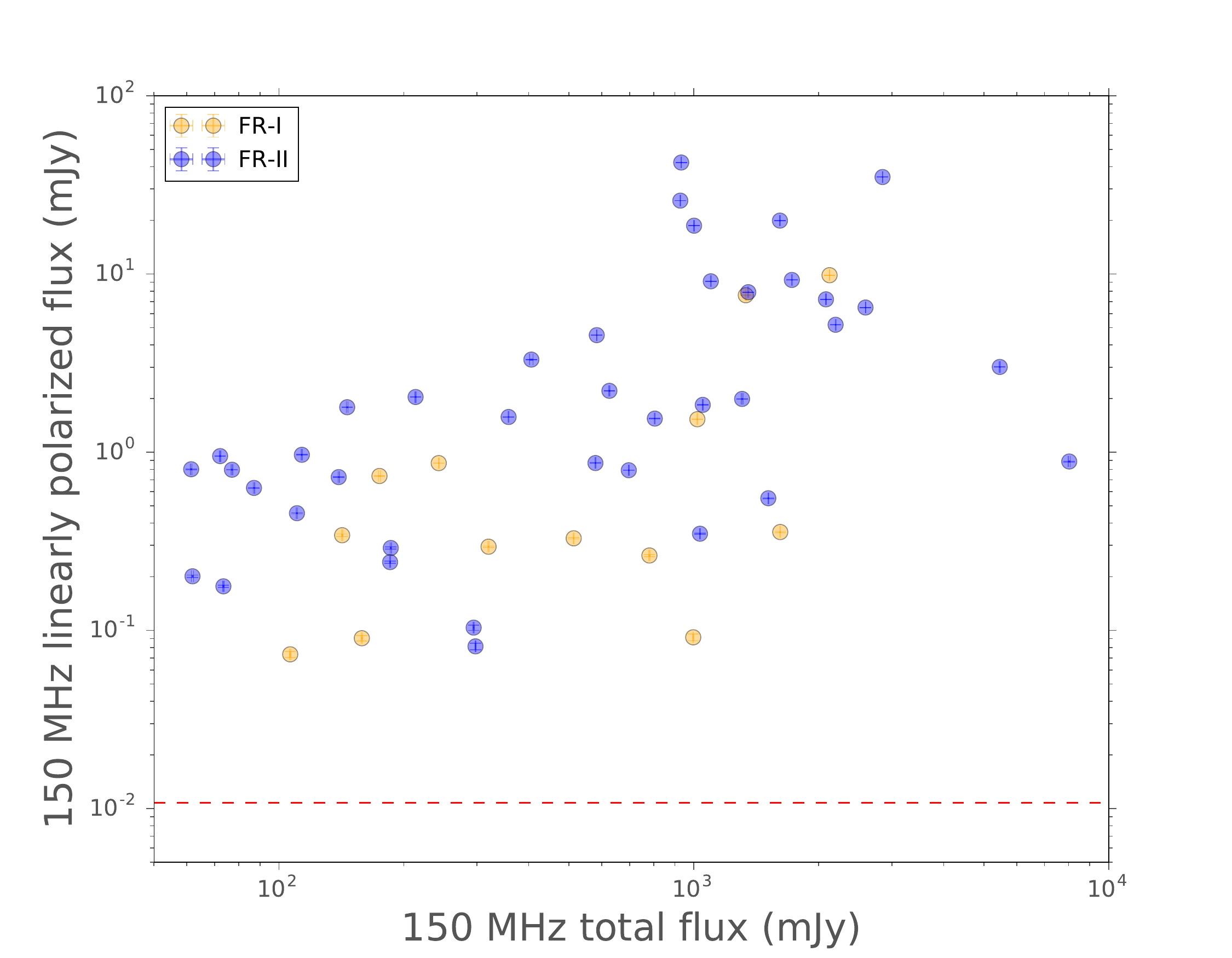}
    \includegraphics[scale=0.41, trim={0.7cm 0 1cm 0}, clip]{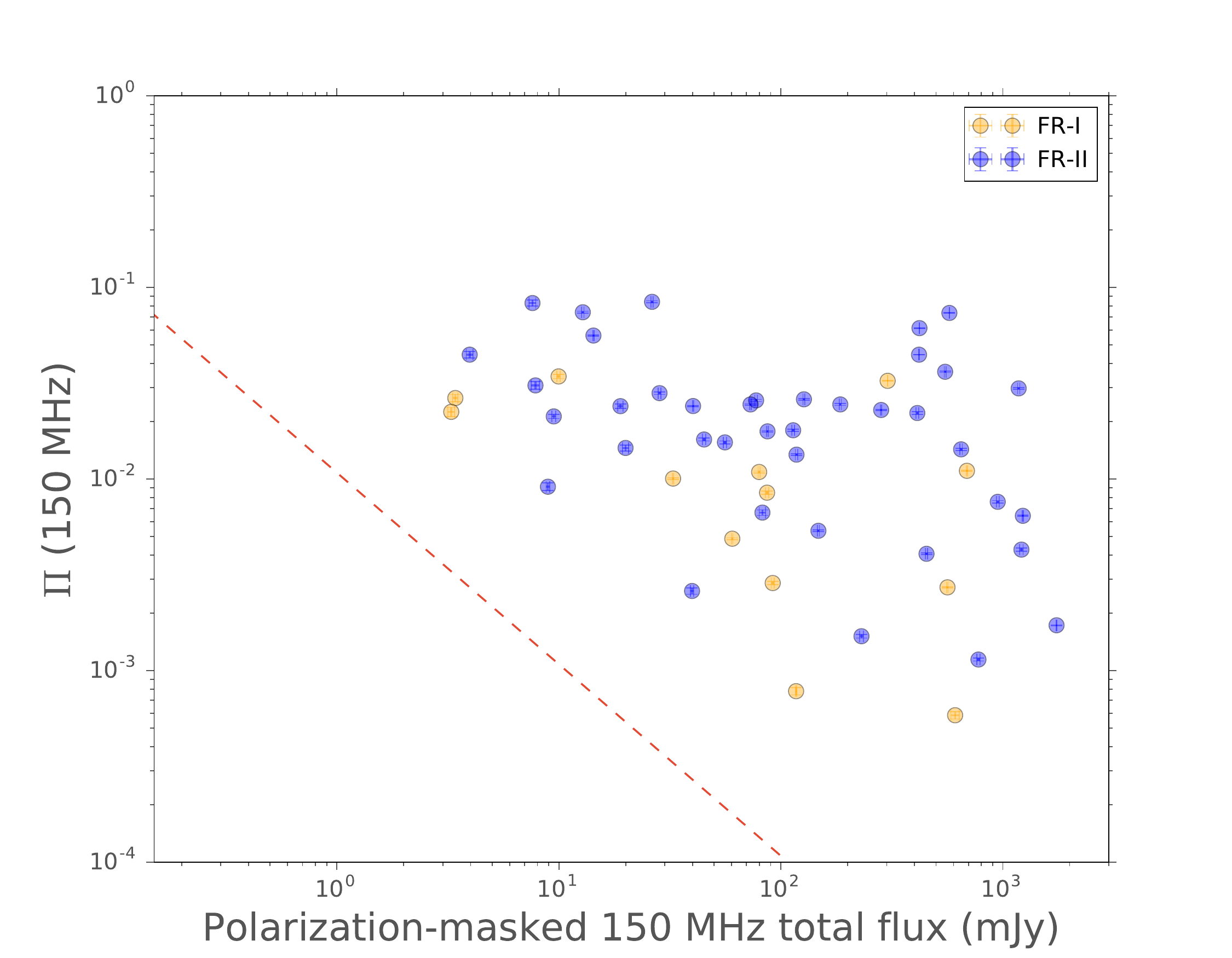}
    \caption{Same as Figure \ref{fig:flux/size_polflux}, for sources classed as FR-I (beige) and FR-II (blue) according to the criteria of \protect\cite{ming19}. Note that indeterminate-morphology sources are not plotted.}
    \label{fig:flux_polflux_morph}
\end{figure*}

It is important to test whether there are correlations within the detected sources due to morphology (i.e different jet or lobe properties). It is well known that FR-I sources are expected to have higher depolarization than FR-II sources for the following reasons: given the same external environment, the bright cores of FR-I sources would typically be associated with higher Faraday depths due to the increasing radial profile towards the centre of the thermal plasma in which they are embedded. On the other hand, FR-II sources are typically polarized at their hotspots (e.g. \citealt{osul18}), which are at larger projected distances from the center of their environments. In addition, FR-I sources are generally found in richer environments than FR-IIs, leading to higher Faraday depths in their line of sight. FR-I sources are also thought to significantly entrain dense material from their surrounding medium \citep[e.g.][]{lain02,cros08,cros14}, which in general would lead to more internal depolarization over that in FR-II sources. Brighter polarization might also be expected in FR-II sources as they tend to be brighter in total flux density in general than FR-I sources -- in our sample polarized FR-Is and FR-IIs have median total flux densities of 513 mJy and 626 mJy, respectively, the difference being statistically significant ($p-$value < 0.05). While FR-II hotspots are the predominant source of polarization at 150 MHz \citep{osul18}, it is important to quantify the detection fractions between FR-Is and FR-IIs, as well as to compare their fractional polarization.

We use the code of \cite{ming19}, which categorizes sources into FR-I and FR-II based on whether the peak radio emission is located close to our away from the centre of the source, respectively, to morphologically categorize the polarized sources in our sample. While visual inspection may be used to classify our sources, an automated classification based on the definition of the FR dichotomy (as used by \citealt{ming19}) represents a systematic morphological stratification of our sample without cognitive bias. The 382 objects in our sample were separated based on the code into three categories; FR-I, FR-II and indeterminate. The latter category represents the case where the code cannot clearly determine the morphology, and this is usually the case for more compact objects or objects with non-symmetric lobes where there may be FR-I-like lobes on one side of the jet and FR-II-like lobes on the other, which are likely due to projection effects in many cases \citep{harw20}. Due to the ability of LOFAR to observe both compact and extended structures of RLAGN, we expected the code to classify a large number of sources which can be visually identified as either FR-I or FR-II as indeterminate. Some of the authors (VHM, MJH and JH) visually checked the 6 arcsec total intensity maps of the indeterminate sources in our sample, and re-classified each source as either FR-I or FR-II where appropriate. Around $\sim 10$ per cent of the indeterminate sample were sources that could be clearly identified as an FR-I or an FR-II, and were moved to those categories. We also checked the sources in the original FR-I and FR-II samples, to conservatively check for obvious contaminants from either class. Only a small number ($\leqslant 5$ per cent of sources) from each category were moved to the other category. In all scenarios, sources were declassified as indeterminate on a conservative basis in order to form robust FR-I and FR-II samples. Our morphological analysis uses the FR-I and FR-II samples only.

Table \ref{table:detectionfractions} lists the detection fractions of our RLAGN sample  with morphology.  
\begin{table}
\centering
    \begin{tabular}{|c|c|c|}
        \hline 
    Sample  & Counts  & Sample detection fraction (\%) \\ 
         \hline
    RLAGN & 382 & 17.5 \\ 

    FR-I & 122 & 3.4 \\ 

    FR-II & 146 & 10.2\\ 
    
    Indeterminate & 114 & 3.9 \\
        \hline 
    \end{tabular}
     \caption{Polarization detection fractions for our RLAGN sample, and the FR-I, FR-II and indeterminate subsets based on morphology. Counts refer to the total number of sources in that sample.}
     \label{table:detectionfractions}
\end{table}
We see that our RLAGN sample of 382 sources is categorized as 122 FR-I sources, 146 FR-II sources and 114 indeterminate-morphology sources. Comparing the polarization statistics, we see that FR-II sources have more than twice the detection fraction of FR-Is (and that of indeterminate sources). These quantities robustly confirm findings by \citetalias{osul18}, that FR-II radio galaxies are much more likely to be detected in polarization than FR-I sources.    

\begin{figure}
    \centering
    \includegraphics[scale=0.42, trim={1.2cm 0 0 0},clip]{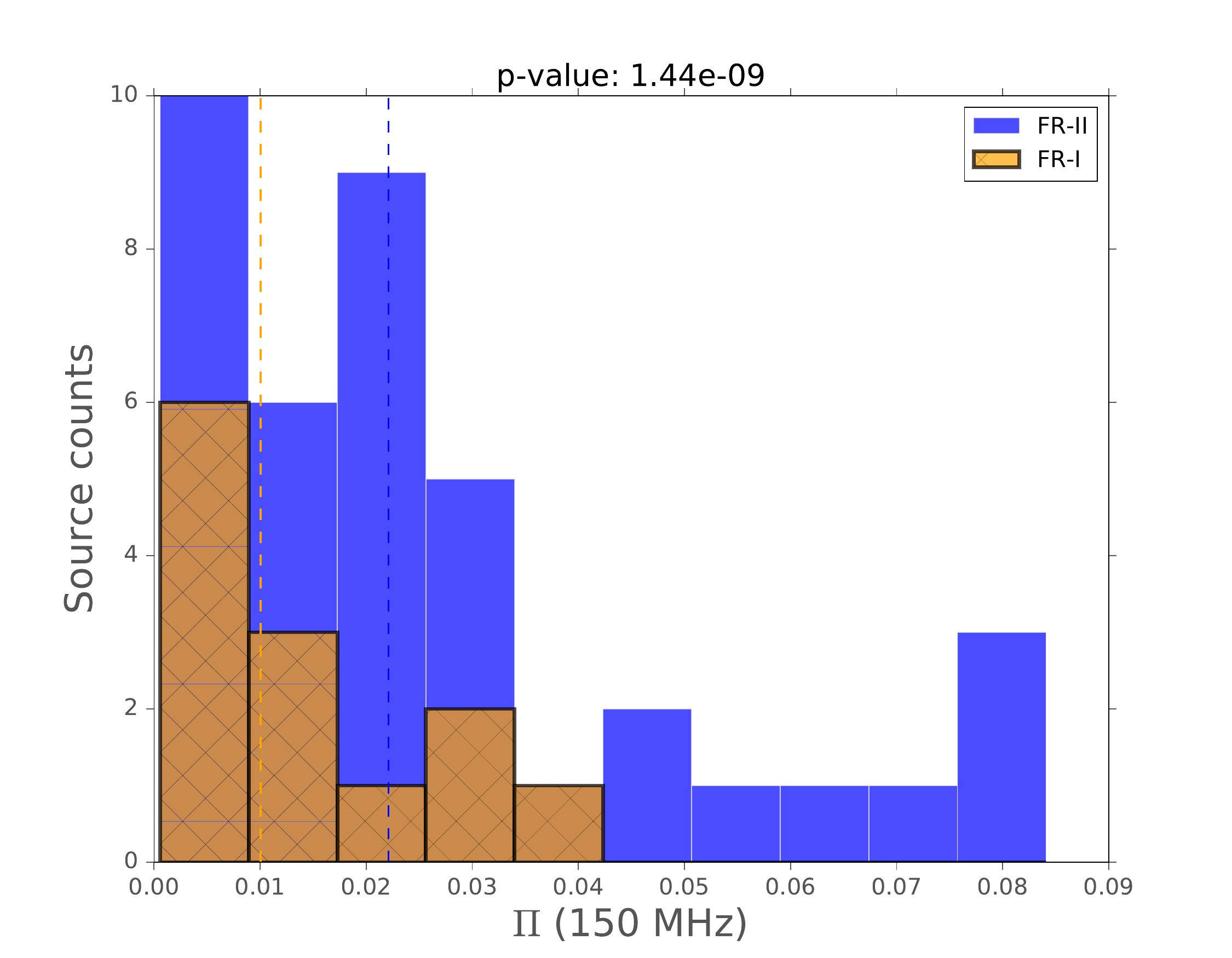}
    \caption{Distribution of fractional polarization between FR-I (hashed orange) and FR-II (blue) radio galaxies. Dashed lines represent median values.}
    \label{fig:Pi_hist}
\end{figure}

We reproduce Figure \ref{fig:flux/size_polflux} in Figure \ref{fig:flux_polflux_morph}, now labelling the sources as FR-I (beige) and FR-II (blue). We see that in general, at a given 150 MHz total flux density, FR-II sources tend to be brighter in polarization than FR-I sources, although with large scatter. In Figure \ref{fig:Pi_hist} we plot the distribution in fractional polarization $\Pi_{150}$ for FR-I and FR-II sources, showing a statistically higher median $\Pi_{150}$ for FR-IIs, where FR-IIs solely dominate at $\Pi_{150}>4$ per cent. This is likely due to the presence of bright hotspots in FR-IIs, that are clearly dominating our statistics, compared to FR-Is mostly with polarized cores. 
\subsection{Host galaxy properties}
The host galaxy properties of RLAGN populations give valuable information on the drivers of AGN activity. It is important to determine if there are differences in detection fractions as a function of host galaxy type. In particular, we test the hypothesis that polarized and depolarized RLAGN at 150 MHz can be driven by the same type of host galaxy.
\begin{figure}
    \centering
    \includegraphics[scale=0.42, trim={0.8cm 0 0 0},clip]{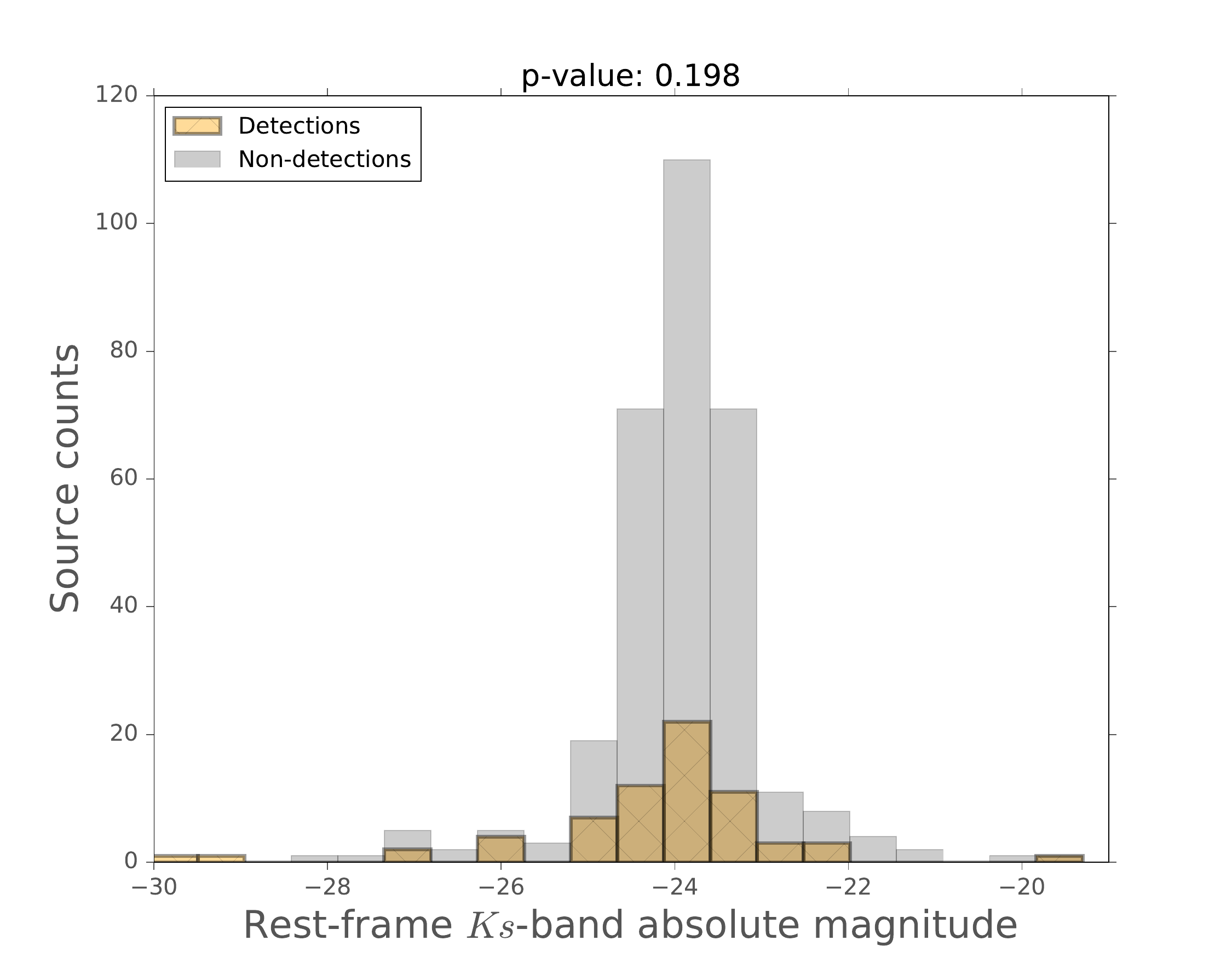}\\
    \includegraphics[scale=0.42, trim={0.8cm 0 0 0},clip]{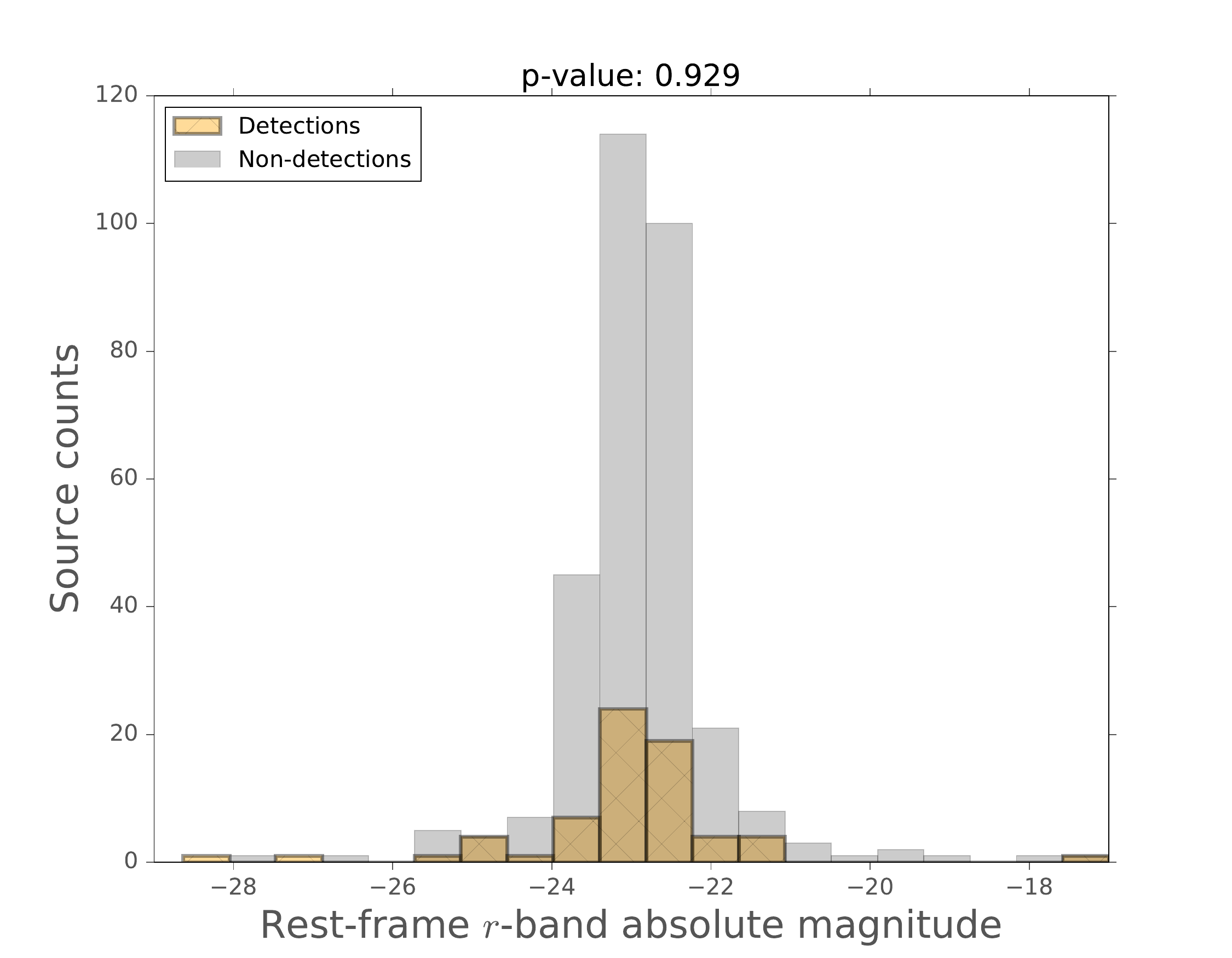}
    \caption{Distributions in rest frame absolute magnitudes in the optical Ks band (top) and r-band (bottom) of the host galaxies of our polarization detected sources (hashed beige) and our non-detected sources (grey).}
    \label{fig:photometry_dist}
\end{figure}

In Figure \ref{fig:photometry_dist} we plot the distributions of host galaxy optical $Ks$ and $r-$ band absolute magnitudes, available in the LoTSS DR1 catalogue, for our polarized (yellow) and depolarized (grey) sources. We see that the distributions in both optical bands are similar -- the $p$ values (quoted above both figures) from a two-sample Kolmogorov-Smirnov (KS) test are both > 0.05, indicating that we cannot reject the hypothesis that both polarized and depolarized subsets have similar distributions, at a confidence level of 95 per cent. The optical and near-IR intrinsic brightness of the host galaxies that drive radio jets in our sample are not in general associated with a detection of polarized emission. 

\begin{figure}
    \centering
    \includegraphics[scale=0.42, trim={0.7cm 0 0 1.2cm},clip]{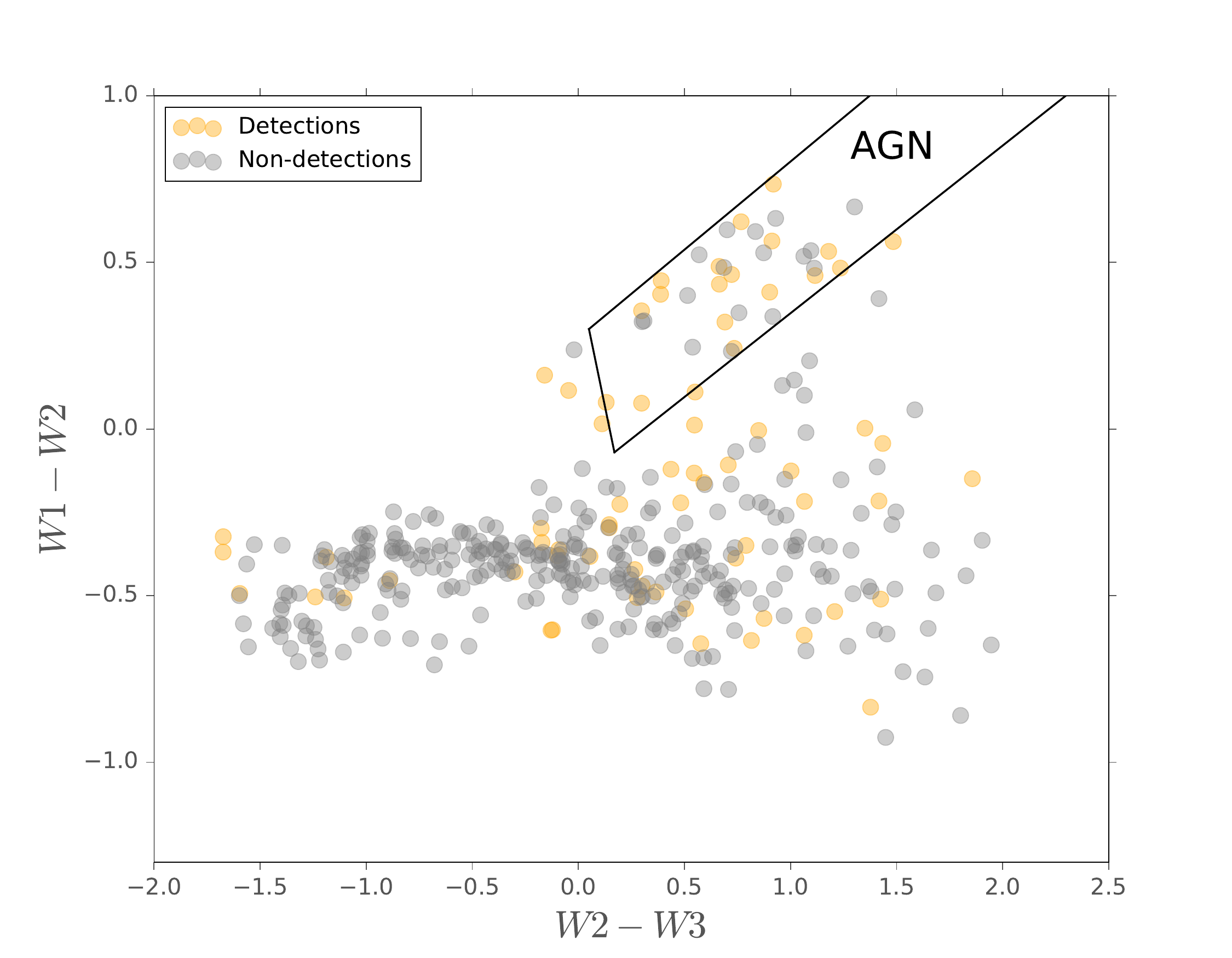}\\
    \includegraphics[scale=0.42, trim={0.7cm 0 0 0.5cm},clip]{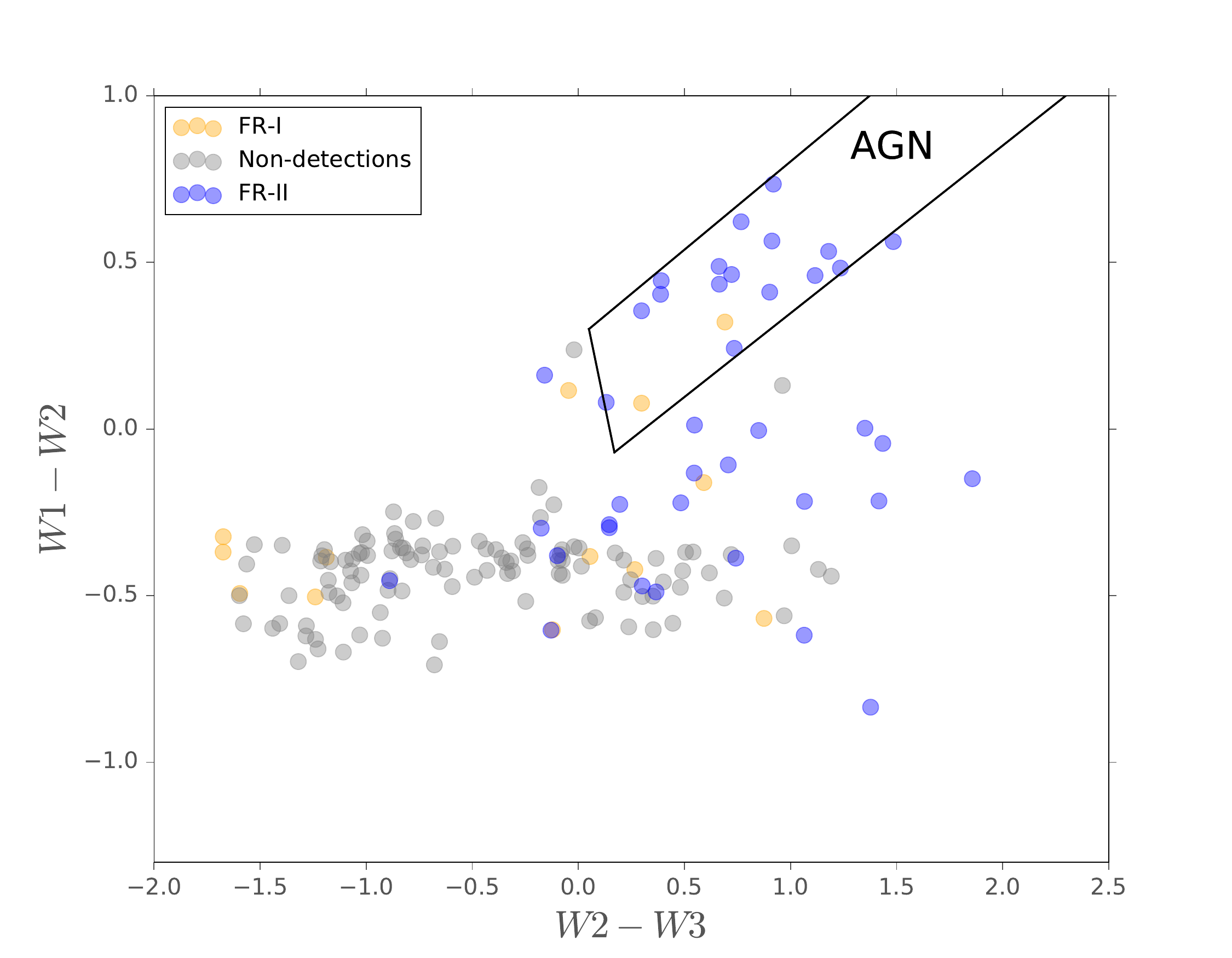}
    \caption{WISE colour-colour diagrams of host galaxies in our sample (top) and in our morphological categories (bottom). Colour coding is the same as for previous figures. Uncertainties are of the order <1\protect{\%} and hence error bars are not shown. The solid lines indicate the `AGN wedge', as defined by \protect\cite{mate12}.}
    \label{fig:wisecolorcolor}
\end{figure}
\begin{figure}
    \centering
    \includegraphics[scale=0.42, trim={0.5cm 0 0 1.2cm},clip]{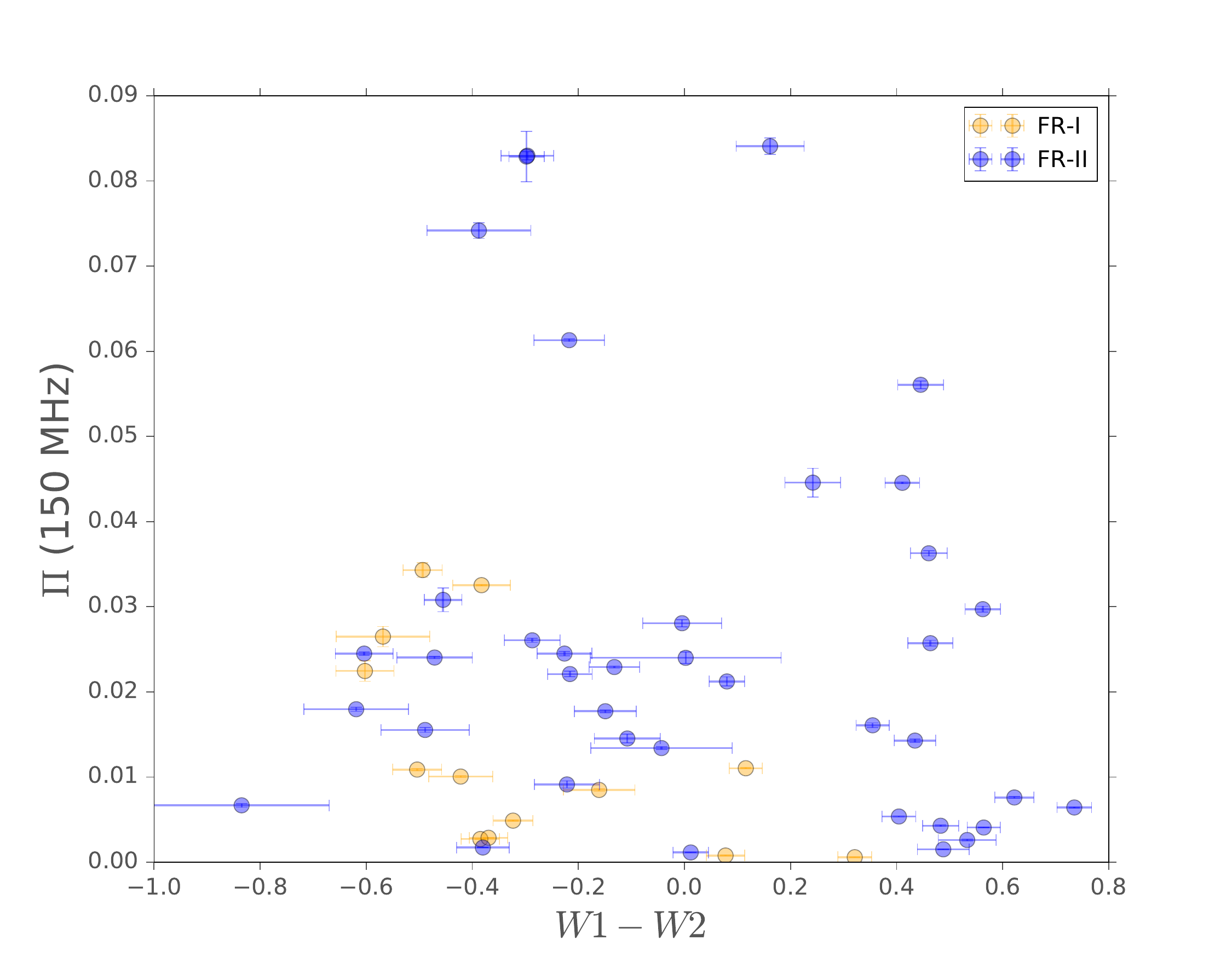}
    \caption{Fractional polarization against host galaxy W1-W2 colour for FR-I (yellow) and FR-II (blue) radio galaxies.}
    \label{fig:w2mw3-pi}
\end{figure}
In terms of polarized radio sources, \cite{banf14} find that quasar-type galaxies typically host sources with lower fractional polarization than quiescent galaxies. \cite{osul15} obtain similar results and find differences in the fractional polarization between High Excitation Radio Galaxies (HERGs) and Low Excitation Radio Galaxies (LERGs), finding that LERGs can achieve higher intrinsic degrees of polarization at GHz frequencies, and \cite{osul17} relate this to LERGs having more intrinsically ordered magnetic fields in the radio plasma. We instead test the low frequency detectability in polarization by comparing the detection rates of HERGs and LERGs using the WISE colour-colour plot, given by WISE mid-infrared apparent magnitudes at 3.4, 4.6, 12, and 22 $\mu$m (W1, W2, W3, W4 bands). While this does not give a direct classification of the HERG and LERG status of a particular galaxy, the majority of LERGs tend to have lower values of W1-W2 and W2-W3, while HERGs tend to have higher values of colour, with higher levels of dust-obscuration (which may arise due to the presence of an optically thick torus surrounding the accretion system). The position of a particular object in the WISE colour-colour plot can therefore give information on the nuclear properties of a galaxy.

In the top panel of Figure \ref{fig:wisecolorcolor} we present the WISE colour-colour plot for our RLAGN sample. We find that the hosts of the polarized sources in our sample have significantly higher values of W1-W2 and W2-W3 than depolarized sources, shown by the larger fraction of polarized objects in the upper-right hand side of the plot (statistics were tested for distributions in W1-W2 and W2-W3 between polarized and depolarized sources, with $p-$values < 0.05). We have over-plotted the `AGN wedge', defined by \cite{mate12}, who select luminous AGN selected in X-rays. This region is typically occupied by obscured AGN and quasars (i.e. HERGs), which, at low redshift, tend to be associated with FR-II radio galaxies \citep{zirb96}. In the bottom panel of Figure \ref{fig:wisecolorcolor} we separate our sample into FR-I and FR-II objects, and it is clearly seen that the majority of the sources with higher values of WISE colours are FR-II sources (while there are still a comparable amount of FR-II sources where FR-Is are situated). Further, in Figure \ref{fig:w2mw3-pi}, we show the fractional polarization $\Pi_{150}$ as a function of the W1-W2 colour. We see that there is a tendency for FR-IIs to have a higher W1-W2 colour than FR-Is, for a given $\Pi_{150}$, though with large overlap. We therefore have insufficient evidence to suggest that quasars with higher values of WISE colour tend to drive radio lobes with higher polarization. Rather, FR-II sources are more likely to be detected due to their more powerful jets than FR-Is, and tend to be hosted by dust-obscured galaxies with a torus, also associated with radiatively efficient accretion in HERGs \citep[][]{lain94,evan06,vand10,gurk14}. While this may present an apparent contradiction with the results of \cite{banf14},\cite{osul15} and \cite{osul17}, in that we have a higher detection rate of polarization for quasar/HERG-type radio galaxies (mostly FR-IIs as seen in Figure \ref{fig:wisecolorcolor}), we emphasise their results are based on the modelled \textit{intrinsic} fractional polarization at high frequencies, while ours is based on observed polarization at low frequencies. Further, and on a related note, it is very likely that LERG FR-Is which may have high intrinsic degrees of polarization are generally depolarized in our sample due to high Faraday dispersions in the centres of their hot gas environments, which are not likely to be detected with LOFAR at 150 MHz (see Section \ref{sect:nvss}). Hence, we suggest that the 150 MHz polarization in radio galaxies is not driven by a particular type of optical brightness or nuclear emission in the host galaxy relative to the population of radio galaxies, but an association exists between polarization and WISE colour due to the association between the FR-class and WISE colour.
\subsection{Frequency dependence}
\label{sect:nvss}
Any frequency dependence of the measured fractional polarization can give evidence for depolarization for any given source. To investigate this, we compared the polarization properties of our sample at 150 MHz with the polarization measured for the same objects with NVSS at 1.4 GHz \citep{tayl09}, as was done by \cite{vane18} for their sample at 4.3 arcmin resolution. \cite{tayl09} present a 1.4 GHz sample of polarized sources in the sky at declination north of $\delta = -40^{\circ}$, which also covers the LoTSS area. The NVSS completeness limit of 2.5 mJy gives an ideal comparison survey, though the restoring beam at 45 arcsec is more than a factor of two larger than that of our sample. \cite{vane18} find that their LOFAR-detected sample contains sources with most of their polarized emission in broad Faraday-thick components (as their sources have higher fractional polarization with NVSS than with LOFAR, with LOFAR only being sensitive to structures $\lesssim$1 rad m$^{-2}$ in Faraday depth). We apply some similar comparisons with our sample to investigate the line-of-sight environments of RLAGN. 

We identify our sources with the \cite{tayl09} sample by using a positional cross-match criterion with a separation limit of 135 arcsec (three NVSS beams). Lower separation limits of one or two NVSS beams resulted in fewer correct cross-matches since our sources are centered on the optical ID, whereas the NVSS polarized sources are centred on the polarized emission, such as hotspots, which can be more than two NVSS beams from the optical ID. To ensure our cross-matching criteria selected the correct NVSS source, we convolved our LOFAR images with a 45 arcsec beam, and compared each image to the cross-matched NVSS source in Stokes I and in polarization, by downloading Stokes IQU cubes using the NRAO postage-stamp server\footnote{\url{https://www.cv.nrao.edu/nvss/postage.shtml}}. We also ensured that sources that did not have an NVSS counterpart were not missed by our cross-matching criteria (i.e. polarized hotspots in NVSS that are more than three NVSS beams from the optical ID). We found that one source (ILTJ105715.33+484108.6) was missed due to its large angular size, and included it as a correct NVSS counterpart. 

Using this method we find 58 NVSS counterparts to our parent RLAGN sample of 382 sources. Out of these sources, 24/58 are detected in polarization in our LOFAR sample, giving approximately a 40\% LOFAR detection rate. However, we note that out of the 67 polarized sources in our sample, there is an absence of polarized NVSS counterparts in 44/67. This is surprising as we would naively expect all 67 LOFAR detections to also be detected with NVSS since the fractional polarization should be higher at higher frequencies and since our sources, being larger than 100 arcsec, are all resolved with NVSS. Beam depolarization could be more prominent in NVSS due to its factor of two larger beam -- comparing the polarized intensity NVSS and LOFAR images at 45 arcsec, we find only one source with significant depolarization in the NVSS images, which we regard as insufficient to explain the lack of NVSS detections. We instead attribute the lack of polarized NVSS counterparts to the fact that LOFAR is more sensitive to steep-spectrum sources relative to NVSS, so that many steep-spectrum sources that suffer little depolarization (so that they are detected at 150 MHz) are undetected in NVSS. As a check, we calculated the expected Faraday dispersion $\sigma_{RM}$ assuming external Faraday rotation using our LOFAR and NVSS polarized counterparts using 
\begin{ceqn}
\begin{equation}
\Pi_{150}/\Pi_{1400} = \exp^{(-2\sigma_{RM}^{2}\lambda_{150}^4 + 2\sigma_{RM}^4\lambda_{1400}^{4})}   
\end{equation}
\end{ceqn}
\citep{soko98}. The average $\sigma_{RM}$ in our sample is 0.24 rad m$^{-2}$. For comparison, the median $\sigma_{RM}$ for 20 double radio galaxies measured by \cite{osul17} is 12.5 rad m$^{-2}$, implying that the sources detected by LOFAR have very little depolarization. Moreover, Equation \ref{equation:polvector} implies that the Faraday dispersion function must be narrow for less depolarization at long wavelengths. 
\begin{figure}
    \centering
    \includegraphics[scale=0.41, trim={1cm 0 0 1cm},clip]{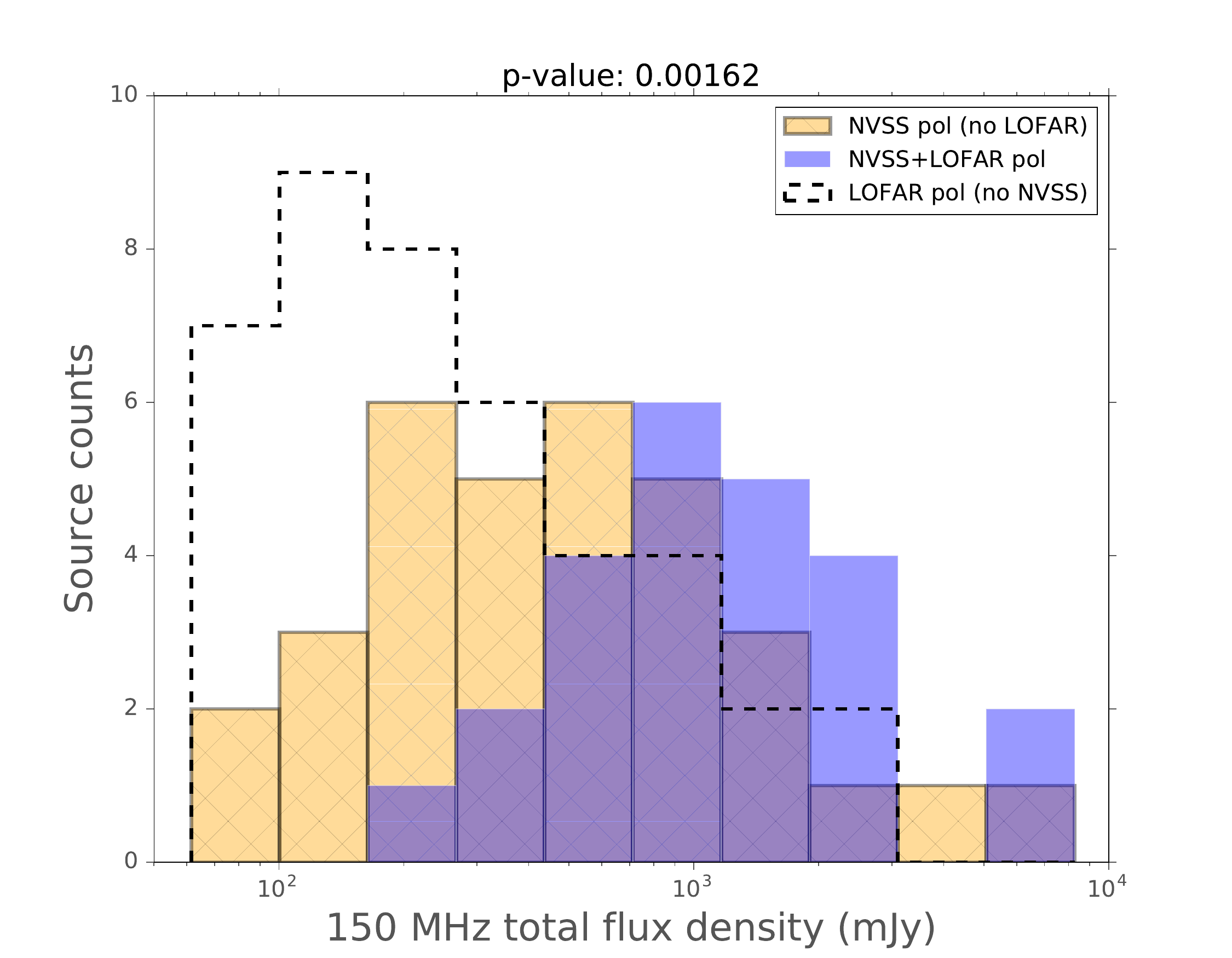}\\
    \includegraphics[scale=0.41, trim={1cm 0 0 0},clip]{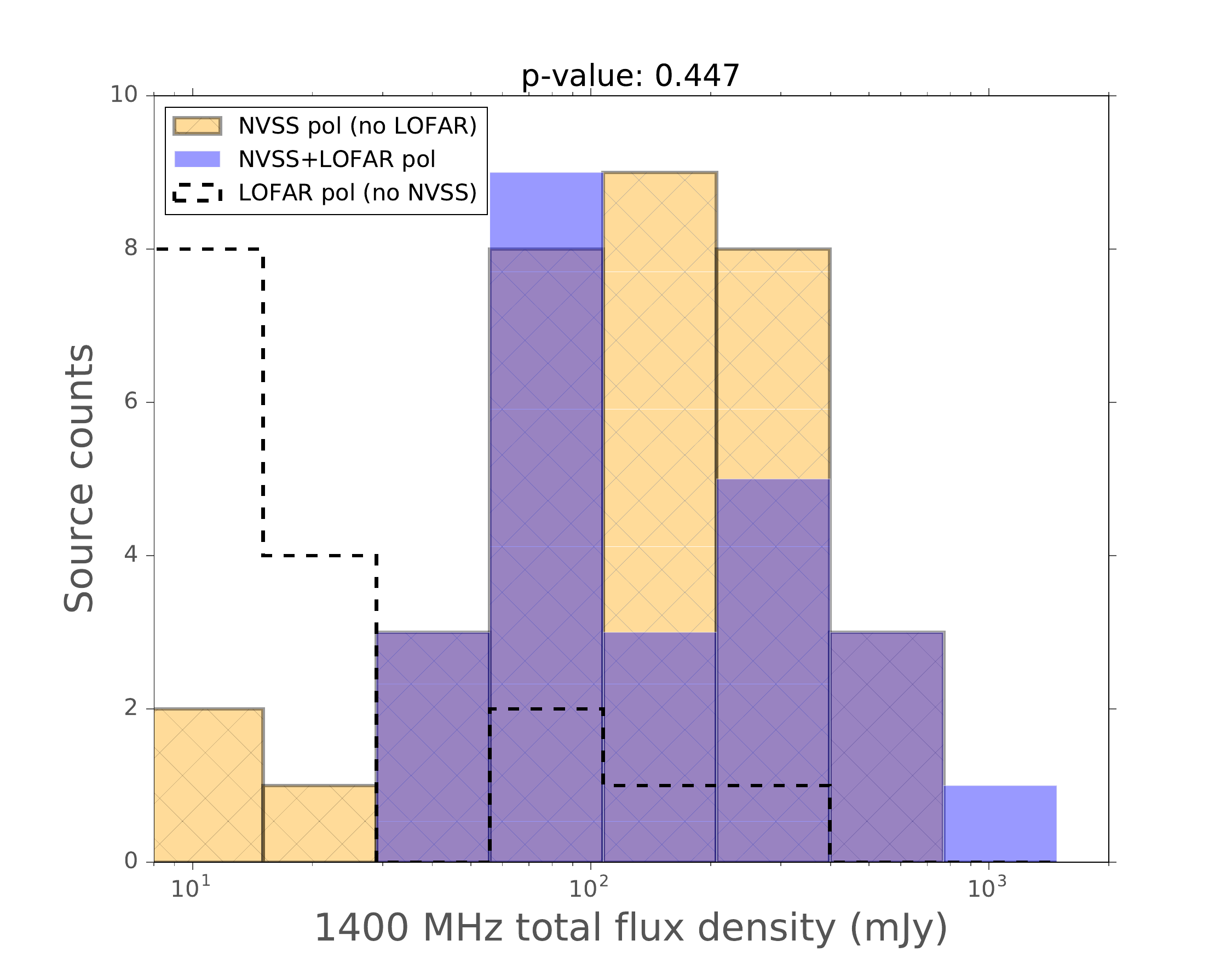}\\
    \includegraphics[scale=0.41,trim={1cm 0.1cm 0 0},clip]{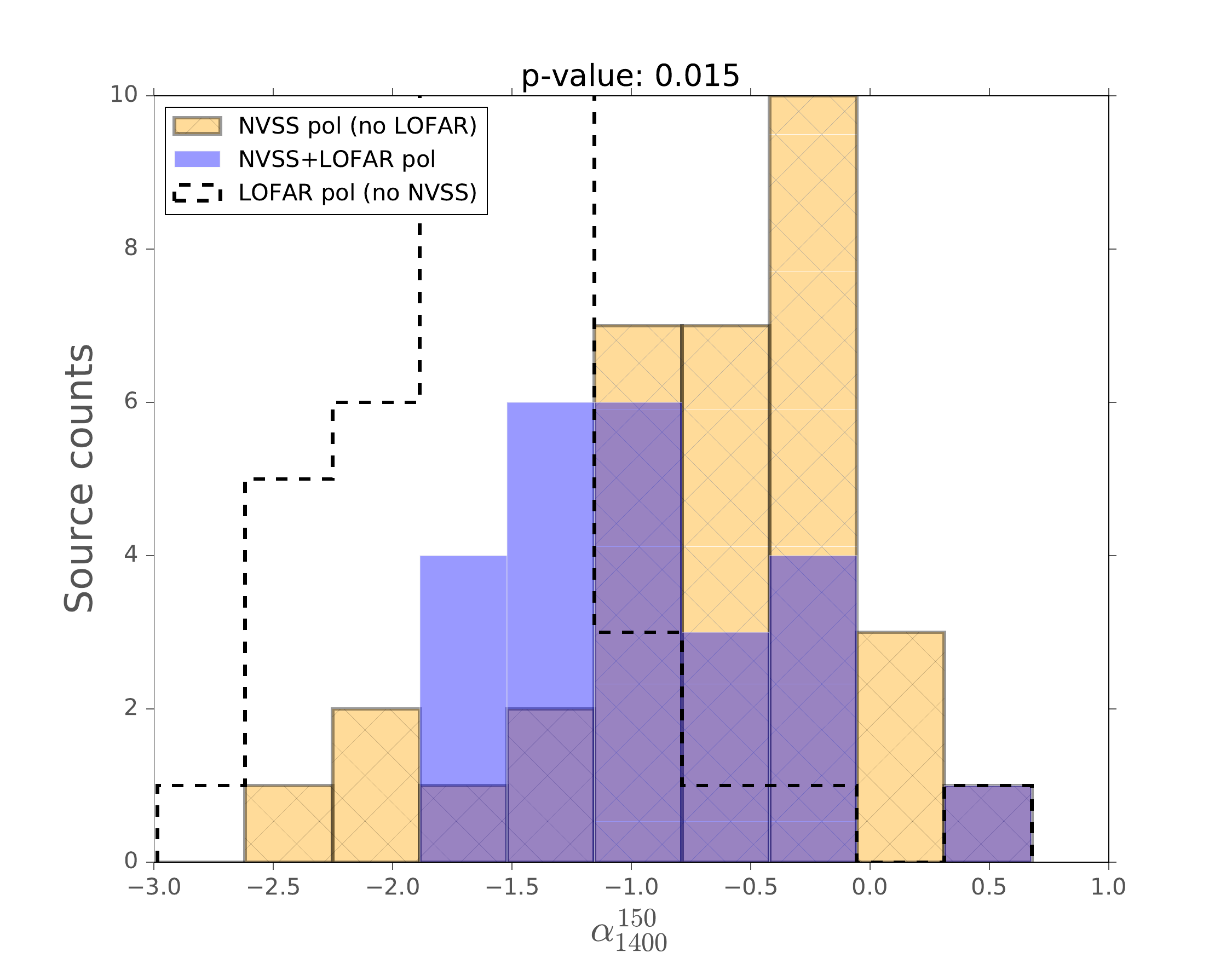}
    \caption{Distributions in total flux density at 150 MHz (top), 1400 MHz (middle) and spectral index (bottom) for our NVSS-LOFAR cross-matched sources. NVSS polarized sources that are not polarized at 150 MHz are shown in hatched beige and those that are polarized at 150 MHz are shown in blue. Sources that are polarized at 150 MHz but not at 1400 MHz are shown in black dashed lines. Note the $p$-values refer to the comparison between blue and hatched beige distributions only. }
    \label{fig:total_flux-nvss-lofar}
\end{figure}

In order to confirm if the lack of polarized NVSS counterparts is due to their steep spectra, we further cross-matched the sources in our sample with no NVSS polarized counterparts (using the same criteria as above) with the NVSS source catalogue \citep{cond98}. In Figure \ref{fig:total_flux-nvss-lofar} we display the distributions in 150 MHz flux density, 1400 MHz flux density and spectral index between those two frequencies (corrected for the difference in beam sizes). The top panel indicates that the polarized NVSS sources which are also polarized with LOFAR (blue) are brighter at 150 MHz than those that are not polarized with LOFAR (beige hatched, note the $p-$value above the figure), as expected. However in the middle panel, there is no statistical evidence for different average flux densities at 1400 MHz, implying LOFAR is more sensitive to steep-spectrum sources at low frequencies. In both panels, the black dashed histogram denotes those sources polarized with LOFAR with no polarized NVSS counterpart, showing that they are fainter than polarized NVSS sources both at 150 MHz and 1400 MHz. In the bottom panel, affirming our earlier inferences, we see that while the average spectral indices of the NVSS polarized sources with LOFAR polarization are steeper than those without LOFAR polarization as expected, the sources polarized with LOFAR but not with NVSS are even steeper (WMU tests performed between all three distributions have $p-$values $\leqslant0.05$). This confirms that the lack of NVSS polarized counterparts is due to their steep spectral indices, so that their flux densities have decreased beyond detection at 1400 MHz, while still being polarized at 150 MHz. Note that the median error in the spectral indices in our sample is $\Delta\alpha^{150}_{1400}=0.11$, taking into account 3$\sigma$ errors on the total flux density measurements.
\begin{figure}
    \centering
    \includegraphics[scale=0.42, trim={0.7cm 0cm 0cm 1.5cm},clip]{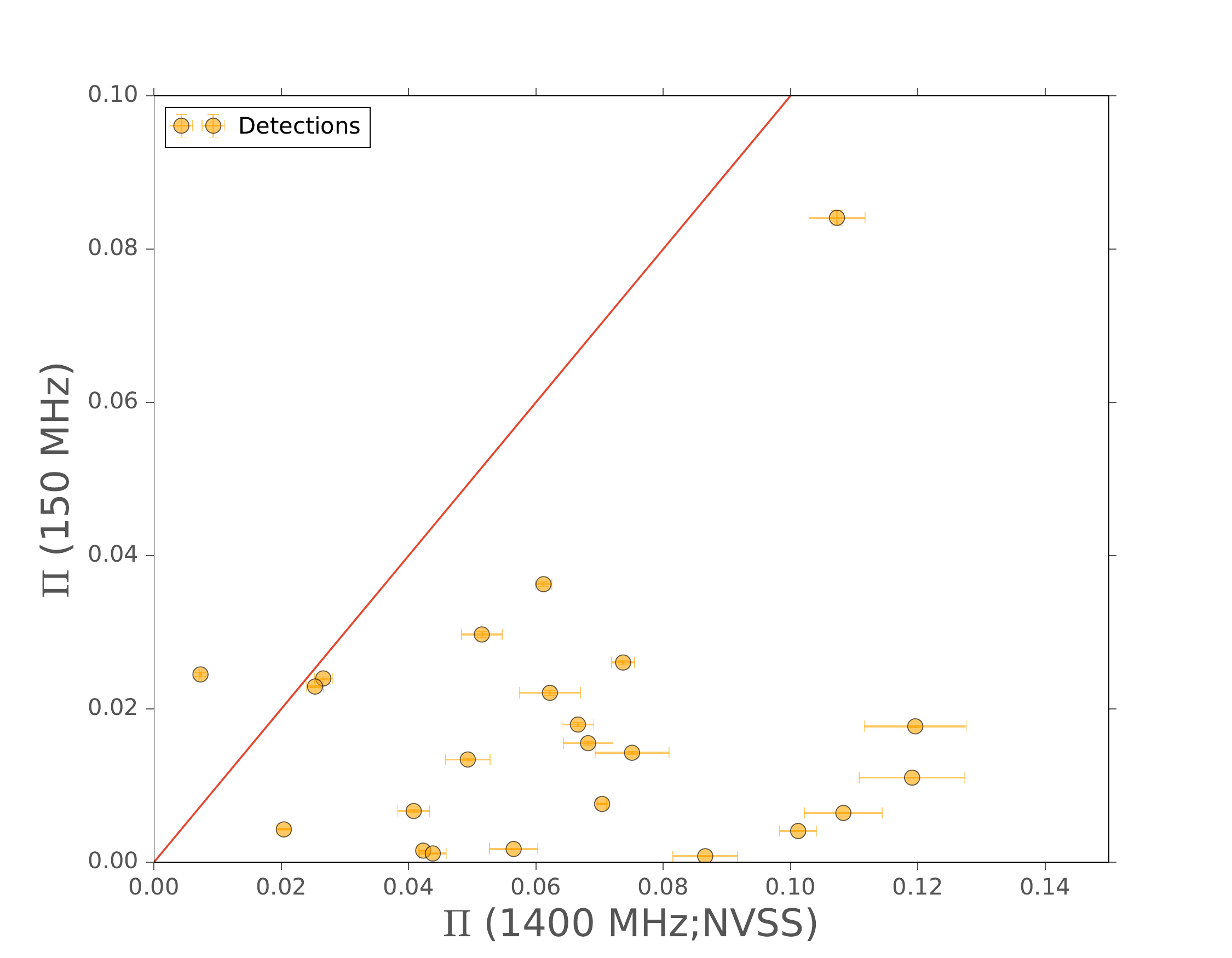}
    \caption{LOFAR fractional polarization at 150 MHz against NVSS fractional polarization at 1400 MHz for 28 sources in our sample with both LOFAR and NVSS detections. The red line shows the line of equality. NVSS error bars are taken from the catalogue of \protect\cite{tayl09}.}
    \label{fig:Pi_1400-Pi_150}
\end{figure}

In Figure \ref{fig:Pi_1400-Pi_150} we plot the fractional polarization at 150 MHz against that at 1400 MHz for our cross-matched sources. We see that all sources, except one, have a lower fractional polarization at 150 MHz, showing depolarization at low frequencies. For the source not depolarized at 150 MHz relative to 1400 MHz (at $\Pi_{150}\sim$0.025), we attribute its higher $\Pi_{150}$ to beam depolarization in NVSS -- upon visual inspection the LOFAR data show resolved components in the lobes which are shown as a single component in the NVSS image.
\subsection{Rotation measure analysis}
In this section we analyse the $RM$s of our sample. The $RM$ arises from the superposition of the line of sight contributions from the Galaxy, the intergalactic medium, the intragroup/intracluster medium and the source itself. We are more interested in the source and local environment properties, and so we subtract the Galactic $RM$ measured at the location of each source in our sample. To do this we positionally cross match the sources in our sample with the Galactic Faraday sky map made by \cite{oppe15}. The pixel size in this map is around 30 arcmin, and hence all our sources are spatially coincident within single pixel regions in this map. We subtract the pixel $RM$ value from the $RM$ value we measure at 150 MHz, as well as from the $RM$s at 1400 MHz from the \cite{tayl09} catalogue, giving Galaxy-subtracted values. For each source we propagate the LOFAR and NVSS $RM$ errors (Section \ref{sect:detection}) with the $RM$ errors catalogued by \cite{oppe15}.

\begin{figure}
    \centering
    \includegraphics[scale=0.42, trim={1cm 0 0 1cm},clip]{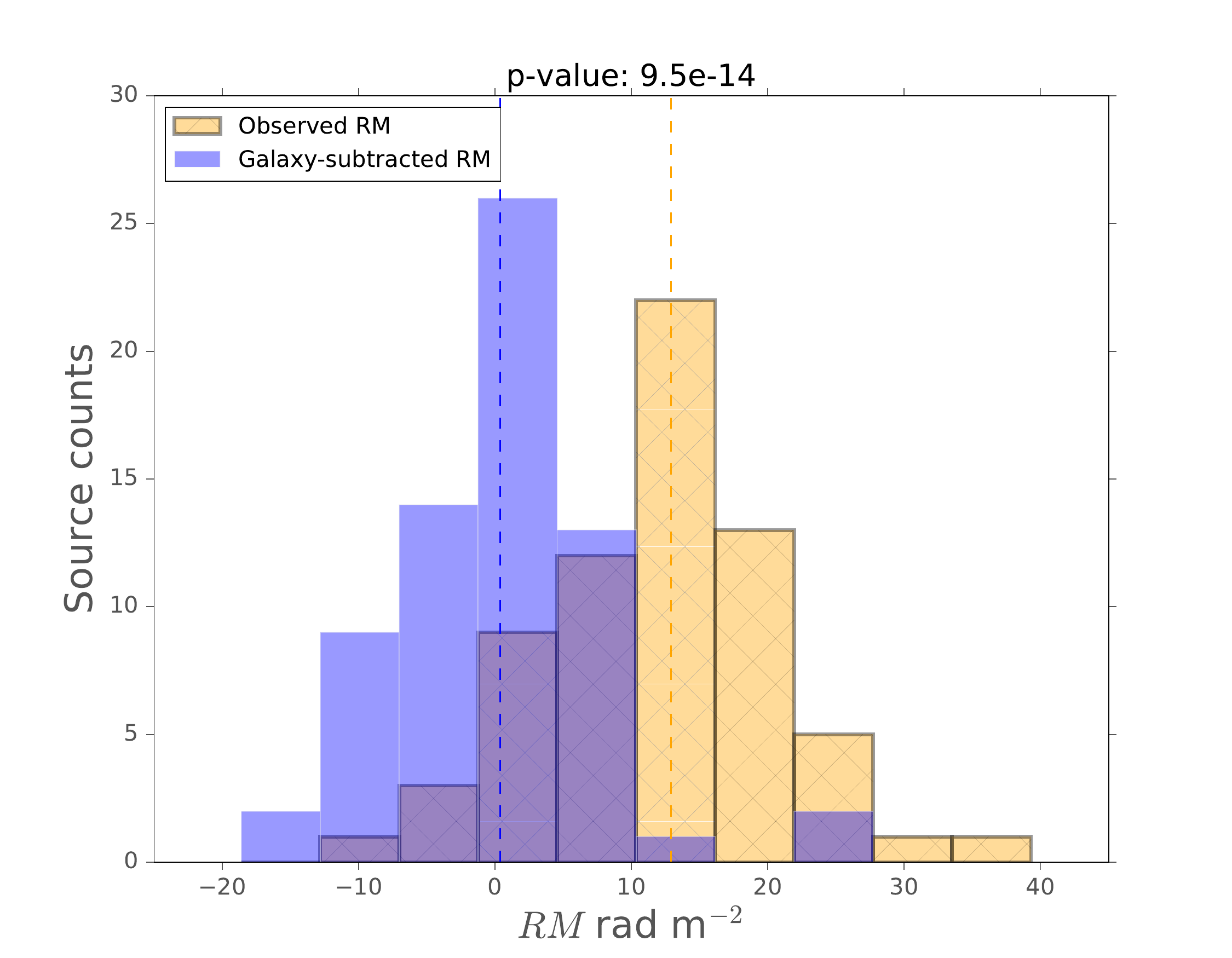}
    \caption{Rotation measure distribution of our sample. The yellow histogram shows our observed rotation measures, and the blue histogram shows the same data after being corrected for the Galactic contribution. Dashed lines represent medians.}
    \label{fig:rm_hist}
\end{figure}

In Figure \ref{fig:rm_hist} we plot the distribution in observed $RM$ at 150 MHz and the Galaxy-subtracted $RM$ for our sample. The average $RM$ in our sample is positive, with a mean\footnote{Uncertainties quoted are the standard errors of the mean} observed value of $+12.6\pm0.15$ rad m$^{-2}$, and the mean Galaxy subtracted value is $+1.0\pm 0.14$ rad m$^{-2}$ (with an rms of 7.56 rad m$^{-2}$). This shows that the bulk of our sample do not have large magnitudes of $RM$ from non-Galactic Faraday screens. It is likely then that the dominant contributor to our $RM$s comes from multiple Faraday screens with multiple magnetic field reversals, which acts to reduce the $RM$ magnitude, or that our sources are preferentially located in low density environments which lead to lower depolarization, or a combination of both factors.  

\begin{figure}
    \centering    
    \includegraphics[scale=0.42, trim={1cm 0 0 0},clip]{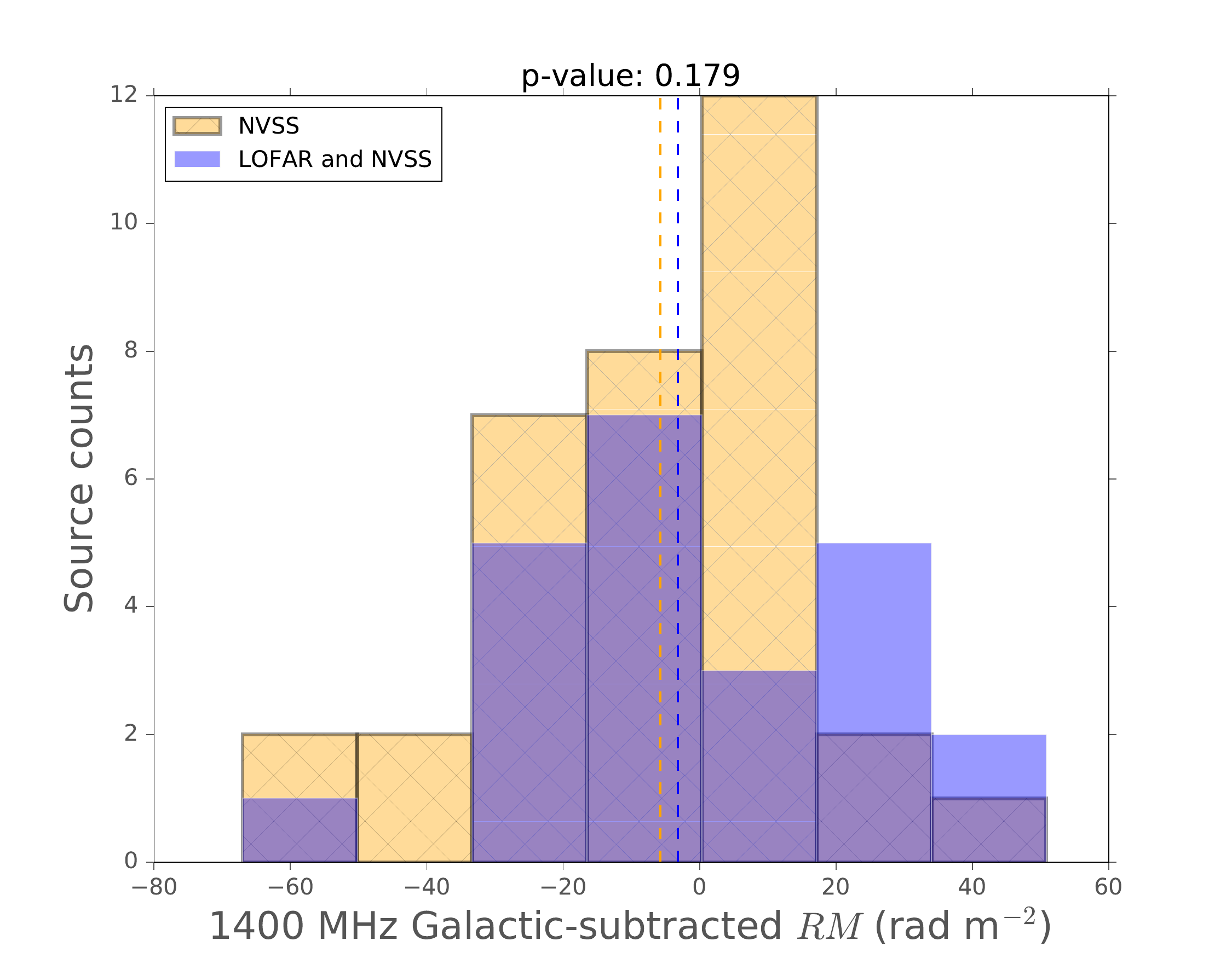}\\
    \includegraphics[scale=0.42, trim={1cm 0 0 0},clip]{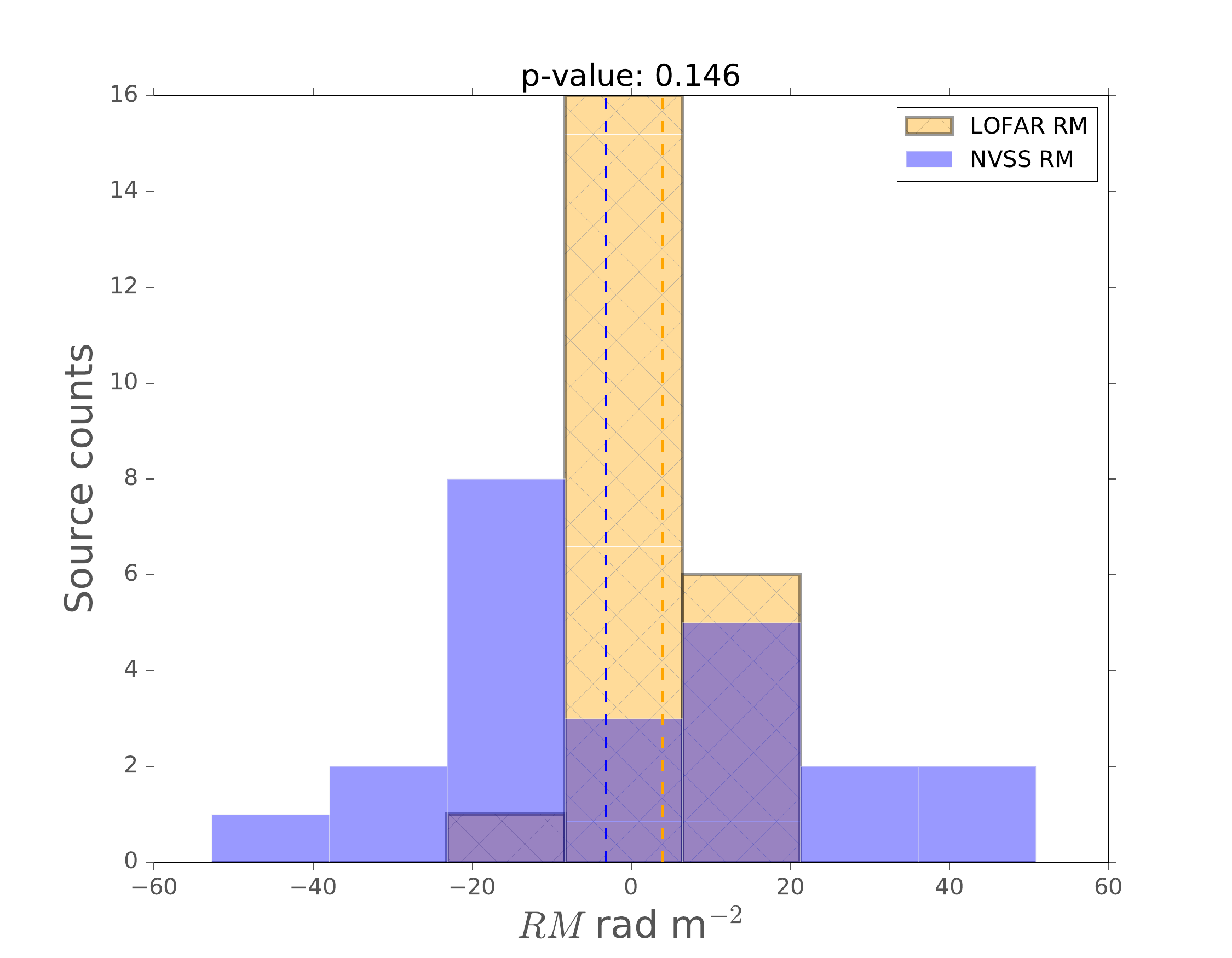}
    \caption{$RM$s for our LOFAR-NVSS cross-matched sample. Top panel shows the 1400 MHz $RM$s between those polarized with LOFAR (blue) and those depolarized with LOFAR (hatched beige). Bottom panel shows the LOFAR (hatched beige) and NVSS (blue) $RM$s for the sources polarized with LOFAR. Medians are shown as dashed lines.}
    \label{fig:rm-nvss-lofar}
\end{figure}
\begin{figure}
    \centering
    \includegraphics[scale=0.42, trim={0.5cm 0 0 0},clip]{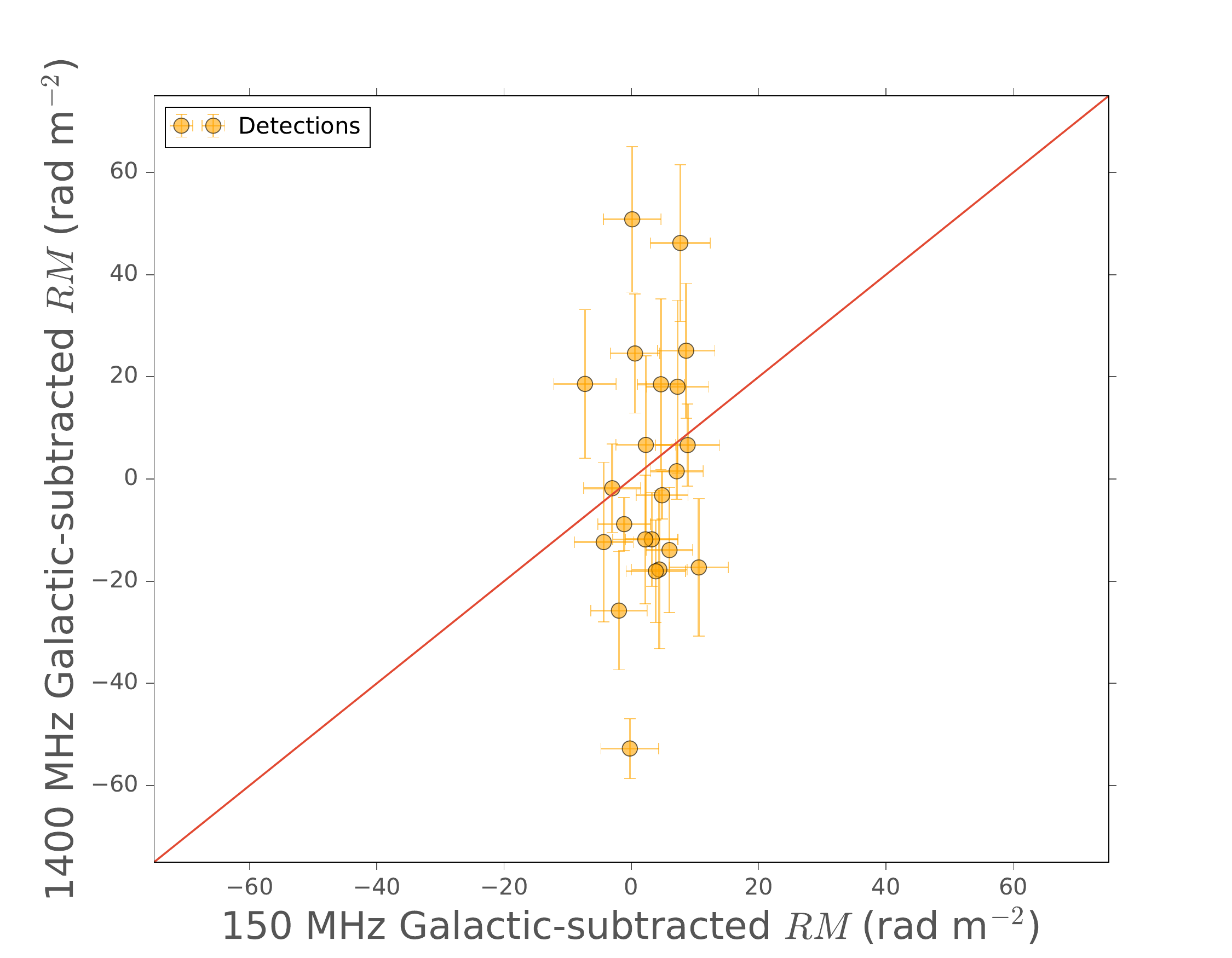}
    \caption{150 MHz $RM$ against 1400 MHz $RM$ for our cross-matched sample, after Galaxy $RM$ subtraction. Calculation of error bars are stated in Section \ref{sect:detection}. The red line is the line of equality, and not a regression line fit to the data.}
    \label{fig:RM_1400-rm_peak}
\end{figure}

We also compare the distributions in Galaxy-subtracted $RM$ at 150 MHz and 1400 MHz in Figure \ref{fig:rm-nvss-lofar}. In the top panel we show the 1400 MHz $RM$s of those sources that are polarized (blue) and depolarized (hatched beige) with LOFAR. We see that there are similar distributions and medians in 1400 MHz subtracted $RM$s between LOFAR polarized and depolarized sources, with a similar range. This suggests, on a statistical level, that the depolarization of RLAGN at low frequencies is not solely driven by the magnitude of Faraday rotation, as expected. Rather, it is the large spatial and/or line of sight dispersion of Faraday rotation that causes depolarization at low frequencies, and since small-scale variations (i.e internal Faraday rotation) will cause depolarization, it is more likely that the sources we detect with little $RM$ variation across the detected regions have significant contributions from the foreground medium local to the source (i.e an ICM). Moreover, as shown by the bottom panel of Figure \ref{fig:rm-nvss-lofar}, the Galaxy subtracted $RM$s have similar peaks between LOFAR and NVSS data for those sources polarized with LOFAR, while there is a much larger spread in NVSS $RM$ as expected due to the lower $RM$ resolution. This also shows a general consistency between Faraday screens in the foreground responsible for depolarization in our sample. These results imply that the absence of polarized emission with LOFAR for these sources is mostly driven by the combination of their total flux density at 150 MHz and their individual Faraday-rotating media. In Figure \ref{fig:RM_1400-rm_peak}, where we plot the $RM$s at 1400 MHz against those at 150 MHz for the cross-matched LOFAR polarized sources, the red diagonal line of equality shows that there is a general lack of a correlation, even with the large errors (calculated using propagation of errors during subtraction). The lack of a clear correlation implies that we may be tracing components of different Faraday screens, rather than different components of the same Faraday screen. Since we infer that our $RM$s are sensitive to external Faraday screens such as the ICM, we may predict $RM$s toward realistic ICM environments around RLAGN and compare them with our observations. 

\subsection{Environment modelling}
X-ray data have long been used to determine environmental properties of radio galaxies \citep[e.g.][]{cros08,cros11,hard16}. However, these are difficult to obtain for large samples, particularly in probing the high-redshift regime ($z>1$). $RM$ maps can provide an independent method of determining the line-of-sight environmental properties that may be used to infer the environments of large samples of radio galaxies. While it is difficult to infer the environment from the $RM$ without information on the electron density, magnetic field strength or the magnetic field reversal scale (the typical physical length between magnetic field direction reversals in the line of sight), we may instead use a model to predict $RM$s based on realistic radio galaxy environments. In particular, assuming purely external Faraday rotation due to a local environment, with plausible assumptions about thermal gas distributions, magnetic field strengths, reversal scales and geometry, we test whether it is possible to reproduce the $RM$ distribution that we observe with LOFAR. If such a distribution is obtained (by way of a two-sample KS test between modelled and observed $RM$ distributions), we may then compare the physical environmental properties of models that LOFAR is sensitive to and those that LOFAR is not sensitive to. We also compare the fraction of models that have $RM$s in our observed range against our polarized detection fraction, and discuss any discrepancies between the two.
\subsubsection{RM prediction model}
\begin{figure}
    \centering
    \includegraphics[scale=0.40, trim={0.5cm 0.1cm 0cm 0.5cm},clip]{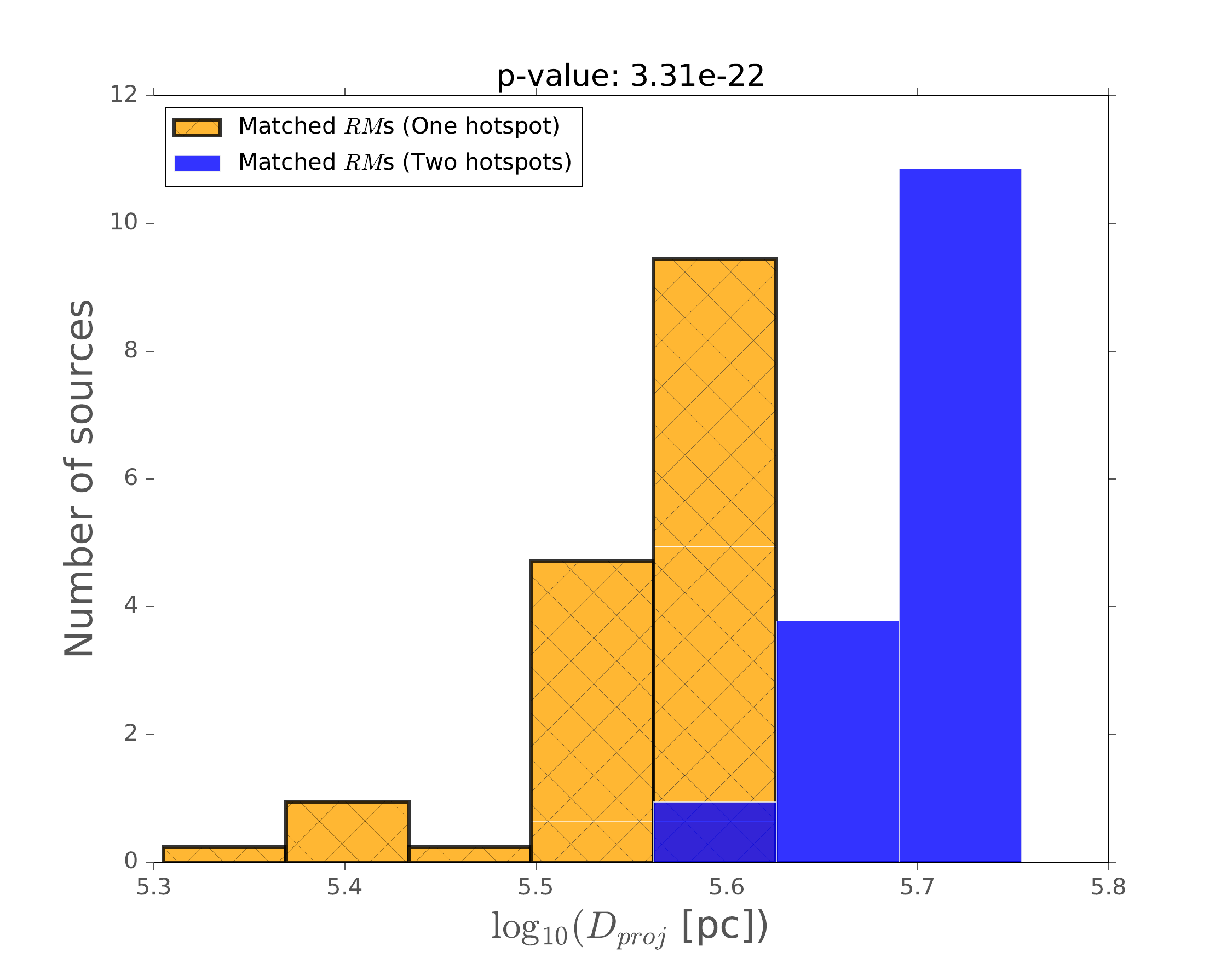}
    \caption{Distribution of projected physical distance of the polarized hotspot(s) from the centre of the ICM/IGM for sources with one polarized hotspot (orange hatched) and polarized hotspots in both lobes (blue).}
    \label{fig:hotspot-hist}
\end{figure}

We create an analytic model which predicts $RM$s using Equation \ref{eq:rm}, which relies on the electron density $n_e$ and the magnetic field $\vec{B}$ through the line of sight toward each polarized source in our sample. A calculation of these properties requires knowledge of the physical environment of each source. Recent studies have shown that the radio lobe properties can be reliable indicators of the ICM pressure at a fixed distance \citep{ines17,cros17}, but such associations based on large samples of the overall RLAGN population have large uncertainties when predicting the properties of any given source. Instead, to determine the physical information needed to predict $RM$s, we draw cluster/group masses from a distribution appropriate for radio galaxy environments. We also include prescriptions for the magnetic field reversal scale (which affects whether the incremental $RM$ is positive or negative) and the orientation of the radio source (which affects the line of sight path length), both of which are unknown, but we use appropriate probability distributions for these input parameters and sample them as a monte carlo simulation. Our $RM$ prediction model is as follows:
\begin{itemize}
    \item We generate distributions of group/cluster masses using the mass function of \cite{gira00}, who show a good agreement between a single Schechter function at $z=0$ and the local mass function for groups and clusters. We generate a Schechter function (with $\alpha = -1.5$) for group/cluster masses in the range $10^{13} - 10^{15} M_{\odot}$, which gives a bias towards masses of groups/clusters which tend to host RLAGN based on optical and X-ray studies \citep{hili91,hard99_environment,best04,ines15}. For each polarized source in our sample, we draw a sample of 1000 values from this function.
    \item We assume an equivalence between the group/cluster mass and $M_{500}$, the mass enclosed in a sphere within which the mean density is 500 times the critical density at $z=0$. For each $M_{500}$ for each source, we determine a radial pressure profile $p(l)$ of the environment, parameterised by $M_{500}$, using the universal pressure profile of \cite{arna10}. The physical size of the pressure profile is determined by calculating the distance from the polarized source at its redshift to $z=0$. 
    \item We determine a density profile $n_{e}(l)$ for each environment by scaling the pressure profile with a single temperature $kT=5.0\times\left(M_{500}[M_{\odot}]/3.84\times10^{14}\right)^{1.0/1.71}$ keV, based on the empirical relationship determined by \cite{arna10}.
    \item For each model we assume a central peak magnetic field strength $\vec{B_0}$, following the prescriptions in the numerical radio galaxy simulations by \cite{hard14}, of $|\vec{B_0}| = 7\sqrt{kT [\text{keV}]/2}$ $\mu$G \citep[calibrated by observations of groups and clusters, see  e.g.][]{guid12}, with a randomly chosen direction (positive or negative). We then determine the magnetic field profile $\vec{B(l)}$ by scaling peak field strength with the density profile using $B(l)\propto n_e(l)^{\gamma}$, where we use $\gamma=0.9$, as found by \cite{dola01} and \cite{dola06} using correlations between rms $RM$s and X-ray surface brightnesses for groups and clusters. We then account for the fact that we only consider the line of sight component of the total magnetic field, so that $\vec{B(l)}_\parallel = \vec{B(l)}/\sqrt{3}$.
    \item We perform the integral $RM_{\text{model}} = 0.812\int^{\text{0}}_{L'} n_e \vec{\bm{B}}_\parallel \text{\textbf{d}}\vec{\bm{l'}}$ incrementally for each model to calculate the total $RM$, where $L'$ is the distance from the polarized emission (either core or hotspot(s) for our sources) to the observer. We visually classified sources as having either; one polarized hotspot, two polarized hotspots or a polarized core. We assume in this model that the AGN is located at the centre of the ICM/IGM, and hence the location of a hotspot is half the linear size of the source in projection from the centre of the ICM/IGM, from where the radial profiles $n_{e}(l')$ and $\vec{B(l')}_\parallel$ are taken. For core-polarized sources we assume a jet with a polarized hotspot at an arbitrarily small distance of 1 pc from the center of the environment, from where the profiles are taken. Due to spherical symmetry of the environment the choice of hotspot (east or west), for core-polarized sources and for sources with one detected hotspot, does not change our results. For models with two polarized hotspots, we calculate the $RM$ from both hotspots and take the mean, as is done for our observations. In Figure \ref{fig:hotspot-hist} we display the distribution of projected physical distances of the polarized emission from the cluster/group centre (i.e half the projected physical size) for the sources in our sample. Note we do not plot the core-polarized sources as their polarized emission has been fixed at 1 pc from the centre of their environments. We immediately see that the sources where both hotspots are polarized are significantly larger (in projection) than sources with a polarized hotspot in one lobe (note the $p$-value from a WMU test in the figure heading). This implies that detecting polarized hotspots in both lobes requires larger sources where, assuming an environment with radially decreasing density (as in our model), the hotspots are located in a less dense medium where the effects of Faraday rotation are less severe. 
    \item The values in the integral above depend on the field reversal scale and orientation $\theta$. We sample a uniform distribution of scales between $10^3$ pc and $10^6$ pc. The choice in field reversal scales were chosen so that they sample the range of scales consistent with observations of groups and clusters \citep[e.g. $\sim 10^4$pc;][]{lain08} and with cosmological magnetohydrodynamic simulations of clusters \citep[e.g. 1 Mpc;][]{dola02}. In each $RM$ integral calculated, the initial sign of $\vec{B}_\parallel$ is changed every time the incremental path length in the integral reaches the  reversal scale. For the orientation angle $\theta$ we draw 1000 values from the distribution $p\left(\theta\right)=1/2\sin\left(\theta\right)$ within the range $0^{\circ}\leqslant\theta\leqslant180^{\circ}$, with $0^{\circ}$ being perpendicular to the plane of the sky and towards the observer and $180^{\circ}$ being away from the observer.
    \item We then truncate each resulting distribution of modelled $RM$s to lie only in the range of those that we sample through $RM$ synthesis: $-150\leqslant RM (\text{rad m}^{-2})\leqslant +150$. This allows the model to return $RM$s that LOFAR can be sensitive to in our analysis.
\end{itemize}
There are caveats to this model which we address before describing our results. The first is in the use of a uniform distribution of field reversal scales, the values of which are relatively unconstrained for the environments around the population of radio galaxies, though we reiterate that they are in approximate agreement with the few studies that do constrain them. Another caveat is that our observations have a finite beam size, whereas we have modelled our sources through single line(s) of sight towards a hotspot(s) with infinite resolution. This is likely to be a very minor affect on our results as the polarized emission we observe is predominantly unresolved, and we take pixel-averaged values as the observed $RM$ within the beam. The final caveat is that we have assumed all sources lie in the geometric centres of their environments. This is supported observationally by studies finding that the majority of X-ray-selected clusters host a central RLAGN \citep[e.g.][]{magl07,best07}, though it is possible that a small number of our sources do not lie in the geometric centers of their environments. In general, while our results (discussed below) are clearly related to our choice of input distributions, we use observationally calibrated results where possible with currently available data. Furthermore our choice of 1000 values drawn from each unknown parameter distribution was made to ensure the models are stochastic.

\begin{figure}
    \centering
    \includegraphics[scale=0.45, trim={0cm 0cm 1cm 0.7cm},clip]{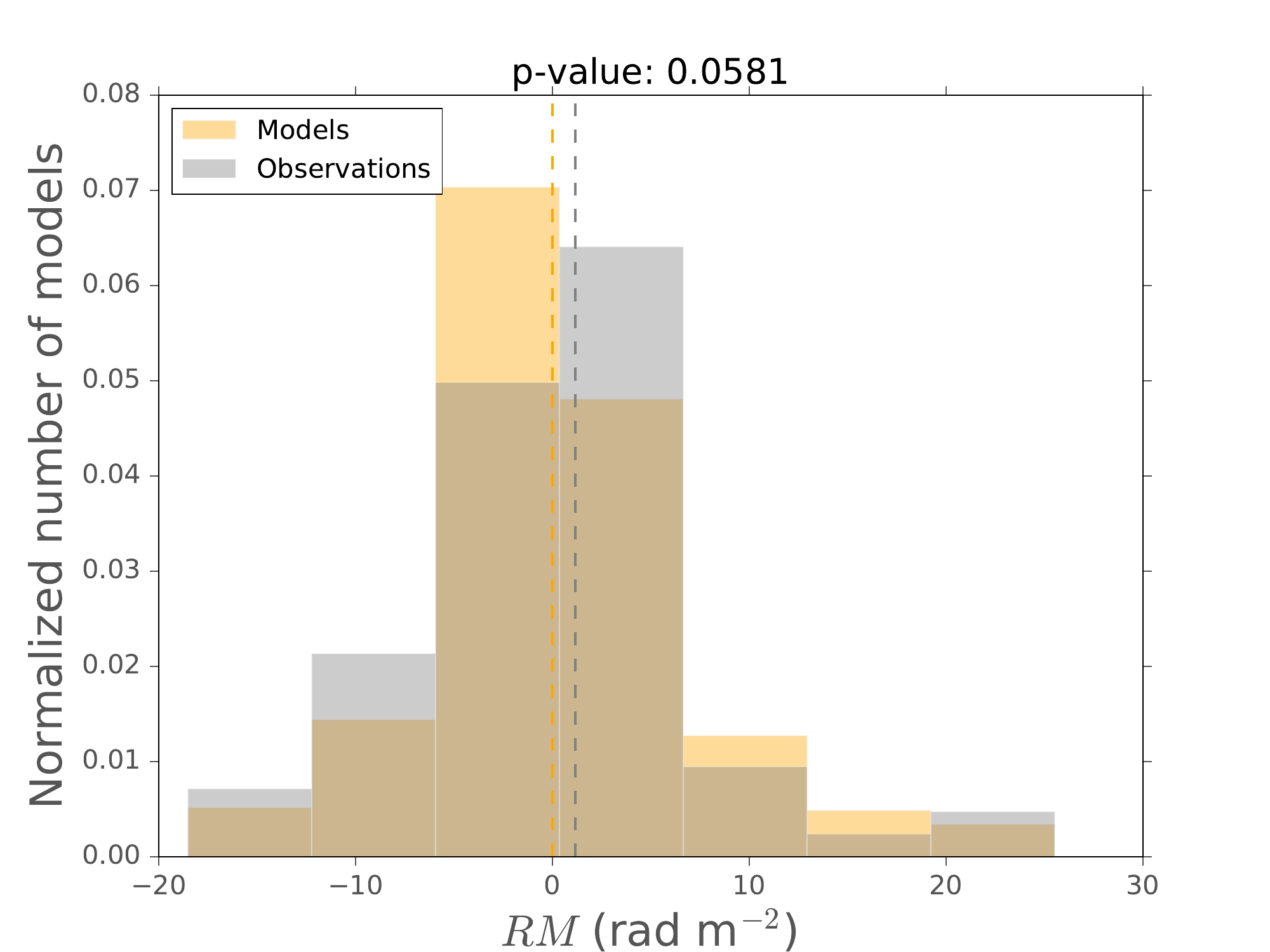}\\
    \includegraphics[scale=0.45, trim={0cm 0cm 1cm 0.7cm},clip]{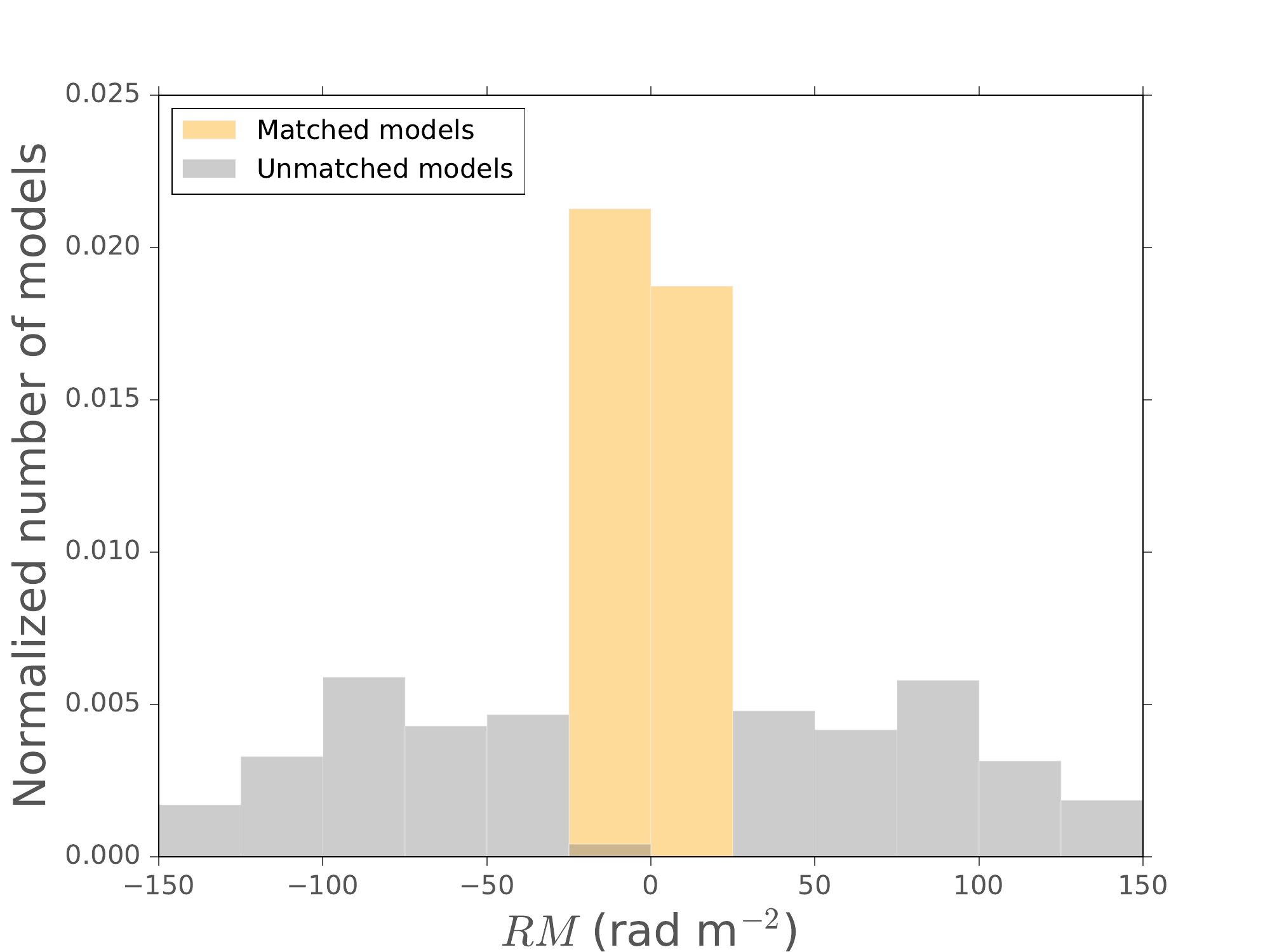}
    \caption{Top: normalised $RM$ distributions of our matched models (yellow) and our observations (grey). Note the superposition of grey and yellow gives a brown colour. Dashed lines  indicate median values. Note that the normalised distribution is such that the integral of the distribution is equal to one. Bottom: normalised $RM$ distributions of our matched (yellow, as in top panel) and unmatched (grey) models.}
    \label{fig:rm_models_obs}
\end{figure}

We separated our models into those which lead to $RM$s within the range of the observed Galaxy-subtracted $RM$ distribution of our sample (`matched' $RM$s; $-20\leqslant RM(\text{rad m}^{-2})\leqslant +25$, see Figure \ref{fig:rm_hist}) and those that are not (`unmatched' $RM$s; outside our observed range but within those that we sample with $RM$ synthesis and that LOFAR is sensitive to, i.e. $-150\leqslant RM (\text{rad m}^{-2})\leqslant +150$). In terms of the model statistics, we have $\sim$61000/67000 models within our $RM$ synthesis range, of which $\sim$32000 are within our observed range, i.e 50 per cent of our models are matched to our observed $RM$ range. Given our observed detection fraction of 18 per cent, the model predicts a factor of three higher detection fraction in the $RM$ range that we observe. We partly associate this with our selection bias for our sample: our models have a matched-$RM$ fraction of $\sim 58$ per cent for core-polarized sources, compared to our observed detection fraction for such sources of 3.4 per cent. While in reality these are mostly based on the FR-I radio galaxies in our sample, such modelled $RM$s would also come from polarized blazars (e.g. as found by \citealt{osul18} in LoTSS), which have been excluded from our sample through our selection of large angular size (>100 arcsec) sources. Hence, the incompleteness in our sample removes polarized sources that we may detect, whereas our models do not take into account any flux or angular size limit. However, more importantly, core-polarized sources in our models that have matched $RM$s produce similar $RM$s based on \textit{all} orientation angles (since the 1 pc distance of their polarized emission at any orientation from the centre of the environment profile does not significantly affect the final aggregated $RM$), our models are already biased towards a very high matched fraction. Further, since FR-I sources (which tend to have polarized cores) live in rich environments relative to FR-IIs, it is possible that the Schechter function we use for all models underestimates the cluster/group masses for FR-Is as a population. 

In top panel of Figure \ref{fig:rm_models_obs} we compare the resulting distribution of $RM$ for matched models and our observations. We see a fairly good agreement between the distributions, with a KS test between both distributions having a $p$-value of $\sim 0.05$. Given this  statistical similarity, we can analyse the observable and physical properties of the matched models to give inferences of the polarization detectability at low frequencies. For completeness we also show the distributions between matched and unmatched $RM$s for our models (bottom panel of Figure \ref{fig:rm_models_obs}), showing a clear peak for models in our observational range and a strong decrease in model counts beyond this range, highlighting that our model inputs are appropriate in predicting $RM$s that we observe with LOFAR.

\begin{figure}
    \centering
    \includegraphics[scale=0.47, trim={0cm 0cm 1cm 1cm},clip]{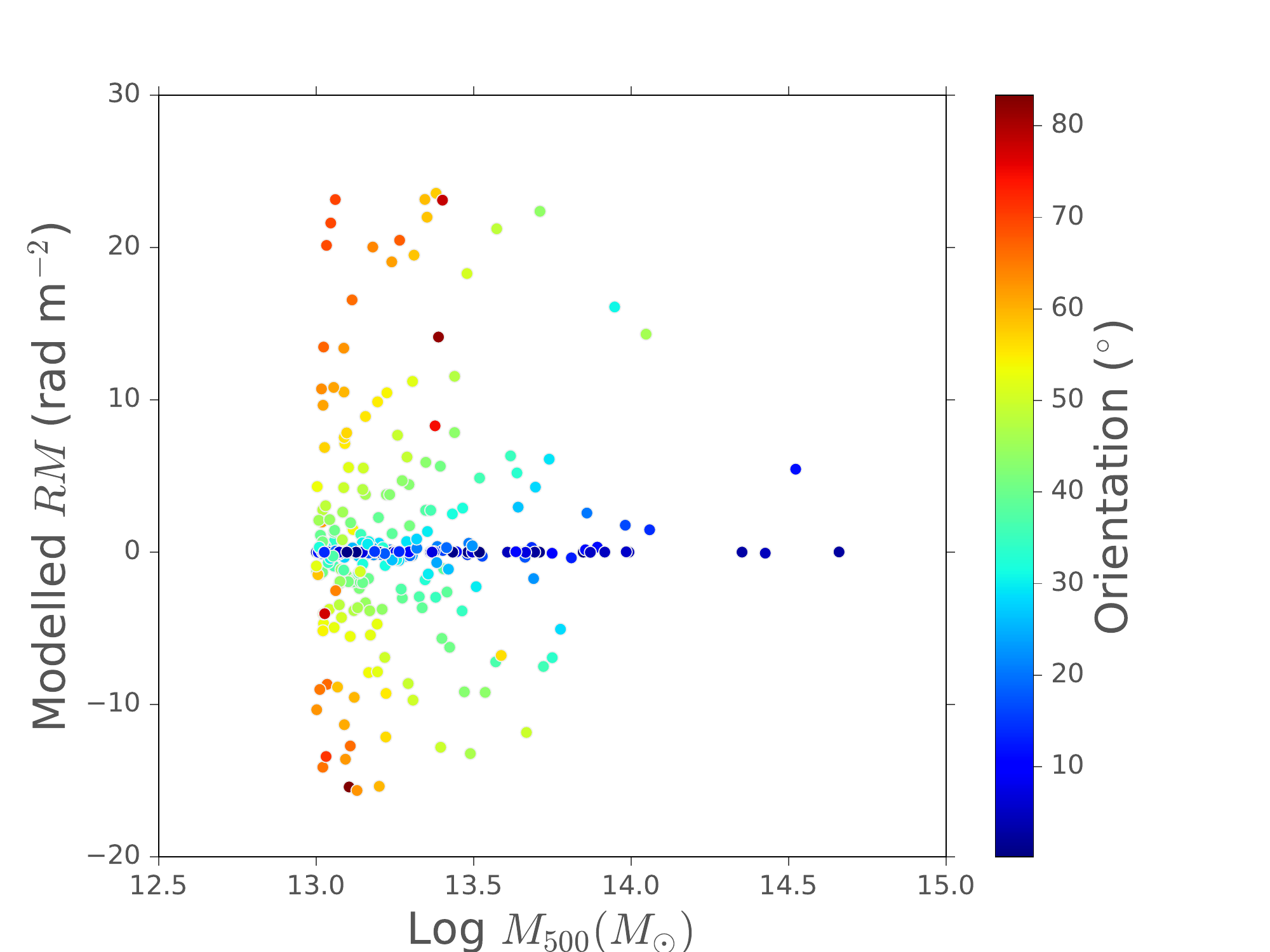}
    \caption{Group/cluster mass against the modelled $RM$ for the polarized source ILTJ112543.21+543903.2 in our sample in an environment with a modelled field reversal scale of $\sim 10^{5.5}$ pc. Models are color-coded by the orientation angle of the jet with the polarized hotspot (0$^{\circ}$ is the jet pointing directly at the observer). Note that for this source angles above $90^{\circ}$ produced $RM$s too extreme to be observed in our sample.}
    \label{fig:model_rm_m500}
\end{figure}
\begin{figure}
    \centering
    \includegraphics[scale=0.42, trim={0.5cm 0cm 0cm 1cm},clip]{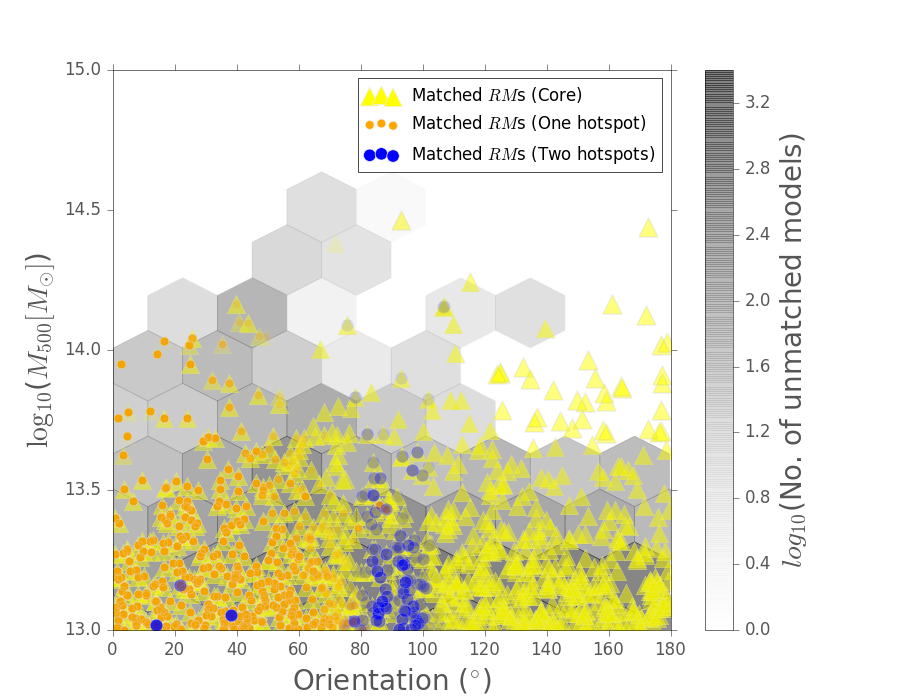}\\
    \includegraphics[scale=0.42, trim={0.5cm 0cm 0cm 1cm},clip]{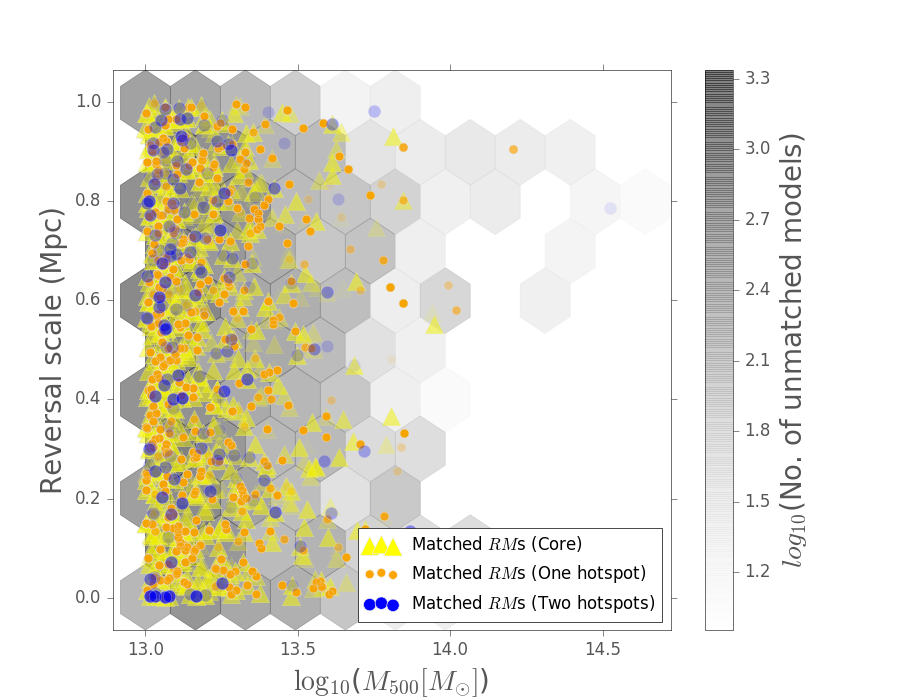}
    \caption{Jet orientation (top) and  reversal scale (bottom) against $M_{500}$ for our $RM$ models. Grey hexagons are our unmatched models (color-coded by density of counts), yellow points are matched models with a polarized core, orange points are matched models with one polarized hotspot and blue points are matched models with polarized hotspot from both lobes.}
    \vspace{-0.4cm}
    \label{fig:mass_orientation}
\end{figure}

\subsubsection{Model properties}
To give an example of the effects of the $RM$ with $M_{500}$ we display our models for ILTJ112543.21+543903.2, a polarized source in our sample with a projected size of 673 kpc and with one polarized hotspot, modelled with a field reversal scale of $10^{5.5}$ pc, color-coded by the orientation angle in each model in Figure \ref{fig:model_rm_m500}. For this type of source we see that there is a preference for the jet with a polarized hotspot to be inclined toward the observer ($\leqslant90^{\circ}$) and as a function of group/cluster mass, meaning that at high cluster/group masses, sources will tend to be depolarized unless one of the jets is orientated towards the observer (as it will experience less Faraday rotation).

In Figure \ref{fig:mass_orientation}, we plot the input physical parameters of our matched and unmatched models. In the top panel we plot the source orientation against $M_{500}$ where we see that for any given orientation, there is a higher matched model fraction at masses below $\sim 10^{13.5}M_{\odot}$, i.e sources tend to be depolarized in higher mass environments as the resulting $|RM|$ would be too high to detect in our observed range (and in reality they would cause higher $RM$ dispersion resulting in depolarization at 150 MHz). As a comparison with sources for which spatially-resolved $RM$ images have been obtained (at high frequencies), the clusters surrounding Cygnus A and Hydra A are at $M_{500}=2.8\times 10^{14} M_{\circ}$ \citep{wils02} and $M_{500}=1.7\times 10^{14} M_{\circ}$ \citep{zhan17} respectively, while a less rich group such as that surrounding 3C31 has a mass of $6.3\times 10^{13} M_{\circ}$ \citep{komo99}. This implies that LOFAR is preferentially sensitive to polarized sources in less rich clusters and poor group environments, and is consistent with our earlier remarks that with LOFAR we are sensitive to polarized radio galaxies with low dispersion in Faraday depth. Interestingly we see that, for sources with a polarized hotspot in one lobe (orange circles), there is a strong preference for angles $\leqslant 90^{\circ}$, meaning that we are seeing the approaching jet which is inclined towards the observer (as seen in Figure \ref{fig:model_rm_m500} for one source). This is due to the fact that such sources will tend to have relatively low $|RM|$ due to smaller path lengths through the line of sight and they experience less Faraday rotation, such that they are within our observed $RM$ range with LOFAR. On the other hand, core-polarized sources (yellow) are populated at all orientation angles, since the jet orientation of their assumed polarized emission at a distance of 1 pc from the centre does not significantly affect the final aggregated $RM$ at $z=0$. Double polarized sources (polarized hotspots in both lobes of an FR-II radio galaxy; blue points) seem to be almost exclusively populated at angles $\sim90^{\circ}$ (i.e on the plane of the sky), as would be expected since the hotspot from the receding jet at a larger angles from the plane of the sky can become more easily depolarized (Laing-Garrington effect; \citealt{lain88,garr88}). Hence, according to our model, LOFAR would tend to only detect both hotspots in polarization if the source is on the plane of the sky.
 
In the bottom panel, showing $M_{500}$ against the field reversal scale, we see no clear correlation and that the range of the reversal scales we sample are equally likely to produce the $RM$s we observe for a given $M_{500}$. Intuitively one expects smaller field reversal scales to produce smaller $RM$s in the range expected for LOFAR data, but many of our sources are very large in physical size (see Figure \ref{fig:hotspot-hist}), so that their hotspots in the periphery of the ICM do not require many reversals to keep the $RM$ low. The core-polarized sources, which are located at the centre of their environment, even with the largest reversal scales, produce the highest $RM$s in our models at the tail of the $RM$ distribution in Figure \ref{fig:rm_models_obs}. This means that our choice of reversal scales and $M_{500}$ are very appropriate for radio galaxies observed with LOFAR, and that more extreme environmental parameters (i.e. $M_{500}\approx10^{15} M_{\odot}$ or reversal scales $\geqslant 10^6$ pc) do not produce the distribution of $RM$s we observe, consistent with the bottom panel of Figure \ref{fig:rm_models_obs}. We note that FR-I sources in reality likely have a different distribution of $M_{500}$, more appropriate for higher mass environments, than is modelled here. While observational evidence for the values of reversal scales at the centres of clusters and groups are unavailable, we note more robust numerical magnetohydrodynamic model are needed to understand the typical field reversal scales required to detect sources at low frequencies.

In summary, we find that, in a model with only external Faraday rotation from the local environment of a source, polarized radio galaxies at low frequencies are predominantly detected in cluster/group masses $\leqslant 10^{13.5} M_{\odot}$, while polarized hotspots are preferentially seen when the jets are on the plane of the sky, otherwise only one hotspot from the approaching jet is seen \citep{lain88,garr88}. We reiterate that our model distributions are a direct product of the assumed input parameter distributions, and in depth analyses of cluster $RM$s are needed to test the robustness of our inputs.
\section{Discussion \& Conclusions}
\label{sect:conclusions}
We have analyzed 20 arcsec resolution polarization data of radio galaxies from part of the upcoming LoTSS DR2. This statistical study of the bulk properties has extended the work of \citetalias{osul18} and \cite{stua20} with the use of an optically identified sample of 382 classified radio galaxies, with 67 detected in polarization. 

We find that at 150 MHz the polarization detection fraction increases with total flux density, as expected; however, the distributions in angular size between detected and non-detected sources are statistically indistinguishable for sources $>100$ arcsec. This trend may be biased due to our selection criteria, and it is possible that the polarized detection fraction for RLAGN increases with smaller angular size due to the presence of blazars. We confirm the conclusions of \citetalias{osul18} that, in terms of resolved sources, the hotspots of FR-II radio galaxies are predominantly detected even at low frequencies. FR-II radio galaxies are not only brighter and more luminous, but they are known to reside in less rich environments than FR-Is, and so physical depolarization due to the ambient medium is less prominent, particularly since the brightest emission is in the hotspots which are far away from the densest part of the IGM/ICM, contrary to the case for FR-Is. The morphologically-classed FR-IIs in our sample generally have a higher polarized flux and fractional polarization than the FR-I sources over the range in total flux density, though with large overlap. 

The comparison of host galaxy photometry between polarized and depolarized sources further highlights the importance of morphology in polarization -- without accounting for morphology, host galaxies with higher values of WISE colour (more AGN-like on a WISE colour-colour diagram) seem to drive RLAGN with a higher detection fraction of polarization. The observed low frequency polarization is related to FR morphology rather than WISE colour, with the more powerful FR-IIs having a high detection fraction, though this case  may be unrelated for intrinsic polarization, for which past studies have found significant differences between HERGs and LERGs (which at low redshift tend to be associated with FR-IIs and FR-Is, respectively). More dense cluster environments contributing to higher internal depolarization via entrainment of the thermal material is a possible explanation for the lack of polarized FR-Is, but our data provide no direct evidence for this hypothesis.

For the sources that have polarized counterparts to the sources in our sample at 1400 MHz, we find that they are de-polarized (weaker but detectable polarization) at 150 MHz for all but one source. This is further confirmed by Figure \ref{fig:total_flux-nvss-lofar}, which shows that the distribution in 150 MHz total flux density is significantly higher for those NVSS sources that are detected with LOFAR, whereas the non-detected sources are not bright enough to produce sufficient polarized emission to be detected. The spectral index distributions imply that the sources detected by LOFAR have significantly steeper spectral indices on average, explaining the lack of polarized NVSS counterparts to the polarized sources in our sample.

Modelling of the environments toward radio galaxies and their subsequently integrated $RM$s shows that, for a range of cluster/group masses, field reversal scales and jet orientation angles, we would expect to preferentially observe polarized hotspots that are inclined towards the observer, for the case where a hotspot from one lobe is detected in polarization. For the case where hotspots in both lobes are detected, our models indicate that the jets are on the plane of the sky, consistent with the Laing-Garrington effect. Core-polarized sources are generally favoured at all orientation angles as a function $M_{500}$ and reversal scale (due to our model assumption that they have a compact polarized component at 1 pc from the centre). Our results generally imply that there is a very low chance of detecting a polarized radio galaxy at 150 MHz if it is in even a moderately rich environment ($M_{500}\geqslant10^{13.5}M_{\odot}$, depending on its orientation angle or physical size (Figure \ref{fig:hotspot-hist}), as the $RM$s would be too high to observe at low frequencies. We reiterate that our results are dependent on our input model parameters, which are based on empirical relationships and observations, but must be validated by more robust numerical modelling.

Our overall results imply that detecting polarized radio galaxies with LOFAR at 150 MHz is related to the combination of total flux density, environment and jet orientation. These results will be useful in determining the properties of polarized sources in the full LoTSS survey, which is expected to contain around ten million radio sources. 
%%%%%%%%%%%%%%%%%%%%%%%%%%%%%%%%%%%%%%%%%%%%%%%%%%
\section*{Acknowledgements}
We thank Aritra Basu, Ettore Carretti, Rainer Beck,  B\l{}a\.{z}ej Nikiel-Wroczy\'{n}ski and the anonymous referee for helpful comments in improving this paper.

VHM thanks the University of Hertfordshire for a research studentship
[ST/N504105/1]. MJH acknowledges support from the UK Science
and Technology Facilities Council [ST/R000905/1].

This paper is based (in part) on data obtained with the International LOFAR Telescope (ILT). LOFAR \citep{vanh13} is the LOw Frequency ARray designed and constructed by ASTRON. It has observing, data processing, and data storage facilities in several countries, which are owned by various parties (each with their own funding sources), and are collectively operated by the ILT foundation under a joint scientific policy. The ILT resources have benefitted from the following recent major funding sources: CNRS-INSU, Observatoire de Paris and Universit{\'e} d'Orl{\'e}ans, France; BMBF, MIWF-NRW, MPG, Germany; Science Foundation Ireland (SFI), Department of Business, Enterprise and Innovation (DBEI), Ireland; NWO, The Netherlands; The Science and Technology Facilities Council, UK; Ministry of Science and Higher Education, Poland.

This research made use of the Dutch national e-infrastructure with support of the SURF
Cooperative (e-infra 180169) and the LOFAR e-infra group. The J{\"u}lich LOFAR Long Term
Archive and the German LOFAR network are both coordinated and operated by the J{\"u}lich
Supercomputing Centre (JSC), and computing resources on the supercomputer JUWELS at JSC
were provided by the Gauss Centre for Supercomputing e.V. (grant CHTB00) through the John
von Neumann Institute for Computing (NIC).

This research made use of the University of Hertfordshire high-performance computing facility
and the LOFAR-UK computing facility located at the University of Hertfordshire and supported
by STFC [ST/P000096/1], and of the Italian LOFAR IT computing infrastructure supported and
operated by INAF, and by the Physics Department of Turin university (under an agreement with
Consorzio Interuniversitario per la Fisica Spaziale) at the C3S Supercomputing Centre, Italy.

\section*{Data Availability}
The data underlying this article will be shared on reasonable request to the corresponding author, after the LoTSS DR2 data is made available to the public in early 2021.

\bibliographystyle{mnras}
\bibliography{references}
% Don't change these lines
%%%%%%%%%%% APPENDIX
\appendix
\section{Faraday spectra}
\begin{figure*}
    \centering
    \includegraphics[scale=0.3, trim={0.2cm 0.3cm 0.1cm 0.2cm}, clip]{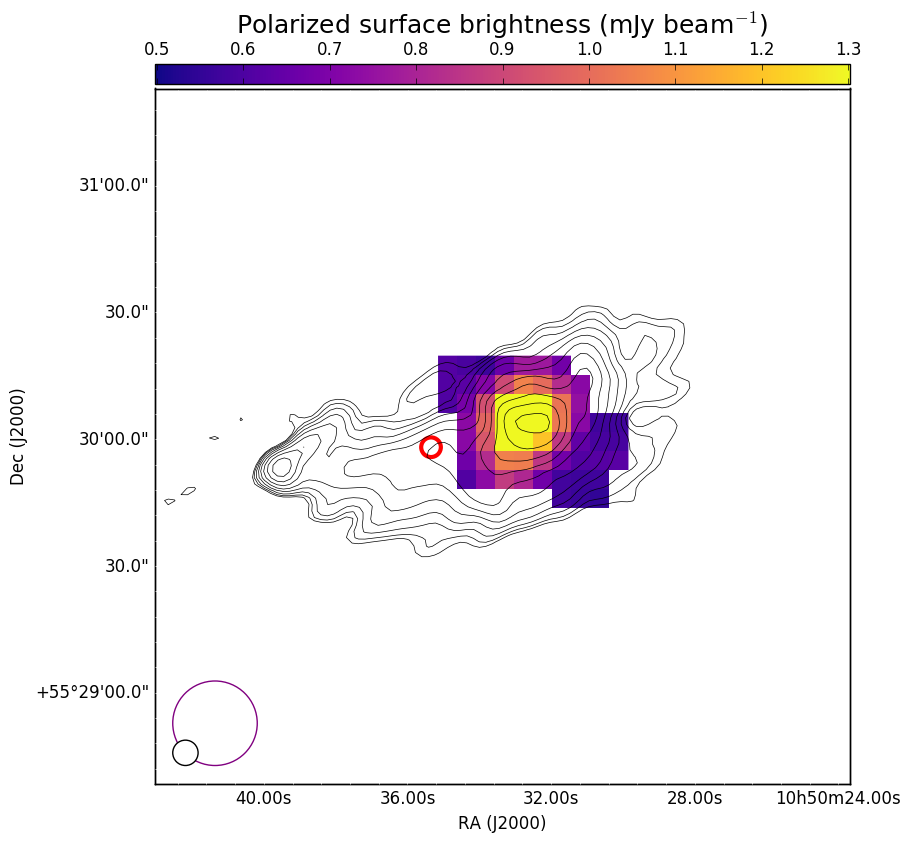}\hspace{0.5cm}
    \includegraphics[scale=0.45, trim={1cm 0.3cm 1cm 0}, clip]{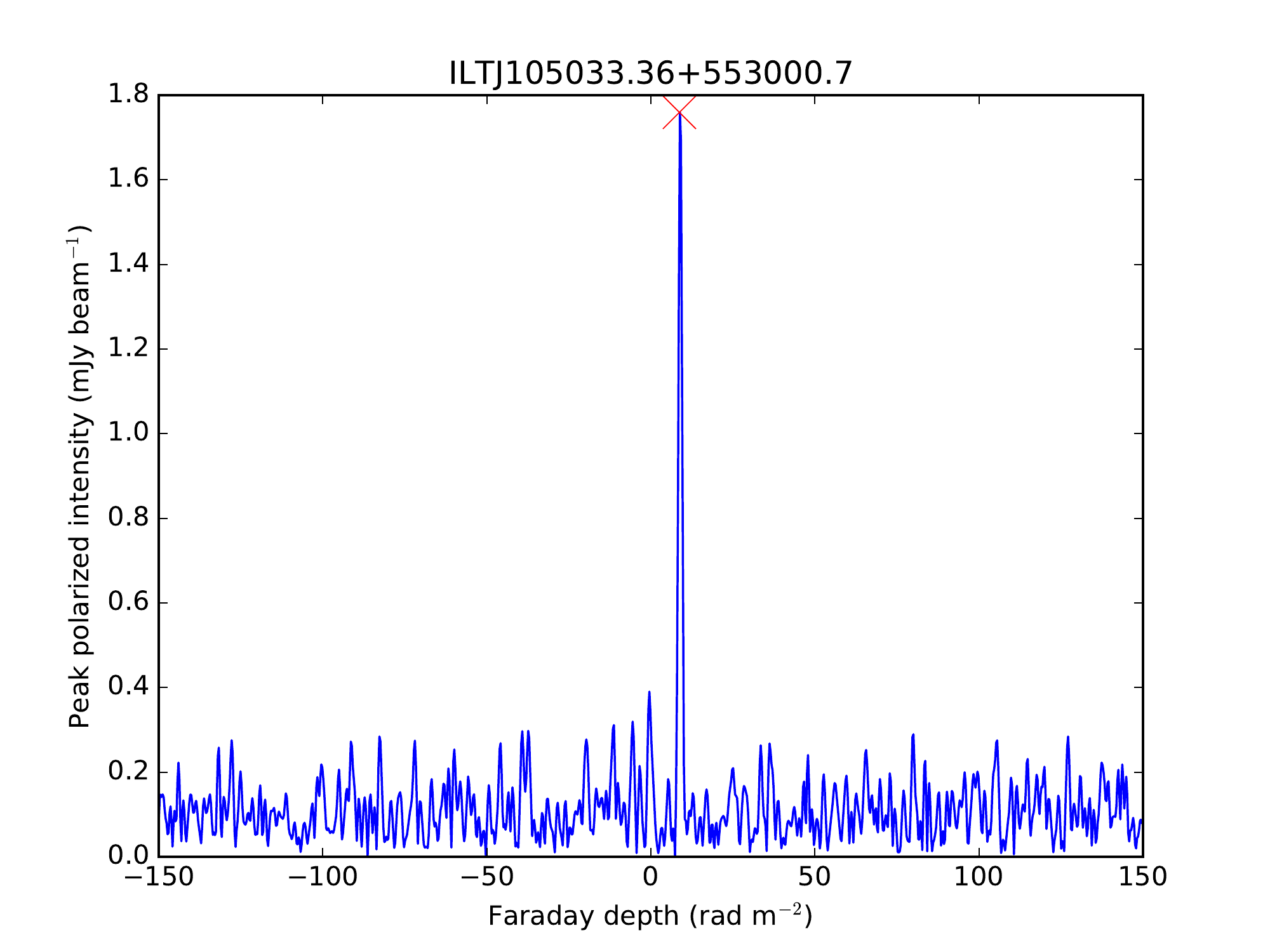}\\
    \includegraphics[scale=0.3, trim={0.2cm 0.3cm 0.1cm 0.2cm}, clip]{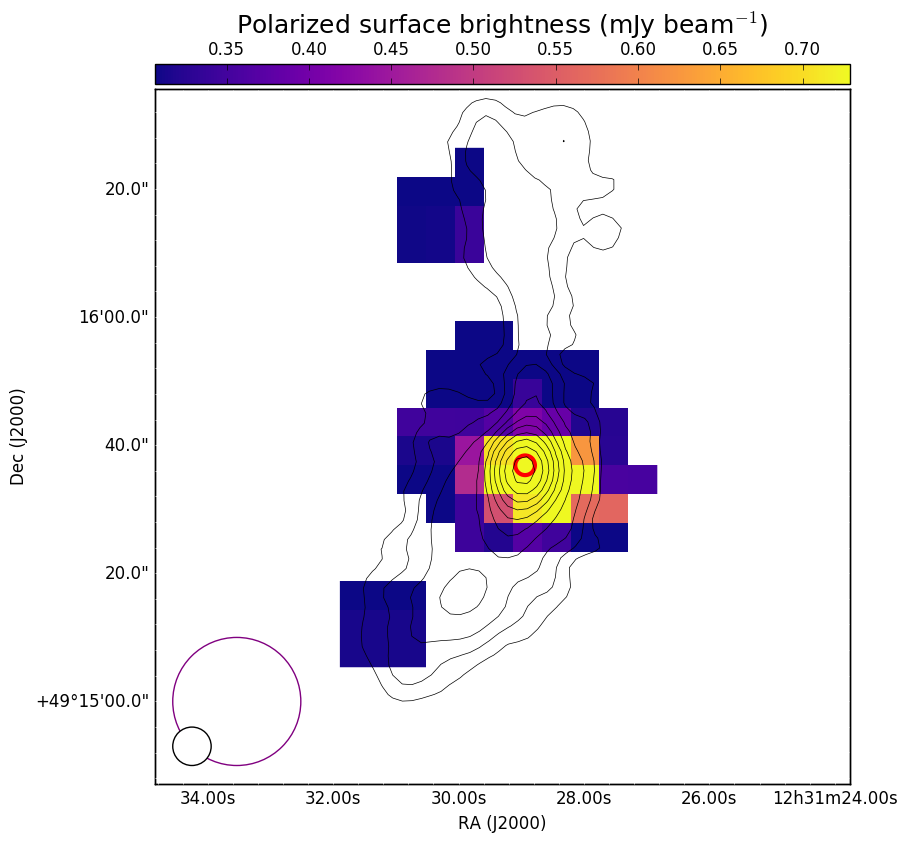}\hspace{0.5cm}
    \includegraphics[scale=0.45, trim={1cm 0.3cm 1cm 0}, clip]{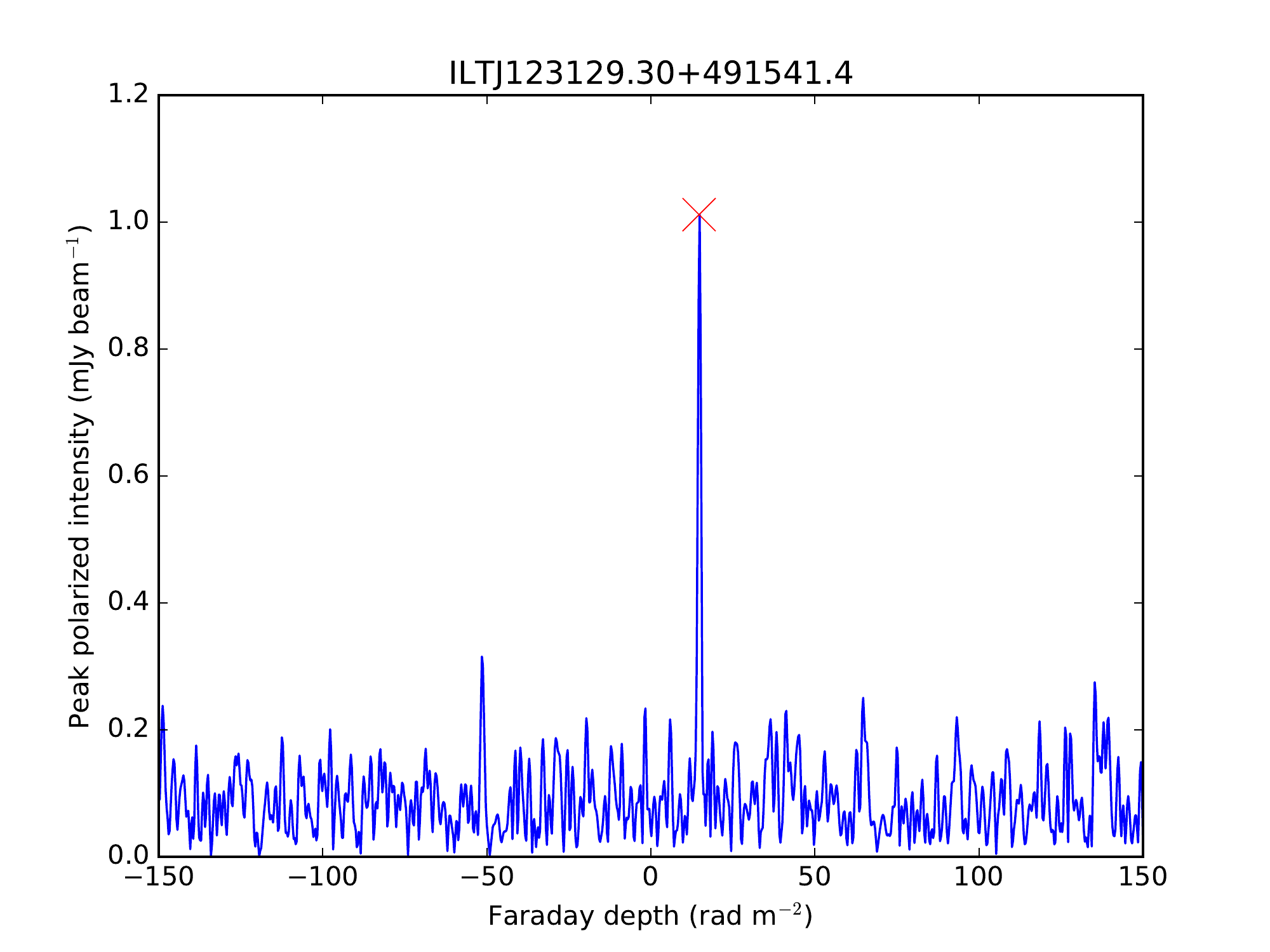}\\
    \includegraphics[scale=0.3, trim={0.2cm 0.3cm 0.1cm 0.2cm}, clip]{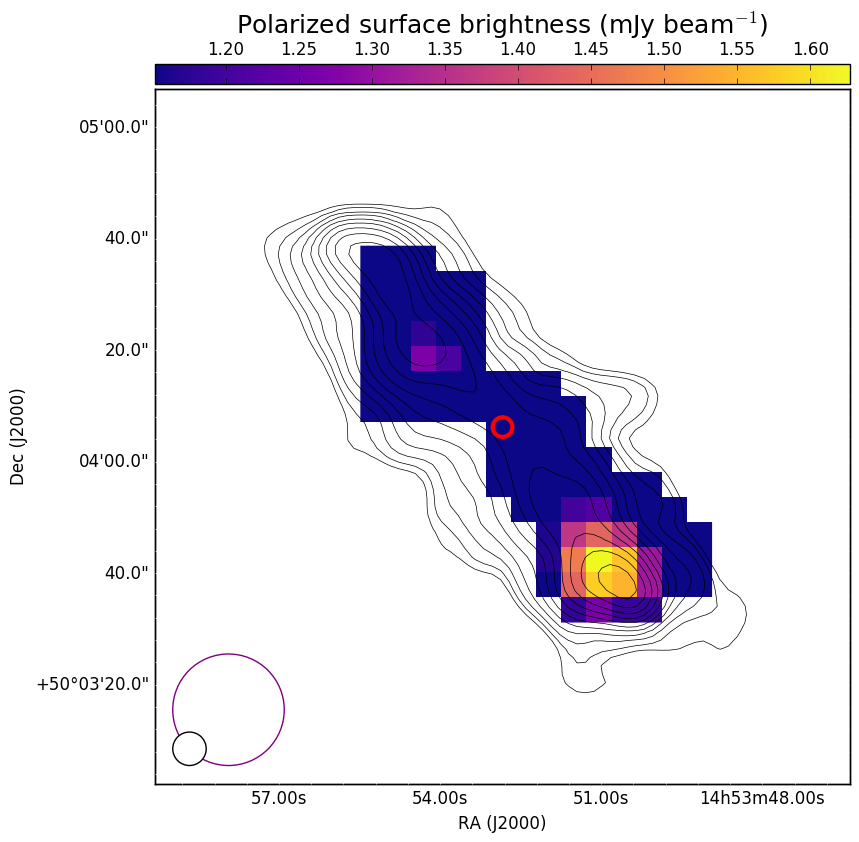}\hspace{0.5cm}
    \includegraphics[scale=0.45, trim={1cm 0.3cm 1cm 0}, clip]{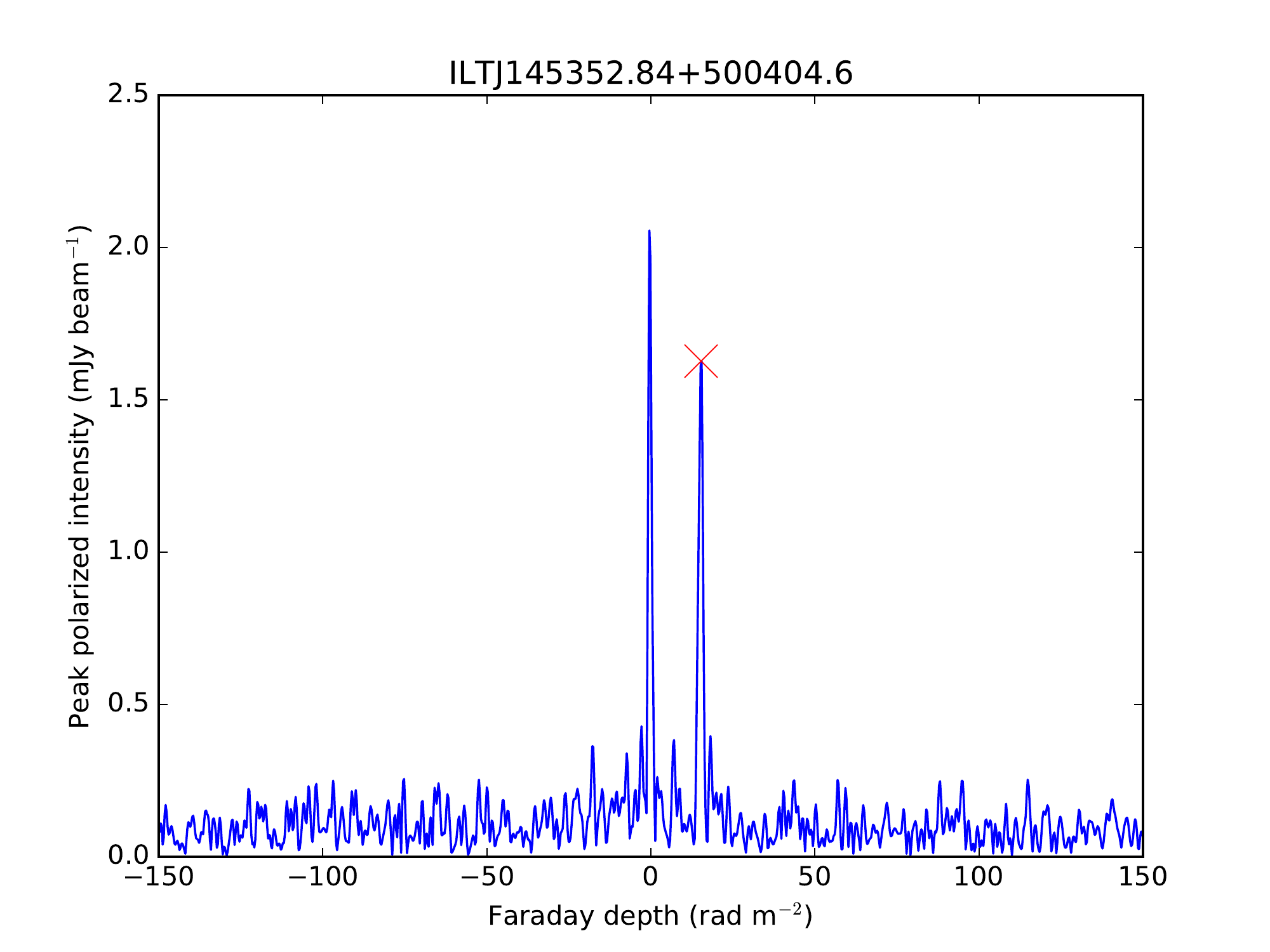}
    \caption{Polarized intensity maps (left; see Figure \ref{fig:pol_fr1} for details) and the Faraday spectrum (right) of their peak polarized intensity pixel. The red cross denotes the $RM$ of the pixel as found using $RM$ synthesis (neglecting $-3\leqslant\phi \text{ (rad m}^{-2})\leqslant1.5$ for leakage signal).}
    \label{fig:fd_spectra}
\end{figure*}
\clearpage
%\bsp	% typesetting comment
\label{lastpage}
\end{document}